\documentstyle[11pt,epsfig]{article}

\def\dag{\dagger}
\def\tr{{\rm tr}}
\def\dim{dimensions }
\def\diml{dimensional }
\def\dslash{\partial \hskip-6pt /}
\def\Dslash{D \hskip-6pt /}
\def\aslash{A \hskip -6pt /}
\def\pslash{p \hskip -6pt /}
\def\kslash{k \hskip -6pt /}

\begin{document}

\title{Aspects of Chern-Simons Theory}

\author{\normalsize{Gerald V. Dunne} \\
\normalsize{Department of Physics}\\
\normalsize{University of Connecticut}\\
\normalsize{Storrs, CT 06269}\\
\normalsize{USA}\\
\normalsize{dunne@hep.phys.uconn.edu}}

\date{}

\maketitle

\begin{abstract}
Lectures at the 1998 Les Houches Summer School: {\it Topological Aspects of
Low Dimensional Systems}. These lectures contain an introduction to various 
aspects of Chern-Simons gauge theory: (i) basics of planar field theory, (ii) 
canonical quantization of Chern-Simons theory, (iii) Chern-Simons vortices, and 
(iv) radiatively induced Chern-Simons terms.
\end{abstract}
\pagebreak

\tableofcontents
\pagebreak

\section{Introduction}
\label{intro}

Planar physics -- physics in two spatial \dim -- presents many interesting surprises, both experimentally and theoretically. The behaviour of electrons and photons [or more generally: fermions and gauge fields] differs in interesting ways from the standard behaviour we are used to in classical and quantum electrodynamics. For example, there exists a new type of gauge theory, completely different from Maxwell theory, in $2+1$ \dim. This new type of gauge theory is known as a ``Chern-Simons theory'' [the origin of this name is discussed below in Section \ref{nabcs} on nonabelian theories]. These Chern-Simons theories are interesting both for their theoretical novelty, and for their practical application for certain planar condensed matter phenomena, such as the fractional quantum Hall effect [see Steve Girvin's lectures at this School].

In these lectures I concentrate on field theoretic properties of Chern-Simons theories. I have attempted to be relatively self-contained, and accessible to someone with a basic knowledge of field theory. Actually, several important new aspects of Chern-Simons theory rely only on quantum mechanics and classical electrodynamics. Given the strong emphasis of this Summer School on condensed matter phenomena, I have chosen, wherever possible, to phrase the discussion in terms of quantum mechanical and solid state physics examples. For example, in discussing the canonical quantization of Chern-Simons theories, rather than delving deeply into conformal field theory, instead I have expressed things in terms of the Landau problem [quantum mechanical charged particles in a magnetic field] and the magnetic translation group. 

In Section \ref{basics}, I introduce the  basic kinematical and dynamical features of planar field theories, such as anyons, topologically massive gauge fields and planar fermions. I also discuss the discrete symmetries ${\cal P}$, ${\cal C}$ and ${\cal T}$, and nonabelian Chern-Simons gauge theories. Section \ref{can} is devoted to the canonical structure and canonical quantization of Chern-Simons theories. This is phrased in quantum mechanical language using a deep analogy between Chern-Simons gauge theories and quantum mechanical Landau levels [which are so important in the understanding of the fractional quantum Hall effect]. For example, this connection gives a very simple understanding of the origin of massive gauge excitations in Chern-Simons theories. In Section \ref{vortex}, I consider the self-dual vortices that arise when Chern-Simons gauge fields are coupled to scalar matter fields, with either relativistic or nonrelativistic dynamics. Such vortices are interesting examples of self-dual field theoretic structures with anyonic properties, and also arise in models for the fractional quantum Hall effect where they correspond to Laughlin's quasipartcle excitations. The final Section concerns Chern-Simons terms that are induced radiatively by quantum effects. These can appear in fermionic theories, in Maxwell-Chern-Simons models and in Chern-Simons models with spontaneous symmetry breaking. The topological nature of the induced term has interesting consequences, especially at finite temperature. 

We begin by establishing some gauge theory notation. The familiar Maxwell (or, in the nonabelian case, Yang-Mills) gauge theory is defined in terms of the fundamental gauge field (connection) $A_\mu=(A_0,\vec{A})$. Here $A_0$ is the scalar potential and $\vec{A}$ is the vector potential. The Maxwell Lagrangian
\begin{equation}
{\cal L}_{\rm M}=-\frac{1}{4} F_{\mu\nu}F^{\mu\nu}-A_\mu J^\mu
\label{maxwell}
\end{equation}
is expressed in terms of the field strength tensor (curvature) $F_{\mu\nu}=
\partial_\mu A_\nu- \partial_\nu A_\mu$, and a matter current $J^{\mu}$ that is
conserved: $\partial_\mu J^\mu =0$. This Maxwell Lagrangian is manifestly
invariant under the gauge transformation $A_\mu\to A_\mu+\partial_\mu\Lambda$;
and, correspondingly, the classical Euler-Lagrange equations of motion
\begin{equation}
\partial_\mu F^{\mu\nu}=J^\nu
\label{maxeqs}
\end{equation}
are gauge invariant. Observe that current conservation $\partial_\nu J^\nu =0$ follows from the antisymmetry of $F_{\mu\nu}$.

Now note that this Maxwell theory could easily be defined in {\it any} space-time dimension $d$ simply by taking the range of the space-time index $\mu$ on the gauge field $A_\mu$ to be $\mu=0,1,2,\dots, (d-1)$ in d-\diml space-time. The field strength tensor is still the antisymmetric tensor $F_{\mu\nu}=\partial_\mu A_\nu- \partial_\nu A_\mu$, and the Maxwell  Lagrangian (\ref{maxwell}) and the equations of motion (\ref{maxeqs}) do not change their form. The only real difference is that the number of independent fields contained in the field strength tensor $F_{\mu\nu}$ is different in different dimensions. [Since $F_{\mu\nu}$ can be regarded as a $d\times d$ antisymmetric matrix, the number of fields is equal to $\frac{1}{2}d(d-1)$.] So at this level, planar (i.e. $2+1$ \diml) Maxwell theory is quite similar to the familiar $3+1$ \diml Maxwell theory. The main difference is simply that the magnetic field is a (pseudo-) scalar $B=\epsilon^{ij}\partial_i A_j$ in $2+1$ \dim, rather than a (pseudo-) vector $\vec{B}=\vec{\nabla}\times\vec{A}$ in $3+1$ \dim. This is just because in $2+1$ \dim the vector potential $\vec{A}$ is a two-\diml vector, and the curl in two dimensions produces a scalar. On the other hand, the electric field $\vec{E}=-\vec{\nabla}A_0-\dot{\vec{A}}$ is a two \diml vector. So the antisymmetric $3\times 3$ field strength tensor has three nonzero field components: two for the electric field $\vec{E}$ and one for the magnetic field $B$.

The real novelty of $2+1$ \dim is that instead of considering this `reduced' form of Maxwell theory, we can also define a completely different type of gauge theory: a Chern-Simons theory. It satisfies our usual criteria for a sensible gauge theory -- it is Lorentz invariant, gauge invariant, and local. The Chern-Simons Lagrangian is
\begin{equation}
{\cal L}_{\rm CS}=\frac{\kappa}{2}\epsilon^{\mu\nu\rho}A_\mu \partial_\nu
A_\rho -A_\mu J^\mu
\label{cs}
\end{equation}
There are several comments to make about this Chern-Simons Lagrangian. First, it does
not {\it look} gauge invariant, because it involves the gauge field $A_\mu$
itself, rather than just the (manifestly gauge invariant) field strength
$F_{\mu\nu}$. Nevertheless, under a gauge transformation, the Chern-Simons Lagrangian
changes by a total space-time derivative
\begin{equation}
\delta {\cal L}_{\rm CS}=\frac{\kappa}{2}\partial_\mu\left(\lambda\,
\epsilon^{\mu\nu\rho} \partial_\nu A_\rho\right).
\label{tot}
\end{equation}
Therefore, if we can neglect boundary terms (later we shall encounter important
examples where this is {\it not} true) then the corresponding Chern-Simons {\it action}, $S_{\rm CS}=\int d^3 x\, {\cal L}_{\rm CS}$, is gauge invariant. This is reflected in the fact that the classical Euler-Lagrange equations
\begin{equation}
\frac{\kappa}{2} \epsilon^{\mu\nu\rho}F_{\nu\rho}=J^\mu; \qquad {\rm
or~equivalently:} \qquad F_{\mu\nu}=\frac{1}{\kappa}
\epsilon_{\mu\nu\rho}J^\rho
\label{cseqs}
\end{equation}
are clearly gauge invariant. Note that the Bianchi identity,
$\epsilon^{\mu\nu\rho}\partial_\mu F_{\nu\rho}=0$, is compatible with current
conservation : $\partial_\mu J^\mu=0$.

A second important feature of the Chern-Simons Lagrangian (\ref{cs}) is that it is {\it first-order} in space-time derivatives. This makes the canonical structure of these theories significantly different from that of Maxwell theory. A related property is that the Chern-Simons Lagrangian is particular to $2+1$ \dim, in the sense that we cannot write down such a term in $3+1$ \dim -- the indices simply do not match up. Actually, it is possible to write down a ``Chern-Simons theory'' in any odd space-time dimension, but it is only in $2+1$ \dim that the Lagrangian is quadratic in the gauge field. For example, the Chern-Simons Lagrangian in five-\diml space-time is 
${\cal L}=\epsilon^{\mu\nu\rho\sigma\tau}A_\mu \partial_\nu A_\rho \partial_\sigma A_\tau$.

At first sight, pure Chern-Simons theory looks rather boring, and possibly trivial, because the source-free classical equations of motion (\ref{cseqs}) reduce to $F_{\mu\nu}=0$, the solutions of which are just pure gauges or ``flat connections''. This is in contrast to pure Maxwell theory, where even the source--free theory has interesting, and physically important, solutions : plane-waves. Nevertheless, Chern-Simons theory can be made interesting and nontrivial in a number of ways:

(i) coupling to dynamical matter fields (charged scalars or fermions)

(ii) coupling to a Maxwell term

(iii) taking the space-time to have nontrivial topology

(iv) nonabelian gauge fields

(v) gravity

I do not discuss $2+1$ \diml gravity in these lectures as it is far from the topic of this School, but I stress that it is a rich subject that has taught us a great deal about both classical and quantum gravity \cite{carlip}. 

\section{Basics of Planar Field Theory}
\label{basics}

\subsection{Chern-Simons Coupled to Matter Fields - ``Anyons''}

\begin{figure}[htb]
\vspace{-1in}
\centering{\epsfig{file=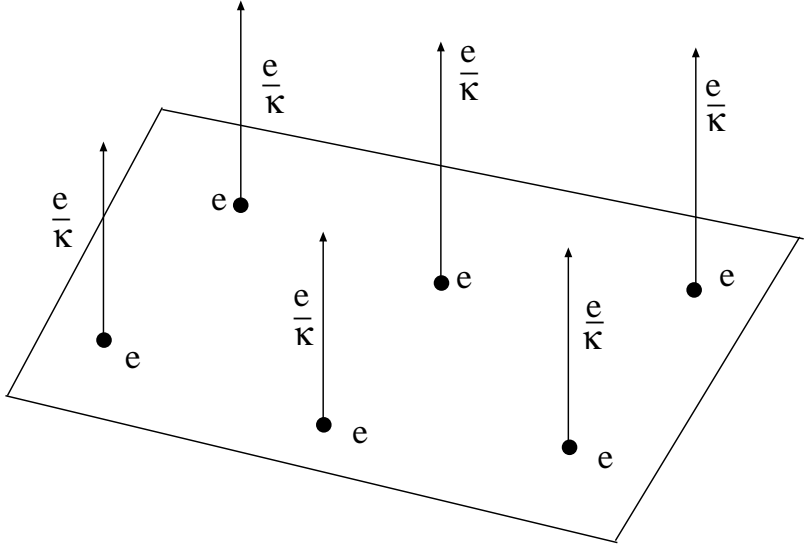,width=5in}}
\vspace{-2.25in}
\caption{A collection of point anyons with charge $e$, and with magnetic flux lines of strength $\frac{e}{\kappa}$ tied to the charges. The charge and flux are tied together throughout the motion of the particles as a result of the Chern-Simons equations (\protect{\ref{cscomps}}).}
\label{fluxes}
\end{figure}

In order to understand the significance of coupling a matter current
$J^\mu=(\rho, \vec{J})$ to a Chern-Simons gauge field, consider the Chern-Simons equations
(\ref{cseqs}) in terms of components :
\begin{eqnarray}
\rho &=& \kappa B \nonumber\\
J^i&=&\kappa \epsilon^{ij}E_j
\label{cscomps}
\end{eqnarray}
The first of these equations tells us that the charge density is locally
proportional to the magnetic field -- thus the effect of a Chern-Simons field is to tie
magnetic flux to electric charge. Wherever there is one, there is the other, and they are locally proportional, with the proportionality constant given by the Chern-Simons coupling parameter $\kappa$. This is illustrated in Figure \ref{fluxes} for a collection of point charges. The second equation in (\ref{cscomps}) ensures that this charge-flux relation is preserved under time evolution because the time derivative of the first equation
\begin{equation}
\dot{\rho}=\kappa \dot{B}=\kappa\epsilon^{ij}\partial_i \dot{A}_j
\label{timeevol}
\end{equation}
together with current conservation, $\dot{\rho}+\partial_i J^i=0$, implies that \begin{equation}
J^i=-\kappa
\epsilon^{ij}\dot{A}_j+\epsilon^{ij}\partial_j \chi
\end{equation}
which is just the second equation in (\ref{cscomps}), with the transverse piece $\chi$ identified with $\kappa A_0$.

Thus, the Chern-Simons coupling at this level is pure constraint -- we can regard the
matter fields as having their own dynamics, and the effect of the Chern-Simons coupling
is to attach magnetic flux to the matter charge density in such a way that it
follows the matter charge density wherever it goes. Clearly, this applies
either to relativistic or nonrelativistic dynamics for the matter fields. (A
word of caution here - although the Chern-Simons term is Lorentz invariant, we can
regard this simply as a convenient shorthand for expressing the constraint
equations (\ref{cscomps}), in much the same way as we can always express a
continuity equation $\dot{\rho}+\partial_i J^i=0$ in a relativistic-looking way
as $\partial_\mu J^\mu=0$. Thus, there is no problem mixing nonrelativistic
dynamics for the matter fields with a `relativistic-looking' Chern-Simons term. The actual dynamics is always inherited from the matter fields.)

This tying of flux to charge provides an explicit realization of ``anyons''
\cite{wilczek,lerda}. (For more details on anyons, see Jan Myrheim's lectures at this school). Consider, for example, nonrelativistic point charged
particles moving in the plane, with magnetic flux lines attached to them. The
charge density
\begin{equation}
\rho(\vec{x},t)=e\,\sum_{a=1}^N\, \delta(\vec{x}-\vec{x}_a(t))
\end{equation}
describes $N$ such particles, with the $a^{th}$ particle following the
trajectory $\vec{x}_a(t)$. The corresponding current density is
$\vec{j}(\vec{x},t)=e\sum_{a=1}^N\,
\dot{\vec{x}}_a(t)\delta(\vec{x}-\vec{x}_a(t))$. The Chern-Simons equations (\ref{cscomps}) attach magnetic flux [see Figure \ref{fluxes}]
\begin{equation}
B(\vec{x},t)=\frac{1}{\kappa}\, e\,\sum_{a=1}^N\,\delta(\vec{x}-\vec{x}_a(t))
\end{equation}
which follows each point particle throughout its motion.

If each particle has mass $m$, the net action is
\begin{equation}
S=\frac{m}{2}\sum_{a=1}^N\int dt\,\vec{v}_a^2+\frac{\kappa}{2}\int d^3x
\epsilon^{\mu\nu\rho} A_\mu \partial_\nu A_\rho -\int d^3x\, A_\mu J^\mu
\label{anylag}
\end{equation}
The Chern-Simons equations of motion (\ref{cseqs}) determine the gauge field $A_\mu
(\vec{x},t)$ in terms of the particle current. The gauge freedom may be fixed
in a Hamiltonian formulation by taking $A_0=0$ and imposing
$\vec{\nabla}\cdot\vec{A}=0$. Then
\begin{equation}
A^i(\vec{x},t)=\frac{1}{2\pi \kappa}\int d^2y\,\epsilon^{ij}{(x^j-y^j)\over
|\vec{x}-\vec{y}|^2}\, \rho(\vec{y},t)=\frac{e}{2\pi\kappa} \sum_{a=1}^N
\epsilon^{ij}{(x^j-x^j_a(t))\over |\vec{x}-\vec{x}_a(t)|^2}
\label{g}
\end{equation}
where we have used the two \diml Green's function
\begin{equation}
\nabla^2\left(\frac{1}{2\pi}\log|\vec{x}-\vec{y}|\right)=\delta^{(2)}
(\vec{x}-\vec{y})
\label{green}
\end{equation}

As an aside, note that using the identity $\partial_i
arg(\vec{x})=-\epsilon_{ij}x^j/|\vec{x}|^2$, where the argument function is
$arg(\vec{x})=\arctan(\frac{y}{x})$, we can express this vector potential
(\ref{g}) as
\begin{equation}
A_i(\vec{x})=\frac{e}{2\pi \kappa}\sum_{a=1}^N\, \partial_i \,
arg(\vec{x}-\vec{x}_a)
\label{sing}
\end{equation}
Naively, this looks like a pure gauge vector potential, which could presumably
therefore be removed by a gauge transformation. However, under such a gauge transformation the corresponding nonrelativistic field $\psi(\vec{x})$ would acquire a phase factor
\begin{equation}
\psi(\vec{x})\to \tilde{\psi}(\vec{x})=\exp\left( -i\frac{e^2}{2\pi\kappa}
\sum_{a=1}^N arg(\vec{x}-\vec{x}_a)\right)\, \psi(\vec{x})
\end{equation}
which makes the field non-single-valued for general values of the Chern-Simons coupling
parameter $\kappa$. This lack of single-valuedness is the nontrivial remnant
of the Chern-Simons gauge field coupling. Thus, even though it looks as though the gauge
field has been gauged away, leaving a `free' system, the complicated
statistical interaction is hidden in the nontrivial boundary conditions for the
non-single-valued field $\tilde{\psi}$.

Returning to the point-anyon action (\ref{anylag}), the Hamiltonian for this
system is
\begin{equation}
H=\frac{m}{2}\sum_{a=1}^N\,\vec{v}_a^2 =
\frac{1}{2m}\sum_{a=1}^N\,[\vec{p}_a-e\vec{A}(\vec{x}_a)]^2
\label{anyham}
\end{equation}
where
\begin{equation}
A^i(\vec{x}_a)= \frac{e}{2\pi\kappa} \sum_{b\neq a}^N
\epsilon^{ij}{(x^j_a-x^j_b)\over |\vec{x}_a-\vec{x}_b|^2}
\label{anygauge}
\end{equation}
The corresponding magnetic field is
\begin{equation}
B(\vec{x}_a)= \frac{e}{\kappa} \sum_{b\neq a}^N \delta(\vec{x}_a-\vec{x}_b)
\label{anymag}
\end{equation}
so that each particle sees each of the $N-1$ others as a point vortex of flux
$\Phi=\frac{e}{\kappa}$, as expected. Note that the gauge field in
(\ref{anygauge}) excludes the self-interaction $a=b$ term, with suitable
regularization \cite{lerda,jackiwpi}.

\begin{figure}[htb]
\vspace{-2cm}
\centering{\epsfig{file=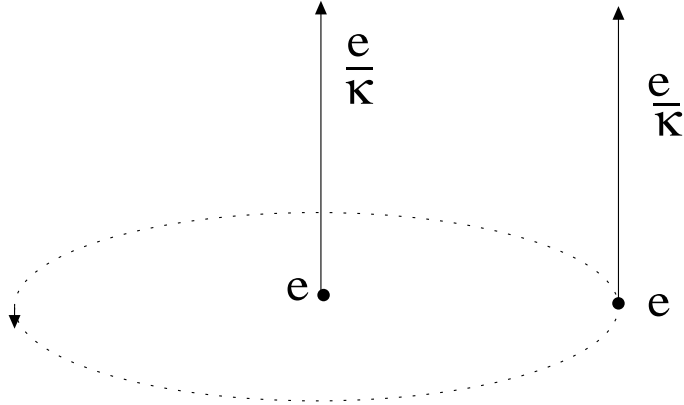,width=4in}}
\vspace{-5cm}
\caption{Aharonov-Bohm interaction between the charge $e$ of an anyon and the flux $\frac{e}{\kappa}$ of another anyon under double-interchange. Under such an adiabatic transport, the multi-anyon wavefunction acquires an Aharonov-Bohm phase (\protect{\ref{abp}}).}
\label{bohm}
\end{figure}

An important consequence of this charge-flux coupling is that it leads to new
Aharonov-Bohm-type interactions. For example, when one such particle moves
adiabatically around another [as shown in Figure \ref{bohm}], in addition to whatever electrical interactions
mediate between them, at the quantum level the nonrelativistic wavefunction
acquires an Aharonov-Bohm phase
\begin{equation}
\exp\left(ie\oint_C \vec{A}\cdot d\vec{x}\right)=\exp\left({ie^2\over
\kappa}\right)
\label{abp}
\end{equation}
If this adiabatic excursion is interpreted as a double interchange of two such
identical particles (each with flux attached), then this gives an ``anyonic''
exchange phase
\begin{equation}
2\pi \Delta\theta={e^2\over 2\kappa}
\label{spinstat}
\end{equation}
which can be tuned to any value by specifying the value of the Chern-Simons coupling
coefficient $\kappa$. This is the origin of anyonic statistics in
point-particle language.

This is a first-quantized description of anyons as point particles. However,
N-anyon quantum mechanics can be treated, in the usual manner of
nonrelativistic many-body quantum mechanics, as the N-particle sector of a
nonrelativistic quantum field theory \cite{lerda,jackiwpi}. In this case, a Chern-Simons field is
required to ensure that the appropriate magnetic flux is always attached to the
(smeared-out) charged particle fields $\varphi(\vec{x},t)$. This, together with
the above-mentioned statistics transmutation, explains the appearance of Chern-Simons
fields in the ``composite boson'' or ``composite fermion'' models for the fractional quantum Hall effect, which involve quasiparticles that have magnetic fluxes attached to charged particles \cite{zhk,jain,zhang,shankar}. In such field theories there is a generalized spin-statistics relation similar to (\ref{spinstat}) -- see later in Eq. (\ref{sp}). By choosing $\kappa$ appropriately, the anyonic exchange phase (\ref{spinstat}) can be chosen so that the particles behave either as fermions or as bosons. An explicit example of this statistical transmutation will be used in Section \ref{zhkm} on the Zhang-Hansson-Kivelson model \cite{zhk} for the fractional quantum Hall effect. 

\subsection{Maxwell-Chern-Simons : topologically massive gauge theory}
\label{tmgt}

Since both the Maxwell and Chern-Simons Lagrangians produce viable gauge theories in $2+1$ \dim, it is natural to consider coupling them together. The result is a surprising new form of gauge field mass generation. Consider the Lagrangian
\begin{equation}
{\cal L}_{\rm MCS}=-\frac{1}{4e^2} F_{\mu\nu}F^{\mu\nu}+
\frac{\kappa}{2}\epsilon^{\mu\nu\rho}A_\mu \partial_\nu A_\rho
\label{mcs}
\end{equation}
The resulting classical field equations are
\begin{equation}
\partial_\mu F^{\mu\nu}+\frac{\kappa e^2}{2}
\epsilon^{\nu\alpha\beta}F_{\alpha\beta} =0
\label{mcseqs}
\end{equation}
which describe the propagation of a single (transverse) degree of freedom with
mass (note that $e^2$ has dimensions of mass in $2+1$ \dim, while $\kappa$ is
dimensionless):
\begin{equation}
m_{\rm MCS}=\kappa e^2
\label{mcsmass}
\end{equation}
This has resulted in the terminology ``topologically massive gauge theory''
\cite{deser1}, where the term ``topological'' is motivated by the nonabelian
Chern-Simons theory (see Section \ref{nabcs}).

The most direct way to see the origin of this mass is to re-write the equation of motion (\ref{mcseqs}) in terms of the pseudovector ``dual'' field
$\tilde{F}^\mu\equiv\frac{1}{2}\epsilon^{\mu\nu\rho}F_{\nu\rho}$:
\begin{equation}
\left[\partial_\mu\partial^\mu+(\kappa e^2)^2\right] \tilde{F}^\nu =0
\label{dualeqs}
\end{equation}
Note that this dual field $\tilde{F}^\mu$ is manifestly gauge invariant, and it also satisfies $\partial_\mu \tilde{F}^\mu=0$.
The MCS mass can also be identified from the corresponding representation
theory of the Poincar\'e algebra in $2+1$-\dim, which also yields the spin of
the massive excitation as
\begin{equation}
s_{\rm MCS}=\frac{\kappa}{|\kappa|}=\pm 1
\label{mcsspin}
\end{equation}
We shall discuss these mass and spin properties further in Section \ref{pa}.

\vskip 1cm
{\bf Exercise 2.2.1 :} Another useful way to understand the origin of the massive gauge excitation is to compute the gauge field propagator in (for example) a covariant gauge with gauge fixing term ${\cal L}_{\rm gf} = -{1\over 2 \xi e^2}\left(\partial_\mu A^\mu\right)^2$. By inverting the quadratic part of the momentum space lagrangian, show that the gauge field propagator is
\begin{equation}
\Delta_{\mu\nu}=e^2\left({p^2g_{\mu\nu}-p_\mu p_\nu
-i\kappa e^2\epsilon_{\mu\nu\rho}p^\rho \over p^2(p^2-\kappa^2 e^4)}+\xi {p_\mu
p_\nu\over
(p^2)^2}\right)
\label{mcsprop}
\end{equation}
This clearly identifies the gauge field mass via the pole at $p^2=(\kappa
e^2)^2$.
\vskip 1cm

I emphasize that this gauge field mass (\ref{mcsmass}) is completely independent of the
standard Higgs mechanism for generating masses for gauge fields through a
nonzero expectation value of a Higgs field. Indeed, we can {\it also} consider
the Higgs mechanism in a Maxwell-Chern-Simons theory, in which case we find {\it two}
independent gauge field masses. For example, couple this Maxwell-Chern-Simons
theory to a complex scalar field $\phi$ with a symmetry breaking potential
$V(|\phi |)$
\begin{equation}
{\cal L}_{\rm MCSH}=-{1\over 4e^2} F_{\mu\nu}F^{\mu\nu}+{\kappa\over
2} \epsilon^{\mu\nu\rho}A_\mu\partial_\nu A_\rho +
\left(D_\mu\phi\right)^*  D^\mu \phi -V(|\phi |)
\label{mcsh}
\end{equation}
where $V(|\phi |)$ has some nontrivial minimum with $<\phi >=v$. In this
broken vacuum there is an additional quadratic term $v^2A_\mu A^\mu$ in the
gauge field Lagrangian which leads to the momentum space propagator (with a
covariant gauge fixing term) \cite{rao}
\begin{eqnarray}
\Delta_{\mu\nu}&=&{e^2(p^2-m_H^2)\over (p^2-m_+^2)(p^2-m_-^2)}\left[
g_{\mu\nu}-{p_\mu p_\nu \over (p^2-\xi m_H^2)}-
i{\kappa e^2 \epsilon_{\mu\nu\rho}p^\rho\over (p^2-m_H^2)}\right]\nonumber\\
&&\qquad\qquad +e^2\xi {p_\mu p_\nu (p^2-\kappa^2 e^4-m_H^2)\over
(p^2-m_+^2)(p^2-m_-^2)(p^2-\xi m_H^2)}
\label{mcshprop}
\end{eqnarray}
where $m_{\rm H}^2=2e^2 v^2$ is the usual Higgs mass scale (squared) and
the other masses are
\begin{equation}
m_\pm^2=m_H^2+{(\kappa e^2)^2\over 2}\pm{\kappa e^2\over 2}
\sqrt{\kappa^2 e^4 +4m_H^2}
\label{pm}
\end{equation}
or
\begin{equation}
m_\pm={m_{\rm MCS}\over 2}\left( \sqrt{1+{4m_H^2\over
m_{\rm MCS}^2}}\pm1\right)
\label{masses}
\end{equation}
{}From the propagator (\ref{mcshprop}) we identify {\it two} physical mass
poles at $p^2=m_\pm^2$.

The counting of degrees of freedom goes as follows. In the unbroken
vacuum, the complex scalar field has two real massive degrees of freedom
and the gauge field has one massive excitation (with mass coming from
the Chern-Simons  term). In the broken vacuum, one component of the scalar
field (the ``Goldstone boson'') combines with the longitudinal part of the
gauge field to produce a new massive gauge degree of freedom. Thus, in the
broken vacuum there is {\it one} real massive scalar degree of freedom (the
``Higgs boson'') and {\it two} massive gauge degrees of freedom.

The Higgs mechanism also occurs, albeit somewhat differently, if the gauge field has just a Chern-Simons term, and no Maxwell term \cite{zyang}. The Maxwell term can be decoupled from the Maxwell-Chern-Simons-Higgs Lagrangian (\ref{mcsh}) by taking the limit
\begin{equation}
e^2\to\infty \hskip 1in \kappa={\rm  fixed}
\label{hagenlimit}
\end{equation}
which leads to the Chern-Simons-Higgs Lagrangian
\begin{equation}
{\cal L}_{\rm CSH}={\kappa\over 2}
\epsilon^{\mu\nu\rho}A_\mu\partial_\nu A_\rho +
\left(D_\mu\phi\right)^\dagger  D^\mu \phi -V(|\phi |)
\label{csh}
\end{equation}

\vskip 1cm
{\bf Exercise 2.2.2 :} Show that the propagator (\ref{mcshprop}) reduces in the limit (\ref{hagenlimit}) to
\begin{equation}
\Delta_{\mu\nu}={1\over
p^2-(\frac{2v^2}{\kappa})^2}\left[\frac{2v^2}{\kappa}g_{\mu\nu}
-\frac{1}{2v^2}p_\mu p_\nu + \frac{i}{\kappa}\epsilon_{\mu\nu\rho}p^\rho\right]
\label{cshprop}
\end{equation}
which has a {\it single} massive pole at $p^2=(\frac{2v^2}{\kappa})^2$.
\vskip 1cm

The counting of degrees of freedom is different in this Chern-Simons-
Higgs model. In the unbroken vacuum the gauge field is nonpropagating, and
so there are just the two real scalar modes of the scalar field $\phi$. In
the broken vacuum, one component of the scalar field (the ``Goldstone
boson'') combines with the longitudinal part of the gauge field to produce a
massive gauge degree of freedom. Thus, in the broken vacuum there is {\it one}
real massive scalar degree of freedom (the ``Higgs boson'') and {\it one}
massive gauge degree of freedom. This may also be deduced from the mass
formulae (\ref{masses}) for the Maxwell-Chern-Simons-Higgs model, which in the limit (\ref{hagenlimit}) tend to
\begin{equation}
m_+\to\infty \qquad\qquad\qquad  m_-\to {2v^2\over\kappa}
\label{masslimit}
\end{equation}
so that one mass $m_+$ decouples to infinity, while the other mass $m_-$ agrees with the mass pole found in (\ref{cshprop}). In Section \ref{csqm} we shall see that there is a simple way to understand these various gauge masses in terms of the characteristic frequencies of the familiar quantum mechanical Landau problem.

\subsection{Fermions in $2+1$-\dim}
\label{ferms}

Fermion fields also have some new and interesting features when restricted to
the plane. The most obvious difference is that the irreducible set of Dirac
matrices consists of $2\times 2$ matrices, rather than $4\times 4$.
Correspondingly, the irreducible fermion fields are 2-component spinors. The
Dirac equation is
\begin{equation}
\left(i\gamma^\mu\partial_\mu-e\gamma^\mu A_\mu-m\right)\psi=0, \qquad {\rm or}
\qquad i{\partial\over \partial t}\psi=\left(-i\vec{\alpha}\cdot\vec{\nabla}
+m\beta\right)\psi
\label{diraceq}
\end{equation}
where $\vec{\alpha}=\gamma^0 \vec{\gamma}$ and $\beta =\gamma^0$.
The Dirac gamma matrices satisfy the anticommutation relations:
$\{\gamma^\mu,\gamma^\nu\}=2g^{\mu\nu}$, where we use the Minkowski metric
$g^{\mu\nu}={\rm diag}(1,-1,-1)$. One natural representation is a `Dirac'
representation:
\begin{eqnarray}
\gamma^0&=&\sigma^3=\left(\matrix{1&0\cr 0&-1}\right)\nonumber\\
\gamma^1&=&i\sigma^1=\left(\matrix{0&i\cr i&0}\right)\nonumber\\
\gamma^2&=&i\sigma^2=\left(\matrix{0&1\cr -1&0}\right)
\label{diracrep}
\end{eqnarray}
while a `Majorana' representation (in which $\beta$ is imaginary while the
$\vec{\alpha}$ are real) is:
\begin{eqnarray}
\gamma^0&=&\sigma^2=\left(\matrix{0&-i\cr i&0}\right)\nonumber\\
\gamma^1&=&i\sigma^3=\left(\matrix{i&0\cr 0&-i}\right)\nonumber\\
\gamma^2&=&i\sigma^1=\left(\matrix{0&i\cr i&0}\right)
\label{majrep}
\end{eqnarray}
These $2\times 2$ Dirac matrices satisfy the identities:
\begin{equation}
\gamma^\mu \gamma^\nu = g^{\mu\nu}{\bf 1}-i\epsilon^{\mu\nu\rho}\gamma_\rho
\end{equation}
\begin{equation}
\tr(\gamma^\mu\gamma^\nu\gamma^\rho)=-2i\,\epsilon^{\mu\nu\rho}
\label{trace}
\end{equation}
Note that in familiar $3+1$ \diml theories, the trace of an odd number of gamma
matrices vanishes. In $2+1$ \dim, the trace of three gamma matrices produces the totally antisymmetric $\epsilon^{\mu\nu\rho}$ symbol. This fact plays a crucial role in the appearance of induced Chern-Simons terms in quantized planar fermion theories, as will be discussed in detail in Section \ref{induced}.  Another important novel feature of $2+1$ \dim is that there is no ``$\gamma^5$'' matrix
that anticommutes with all the Dirac matrices - note that $i\gamma^0 \gamma^1 \gamma^2={\bf 1}$. Thus, there is no notion of
chirality in the usual sense.

\subsection{Discrete Symmetries: ${\cal P}$, ${\cal C}$ and ${\cal T}$}
\label{pct}

The discrete symmetries of parity, charge conjugation and time reversal act
very differently in $2+1$-\dim. Our usual notion of a parity transformation is
a reflection $\vec{x}\to -\vec{x}$ of the spatial coordinates. However, in the
plane, such a transformation is equivalent to a rotation (this Lorentz
transformation has $\det(\Lambda)=(-1)^2=+1$ instead of $\det(\Lambda)= (-1)^3=-1$). So the
improper discrete `parity' transformation should be taken to be reflection in
just one of the spatial axes (it doesn't matter which we choose):
\begin{eqnarray}
x^1&\to& -x^1\nonumber\\
x^2&\to& x^2
\label{par}
\end{eqnarray}
{}From the kinetic part of the Dirac Lagrangian we see that the spinor field
$\psi$ transforms as
\begin{equation}
\psi\to \gamma^1 \psi
\end{equation}
(where we have suppressed an arbitrary unimportant phase). But this means that
a fermion mass term breaks parity
\begin{equation}
\bar{\psi}\psi\to - \bar{\psi}\psi
\label{massterm}
\end{equation}
Under ${\cal P}$, the gauge field transforms as
\begin{equation}
A^1\to -A^1, \qquad A^2\to A^2,  \qquad A^0\to A^0
\end{equation}
which means that while the standard Maxwell kinetic term is
${\cal P}$-invariant, the Chern-Simons term changes sign under ${\cal P}$:
\begin{equation}
\epsilon^{\mu\nu\rho}A_\mu \partial_\nu A_\rho\to - \epsilon^{\mu\nu\rho}A_\mu
\partial_\nu A_\rho
\end{equation}

Charge conjugation converts the ``electron'' Dirac equation (\ref{diraceq})
into the ``positron'' equation:
\begin{equation}
\left(i\gamma^\mu\partial_\mu+e\gamma^\mu A_\mu-m\right)\psi_c=0
\end{equation}
As is standard, this is achieved by the definition $\psi_c\equiv C\gamma^0
\psi^*$, where the charge conjugation matrix $C$ must satisfy
\begin{equation}
(\gamma^\mu)^T=-C^{-1}\gamma^\mu C
\label{chargeconj}
\end{equation}
In the Dirac representation (\ref{diracrep}) we can choose $C=\gamma^2$. Note then that the
fermion mass term is invariant under ${\cal C}$ (recall the anticommuting
nature of the fermion fields), as is the Chern-Simons term for the gauge field.

Time reversal is an anti-unitary operation (${\cal T}:i\to -i$) in order to
implement $x^0\to -x^0$ without taking $P^0\to -P^0$. The action on spinor and
gauge fields is [using the Dirac representation (\ref{diracrep})]
\begin{equation}
\psi\to \gamma^2\psi, \qquad \vec{A}\to -\vec{A}, \qquad A^0\to A^0
\end{equation}
{}From this we see that both the fermion mass term and the gauge field Chern-Simons term
change sign under time reversal.

The fact that the fermion mass term and the Chern-Simons term have the same
transformation properties under the discrete symmetries of ${\cal P}$, ${\cal
C}$ and ${\cal T}$ will be important later in Section \ref{induced} when we consider radiative
corrections in planar gauge and fermion theories. One way to understand this
connection is that these two terms are supersymmetric partners in $2+1$ \dim \cite{gates}.

\subsection{Poincar\'e Algebra in $2+1$-\dim}
\label{pa}

The novel features of fermion and gauge fields in $2+1$-\dim, as well as the
anyonic fields, can be understood better by considering the respresentation
theory of the Poincar\'e algebra. Our underlying guide is Wigner's Principle:
that in quantum mechanics the relativistic single-particle states should carry
a unitary, irreducible representation of the universal covering group of the
Poincar\'e group \cite{sw}.

The Poincar\'e group $ISO(2,1)$ combines the proper Lorentz group $SO(2,1)$
with space-time translations \cite{binegar,carlo}. The Lorentz generators
$L^{\mu\nu}$ and translation generators $P^\mu$ satisfy the standard Poincar\'e
algebra commutation relations, which can be re-expressed in $2+1$-\dim as
\begin{eqnarray}
[J^\mu , J^\nu ]&=&i\epsilon^{\mu\nu\rho}J_\rho \nonumber\cr
[J^\mu , P^\nu ]&=&i\epsilon^{\mu\nu\rho}P_\rho \nonumber\cr
[P^\mu , P^\nu ]&=&0
\label{poincare}
\end{eqnarray}
where the pseudovector generator $J^\mu$ is
$J^\mu=\frac{1}{2}\epsilon^{\mu\nu\rho}L_{\nu\rho}$. Irreducible
representations of this algrebra may be characterized by the eigenvalues of the
two Casimirs:
\begin{equation}
P^2=P_\mu P^\mu , \qquad\qquad W=P_\mu J^\mu
\label{lubanski}
\end{equation}
Here, $W$ is the Pauli-Lubanski pseudoscalar, the $2+1$ \diml analogue of the
familiar Pauli-Lubanski pseudovector in $3+1$ \dim. We define single-particle
representations $\Phi$ by
\begin{equation}
P^2\Phi=m^2\Phi\qquad\qquad W\Phi=-s m \Phi
\label{egg}
\end{equation}
defining the mass $m$ and spin $s$.

For example, a spin $0$ scalar field may be represented by a momentum space
field $\phi(p)$ on which $P^\mu$ acts by multiplication and $J^\mu$ as an
orbital angular momentum operator:
\begin{equation}
P^\mu \phi=p^\mu\phi\qquad\qquad J^\mu
\phi=-i\epsilon^{\mu\nu\rho}p_\nu{\partial\over \partial p^\rho}\phi
\label{scalar}
\end{equation}
Then the eigenvalue conditions (\ref{egg}) simply reduce to the Klein-Gordon equation $(p^2-m^2)\phi=0$ for a spin $0$ field since $P\cdot J\phi=0$.

For a two-component spinor field $\psi$ we take
\begin{equation}
J^\mu=-i\epsilon^{\mu\nu\rho}p_\nu{\partial\over \partial p^\rho}{\bf 1}
-\frac{1}{2}\gamma^\mu
\label{spinor}
\end{equation}
so the eigenvalue conditions (\ref{egg}) become a Dirac equation of motion $(i\gamma^\mu\partial_\mu-m)\psi=0$, corresponding to spin $s=\pm\frac{1}{2}$.

For a vector field $A_\mu$, whose gauge invariant content may be represented
through the pseudovector dual
$\tilde{F}^\mu=\frac{1}{2}\epsilon^{\mu\nu\rho}F_{\nu\rho}$, we take
\begin{equation}
(J^\mu)_{\alpha\beta}=-i\epsilon^{\mu\nu\rho}p_\nu{\partial\over \partial
p^\rho}\delta_{\alpha\beta}+i\epsilon^{\mu}_{\;\alpha\beta}
\label{vector}
\end{equation}
Then the eigenvalue condition
\begin{equation}
(P\cdot J)_{\alpha\beta}\tilde{F}^\beta=i\epsilon^\mu_{\:\alpha\beta}p_\mu
\tilde{F}^\beta =-s m \tilde{F}_\alpha
\end{equation}
has the form of the topologically massive gauge field equation of motion
(\ref{mcseqs}). We therefore deduce a mass $m=\kappa e^2$ and a spin $s=sign(\kappa)=\pm 1$. This agrees with the Maxwell-Chern-Simons mass found earlier in (\ref{mcsmass}), and is the source of the Maxwell-Chern-Simons spin quoted in (\ref{mcsspin}).

In general, it is possible to modify the standard ``orbital'' form of $J^\mu$
appearing in the scalar field case (\ref{scalar}) without affecting the
Poincar\'e algebra:
\begin{equation}
J^\mu=-i\epsilon^{\mu\nu\rho}p_\nu{\partial\over \partial
p^\rho}-s\left({p^\mu+m\eta^\mu\over p\cdot \eta+m}\right),\qquad
\eta^\mu=(1,0,0)
\label{general}
\end{equation}
It is easy to see that this gives $W=P\cdot J=-s m$, so that the spin can be
arbitrary. This is one way of understanding the possibility of anyonic spins in
$2+1$ \dim. Actually, the real question is how this form of $J^\mu$ can be
realized in terms of a local equation of motion for a field. If $s$ is an
integer or a half-integer then this can be achieved with a $(2s+1)$-component
field, but for arbitrary spin $s$ we require infinite component fields
\cite{jackiwnair}.

\subsection{Nonabelian Chern-Simons Theories}
\label{nabcs}

It is possible to write a nonabelian version of the Chern-Simons Lagrangian (\ref{cs}):
\begin{equation}
{\cal L}_{CS}= \kappa \epsilon^{\mu\nu\rho} tr \left ( A_\mu\partial_\nu A_\rho
+{2\over 3} A_\mu A_\nu A_\rho \right )
\label{cslag}
\end{equation}
The gauge field $A_\mu$ takes values in a finite dimensional representation of
the (semi-simple) gauge Lie algebra ${\cal G}$. In these lectures we take ${\cal G}=su(N)$. In an abelian theory, the gauge fields
$A_\mu$ commute, and so the trilinear term in (\ref{cslag}) vanishes due to the
antisymmetry of the $\epsilon^{\mu\nu\rho}$ symbol. In the nonabelian case [just as in Yang-Mills theory] we write $A_\mu=A_\mu^a T^a$ where the $T^a$ are the generators of ${\cal G}$ [for $a=1,\dots dim({\cal G})$], satisfying the commutation relations $[ T^a , T^b ]=f^{abc}T^c$, and the normalization $\tr(T^a T^b)=-\frac{1}{2}\delta^{ab}$.
\vskip 1cm
{\bf Exercise 2.6.1 :} 
Show that under infinitesimal variations $\delta A_\mu$ of the gauge field the change in the nonabelian Chern-Simons Lagrangian is 
\begin{equation}
\delta {\cal L}_{CS} = \kappa \epsilon^{\mu\nu\rho} \tr\left(\delta A_\mu
F_{\nu\rho}\right)
\label{var}
\end{equation}
where $F_{\mu\nu}=\partial_\mu A_\nu - \partial_\nu A_\mu+[A_\mu, A_\nu ]$ is
the nonabelian field strength. 
\vskip 1cm
From the variation (\ref{var}) we see that the nonabelian equations of motion have the same form as the abelian ones: $\kappa
\epsilon^{\mu\nu\rho}F_{\nu\rho}=J^\mu$. Note also that the Bianchi identity,
$\epsilon^{\mu\nu\rho}D_\mu F_{\nu\rho}=0$, is compatible with covariant
current conservation: $D_\mu J^\mu=0$. The source-free equations are once again
$F_{\mu\nu} =0$, for which the solutions are pure gauges (flat connections)
$A_\mu=g^{-1}\partial_\mu g$, with $g$ in the gauge group.

An important difference, however, lies in the behaviour of the nonabelian Chern-Simons
Lagrangian (\ref{cslag}) under a gauge transformation. The nonabelian gauge
transformation $g$ (which is an element of the gauge group) tarnsforms the gauge field as
\begin{equation}
A_\mu \to A_\mu^g\equiv g^{-1}A_\mu g +g^{-1}\partial_\mu g
\label{nabgt}
\end{equation}
\vskip 1cm
{\bf Exercise 2.6.2 :} Show that under the gauge transformation (\ref{nabgt}), the Chern-Simons Lagrangian ${\cal L}_{CS}$ in (\ref{cslag}) transforms as
\begin{equation}
{\cal L}_{CS} \to {\cal L}_{CS} - \kappa \epsilon^{\mu\nu\rho}\partial_\mu \tr
\left( \partial_\nu g\,g^{-1}\, A_\rho\right) -{\kappa\over 3} \epsilon^{\mu\nu\rho} \tr \left( g^{-1}\partial_\mu g g^{-1}\partial_\nu g g^{-1}\partial_\rho g\right)
\label{change}
\end{equation}
\vskip 1cm
We recognize, as in the abelian case, a total space-time derivative term, which
vanishes in the action with suitable boundary conditions. However, in the
nonabelian case there is a new term in (\ref{change}), known as the winding
number density of the group element $g$ :
\begin{equation}
w(g)={1\over 24\pi^2} \epsilon^{\mu\nu\rho} \tr \left( g^{-1}\partial_\mu g
g^{-1}\partial_\nu g g^{-1}\partial_\rho g \right)
\label{windingdensity}
\end{equation}
With appropriate boundary conditions, the integral of $w(g)$ is an integer - see Exercise 2.6.3. Thus, the Chern-Simons action changes by an additive constant under a large gauge transformation ({\it i.e.}, one with nontrivial winding number $N$):
\begin{equation}
{\cal S}_{\rm CS}\to {\cal S}_{\rm CS} -8\pi^2 \kappa N
\label{shs}
\end{equation}
This has important implications for the development of a quantum nonabelian
Chern-Simons theory. To ensure that the quantum amplitude $\exp(i\,S)$ remains
gauge invariant, the Chern-Simons coupling parameter $\kappa$ must assume discrete
values \cite{deser1}
\begin{equation}
\kappa = {{\rm integer}\over 4\pi}
\label{dirac}
\end{equation}
This is analogous to Dirac's quantization condition for a magnetic monopole
\cite{dirac}. We shall revisit this Chern-Simons discreteness condition in more detail
in later Sections.
\vskip 1cm
{\bf Exercise 2.6.3:} In three \diml Euclidean space, take the $SU(2)$ group element
\begin{equation}
g=\exp\left(i\pi N{\vec{x}\cdot\vec{\sigma}\over \sqrt{\vec{x}^2+R^2}}\right)
\end{equation}
where $\vec{\sigma}$ are the Pauli matrices, and $R$ is an arbitrary scale
parameter. Show that the winding number for this $g$ is equal to $N$. Why must
$N$ be an integer?
\vskip 1cm

To conclude this brief review of the properties of nonabelian Chern-Simons terms, I
mention the original source of the name ``Chern-Simons''. S. S. Chern and J.
Simons were studying a combinatorial approach to the Pontryagin density $\epsilon^{\mu\nu\rho\sigma}\tr\left(F_{\mu\nu}F_{\rho\sigma}\right)$ in four
dimensions and noticed that it could be written as a total derivative:
\begin{equation}
\epsilon^{\mu\nu\rho\sigma}\tr\left(F_{\mu\nu}F_{\rho\sigma}\right) = 4 \,
\partial_\sigma \left[\epsilon^{\mu\nu\rho\sigma} tr \left ( A_\mu\partial_\nu
A_\rho +{2\over 3} A_\mu A_\nu A_\rho \right )\right]
\label{pontryagin}
\end{equation}
Their combinatorial approach ``got stuck by the emergence of a boundary term
which did not yield to a simple combinatorial analysis. The boundary term
seemed interesting in its own right, and it and its generalizations are the
subject of this paper'' \cite{chern}. We recognize this interesting boundary
term as the Chern-Simons Lagrangian (\ref{cslag}).

\section{Canonical Quantization of Chern-Simons Theories}
\label{can}

There are many ways to discuss the quantization of Chern-Simons theories. Here I focus on canonical quantization because it has the most direct relationship with the condensed matter applications which form the primary subject of this
School. Indeed, the well-known Landau and Hofstadter problems of solid state
physics provide crucial physical insight into the canonical quantization of
Chern-Simons theories.

\subsection{Canonical Structure of Chern-Simons Theories}

In this Section we consider the classical canonical structure and Hamiltonian
formulation of Chern-Simons theories, in preparation for a discussion of their
quantization. We shall discover an extremely useful quantum mechanical analogy
to the classic Landau problem of charged electrons moving in the plane in the
presence of an external uniform magnetic field perpendicular to the plane. I
begin with the abelian theory because it contains the essential physics, and
return to the nonabelian case later.

The Hamiltonian formulation of Maxwell (or Yang-Mills) theory is standard. In
the Weyl gauge ($A_0=0$) the spatial components of the gauge field $\vec{A}$
are canonically conjugate to the electric field components $\vec{E}$, and
Gauss's law $\vec{\nabla}\cdot\vec{E}=\rho$ appears as a constraint, for which
the nondynamical field $A_0$ is a Lagrange multiplier. [If you 
wish to remind yourself of the Maxwell case, simply set the Chern-Simons coupling $\kappa$ to zero in the following equations (\ref{mcsl}) -
(\ref{electriccrs})].

Now consider instead the canonical structure of the Maxwell-Chern-Simons theory with Lagrangian (\ref{mcs}):
\begin{equation}
{\cal L}_{\rm MCS}=\frac{1}{2e^2}E_i^2-\frac{1}{2e^2}B^2 +
\frac{\kappa}{2} \epsilon^{ij}\dot{A}_i A_j+
\kappa A_0 B
\label{mcsl}
\end{equation}
The $A^0$ field is once again nondynamical, and can be regarded as a Lagrange
multiplier enforcing the Gauss law constraint
\begin{equation}
\partial_i F^{i0}+\kappa e^2 \epsilon^{ij}\partial_i A_j=0
\label{mcsgauss}
\end{equation}
This is simply the $\nu=0$ component of the Euler-Lagrange equations
(\ref{mcseqs}). In the $A_0=0$ gauge we identify the $A_i$ as `coordinate'
fields, with corresponding `momentum' fields
\begin{equation}
\Pi^i\equiv {\partial {\cal L}\over \partial \dot{A_i}}=
\frac{1}{e^2}\dot{A}_i+\frac{\kappa}{2}\epsilon^{ij}A_j
\label{momenta}
\end{equation}
The Hamiltonian is obtained from the Lagrangian by a Legendre transformation
\begin{eqnarray}
{\cal H}_{\rm MCS}&=&\Pi^i \dot{A}_i- {\cal L}\nonumber\\
&=&\frac{e^2}{2}\left(\Pi^i-\frac{\kappa}{2}\epsilon^{ij} A_j\right)^2
+\frac{1}{2e^2}B^2+ A_0\left(\partial_i \Pi^i+\kappa B\right)
\label{mcsham}
\end{eqnarray}
At the classical level, the fields $A_i(\vec{x},t)$ and $\Pi^i(\vec{x},t)$
satisfy canonical equal-time Poisson brackets. These become equal-time
canonical commutation relations in the quantum theory:
\begin{equation}
[A_i(\vec{x}),\Pi^j(\vec{y})]=i\,\delta_i^{\,j}\delta(\vec{x}-\vec{y})
\label{mcscrs}
\end{equation}
Notice that this implies that the electric fields do not commute (for
$\kappa\neq 0$)
\begin{equation}
[E_i(\vec{x}),E_j(\vec{y})]=-i\,\kappa e^4\,
\epsilon_{ij}\delta(\vec{x}-\vec{y})
\label{electriccrs}
\end{equation}
The Hamiltonian (\ref{mcsham}) still takes the standard Maxwell form ${\cal
H}=\frac{1}{2e^2}(\vec{E}^2+B^2)$ when expressed in terms of the electric and
magnetic fields. This is because the Chern-Simons term does not modify the energy -- it is, after all, first order in time derivatives. But it does modify the relation between momenta and velocity fields. This is already very suggestive of the effect of an external magnetic field on the dynamics of a charged particle.

Now consider a {\it pure} Chern-Simons theory, with {\it no} Maxwell term in the Lagrangian.
\begin{equation}
{\cal L}_{\rm CS}=\frac{\kappa}{2}\epsilon^{ij}\dot{A}_i A_j +\kappa A_0 B
\label{pure}
\end{equation}
Once again, $A_0$ is a Lagrange multiplier field, imposing the Gauss law:
$B=0$. But the Lagrangian is first order in time derivatives, so it is already
in the Legendre transformed form $L=p\dot{x}-H$, with $H=0$. So there is no
dynamics -- indeed, the only dynamics would be inherited from coupling to
dynamical matter fields. Another way to see this is to notice that the pure Chern-Simons energy momentum tensor
\begin{equation}
T^{\mu\nu}\equiv{2\over \sqrt{detg}}{\delta S_{\rm CS}\over \delta g_{\mu\nu}}
\label{enmom}
\end{equation}
vanishes identically because the Chern-Simons action is independent of the metric, since the Lagrange density is a three-form 
${\cal L}=\tr(AdA+\frac{2}{3}AAA)$.

Another important fact about the pure Chern-Simons system (\ref{pure}) is that the components of the gauge field are canonically conjugate to one another:
\begin{equation}
[A_i(\vec{x}),A_j(\vec{y})]=\frac{i}{\kappa}\epsilon_{ij}
\delta(\vec{x}-\vec{y})
\label{purecrs}
\end{equation}
This is certainly very different from the Maxwell theory, for which the
components of the gauge field commute, and it is the $A_i$ and $E_i$ fields that are canonically conjugate. So pure Chern-Simons is a strange new type of gauge theory, with the components $A_i$ of the gauge field not commuting with one another.

We can recover this noncommutativity property from the Maxwell-Chern-Simons case by
taking the limit $e^2\to\infty$, with $\kappa$ kept fixed. Then, from the
Hamiltonian (\ref{mcsham}) we see that we are forced to impose the constraint
\begin{equation}
\Pi^i=\frac{\kappa}{2} \epsilon^{ij}A_j
\label{con}
\end{equation}
then the Maxwell-Chern-Simons Hamiltonian (\ref{mcsham}) vanishes and the Lagrangian
(\ref{mcsl}) reduces to the pure Chern-Simons Lagrangian (\ref{pure}). The canonical
commutation relations (\ref{purecrs}) arise because of the constraints
(\ref{con}), noting that these are second-class constraints so we must use
Dirac brackets to find the canonical relations between $A_i$ and $A_j$
\cite{dunne2}.

\subsection{Chern-Simons Quantum Mechanics}
\label{csqm}

\begin{figure}[htb]
\vspace{-1cm}
\centering{\epsfig{file=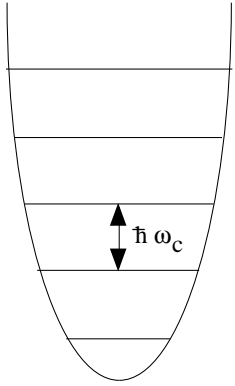,width=3in}}
\vspace{-3cm}
\caption{The energy spectrum for charged particles in a uniform magnetic field consists of equally spaced `Landau levels', separated by $\hbar \omega_c$ where $\omega_c$ is the cyclotron frequency. Each Landau level has degeneracy given by the total magnetic flux through the sample.}
\label{llevels}
\end{figure}

To understand more deeply this somewhat unusual projection from a Maxwell-Chern-Simons theory to a pure Chern-Simons theory we appeal to the following quantum mechanical analogy \cite{kogan,dunne2}. Consider the long wavelength limit of the Maxwell-Chern-Simons Lagrangian, in which we drop all spatial derivatives. (This is sufficient for identifying the masses of excitations.) Then the resulting Lagrangian
\begin{equation}
L=\frac{1}{2e^2}\dot{A}_i^2+\frac{\kappa}{2}\epsilon^{ij}\dot{A}_i A_j
\end{equation}
has exactly the same form as the Lagrangian for a nonrelativistic charged
particle moving in the plane in the presence of a uniform external magnetic
field $b$ perpendicular to the plane
\begin{equation}
L=\frac{1}{2}m\dot{x}_i^2+\frac{b}{2}\epsilon^{ij}\dot{x}_ix_j
\label{ml}
\end{equation}
The canonical analysis of this mechanical model is a simple undergraduate
physics exercise. The momenta
\begin{equation}
p_i={\partial L\over \partial\dot{x}_i}=m\dot{x}_i+\frac{b}{2}\epsilon^{ij}x_j
\end{equation}
are shifted from the velocities and the Hamiltonian is
\begin{equation}
H=p_i\dot{x}_i-L=\frac{1}{2m}(p_i-\frac{b}{2}\epsilon^{ij}x_j)^2=
\frac{m}{2}v_i^2
\label{mechham}
\end{equation}
At the quantum level the canonical commutation relations,
$[x_i,p_j]=i\delta_{ij}$, imply that the velocities do not commute:
$[v_i,v_j]=-i\frac{b}{m^2}\epsilon_{ij}$. It is clear that these features of
the Landau problem mirror precisely the canonical structure of the Maxwell-Chern-Simons
system, for both the Hamiltonian (\ref{mcsham}) and the canonical commutation
relations (\ref{mcscrs}) and (\ref{electriccrs}).
\begin{eqnarray}
{\rm MCS~field~theory} &\longleftrightarrow &  {\rm Landau~problem}\nonumber\cr
e^2&\longleftrightarrow & \frac{1}{m}\nonumber\cr
\kappa &\longleftrightarrow & b\nonumber\cr
m_{\rm MCS} =\kappa e^2 &\longleftrightarrow & \omega_c=\frac{b}{m}
\label{corr}
\end{eqnarray}
This correspondence is especially useful because the quantization of the Landau
system is well understood. The quantum mechanical spectrum consists of equally
spaced energy levels (Landau levels), spaced by $\hbar \omega_c$ where the
cyclotron frequency is $\omega_c=\frac{b}{m}$. See Figure \ref{llevels}. Each Landau level is infinitely degenerate in the open plane, while for a finite area the degeneracy is related to the net magnetic flux
\begin{equation}
N_{\rm deg}={b A\over 2\pi}
\label{landau}
\end{equation}
where $A$ is the area \cite{landau,aharonov}.

The pure Chern-Simons limit is when $e^2\to\infty$, with $\kappa$ fixed. In the quantum
mechanical case this corresponds to taking the mass $m\to 0$, with $b$ fixed.
Thus, the cyclotron frequency $\omega_c=\frac{b}{m}$ becomes infinite and so
the energy gap between Landau levels becomes infinite, isolating each level
from the others. We therefore have a formal projection onto a highly degenerate
ground state - the lowest Landau level (LLL). Interestingly, this is exactly
the type of limit that is of physical interest in quantum Hall systems -- see
Steve Girvin's lectures at this School for more details on the importance of
the lowest Landau level. In the limit $m\to 0$ of projecting to the lowest Landau
level, the Lagrangian (\ref{ml}) becomes 
$L=\frac{b}{2}\epsilon^{ij}\dot{x}_ix_j$. This is first order in time
derivatives, so the two coordinates $x_1$ and $x_2$ are in fact canonically
conjugate to one another, with commutation relations [compare with
(\ref{purecrs})]:
\begin{equation}
[x_i,x_j]=\frac{i}{b} \epsilon_{ij}
\label{noncomm}
\end{equation}
Thus, the two-dimensional coordinate space has become [in the LLL projection
limit] a two-dimensional phase space, with $\frac{1}{b}$ playing the role of
``$\hbar$''. Applying the Bohr-Sommerfeld estimate of the number of quantum
states in terms of the area of phase space, we find
\begin{equation}
N_{\rm deg} \approx \frac{A}{``h"}=\frac{bA}{2\pi}
\end{equation}
which is precisely Landau's estimate (\ref{landau}) of the degeneracy of the
lowest Landau level.

This projection explains the physical nature of the pure Chern-Simons theory. The pure
Chern-Simons theory can be viewed as the $e^2\to\infty$ limit of the topologically
massive Maxwell-Chern-Simons theory, in which one truncates the Hilbert space onto the
ground state by isolating it from the rest of the spectrum by an infinite gap.
The Chern-Simons analogue of the cyclotron frequency $\omega_c$ is the Chern-Simons mass $\kappa
e^2$. So the inclusion of a Chern-Simons term in a gauge theory Lagrangian is analogous
to the inclusion of a Lorentz force term in a mechanical system. This explains
how it was possible to obtain a mass (\ref{mcsmass}) for the gauge field in the
Maxwell-Chern-Simons theory without the Higgs mechanism - in the mechanical analogue,
the Higgs mechanism corresponds to introducing a harmonic binding term
$\frac{1}{2}m\omega^2 \vec{x}^2$, which gives a characteristic frequency in the
most obvious way. But the Landau system shows how to obtain a characteristic
frequency (the cyclotron frequency) {\it without} introducing a harmonic
binding term. We can view the Chern-Simons theory as a gauge field realization of this
mechanism.

To clarify the distinction between these two different mass generation
mechanisms for the gauge field, consider them both acting together, as we did
in Section \ref{tmgt}. That is, consider the broken (Higgs) phase of a
Maxwell-Chern-Simons theory coupled to a scalar field (\ref{mcsh}). If we are only
interested in the {\it masses} of the excitations it is sufficient to make a
zeroth-order (spatial) derivative expansion, neglecting all spatial derivatives,
in which case the functional Schr\"odinger representation reduces to the familiar
Schr\"odinger representation of quantum mechanics. Physical masses of the field
theory appear as physical frequencies of the corresponding quantum mechanical
system. In the Higgs phase, the quadratic Lagrange density becomes
\begin{equation}
L={1\over 2 e^2}\dot{A}_i^2+{\kappa\over 2}\epsilon^{ij}\dot{A}_i
A_j - v^2A_iA_i
\label{qmlimit}
\end{equation}
This is the Maxwell-Chern-Simons Lagrangian with a Proca mass term $v^2 A_i^2$. In the
analogue quantum mechanical system this corresponds to a charged particle of mass
$1\over e^2$ moving in a uniform magnetic field of strength $\kappa$, and a
harmonic potential well of frequency $\omega=\sqrt{2}ev$.  Such a quantum
mechanical model is exactly solvable, and is well-known [see Exercise 3.2.1] to
separate into two distinct harmonic oscillator systems of characteristic
frequencies
\begin{equation}
\omega_\pm={\omega_c\over 2}\left(\sqrt{1+{4\omega^2\over
\omega_c^2}}\pm
1\right)
\label{frequencies}
\end{equation}
where $\omega_c$ is the cyclotron frequency corresponding to the
magnetic field and $\omega$ is the harmonic well frequency. Taking
$\omega_c=\kappa e^2$ and $\omega=\sqrt{2}ev$, we see that these characteristic
frequencies are exactly the mass poles $m_\pm$ in (\ref{masses}) of the
Maxwell-Chern-Simons Higgs system, identified from the covariant gauge propagator. The
pure Chern-Simons Higgs limit corresponds to the physical limit in which the
cyclotron frequency dominates, so that
\begin{equation}
\omega_-\to {\omega^2\over \omega_c}={2v^2\over\kappa}=m_- \hskip 1.5in
\omega_+\to\infty
\label{freq-limit}
\end{equation}
The remaining finite frequency $\omega_-$ is exactly the mass $m_-$ found in
the covariant propagator (\ref{cshprop}) for the Higgs phase of a pure Chern-Simons
Higgs theory. 

So we see that in $2+1$ \dim, the gauge field can acquire one massive mode
via the standard Higgs mechanism [no Chern-Simons term], or via the Chern-Simons-Higgs mechanism
[no Maxwell term]; or the gauge field can acquire two massive modes [both Chern-Simons 
and Maxwell term].
\vskip 1cm
{\bf Exercise 3.2.1 :} Consider the planar quantum mechanical system with
Hamiltonian
$H=\frac{1}{2m}(p^i+\frac{b}{2}\epsilon^{ij}x^j)^2
+\frac{1}{2}m\omega^2\vec{x}^2$. Show that the definitions
\begin{eqnarray}
p_\pm&=&\sqrt{\omega_\pm\over 2m\Omega}p^1 \pm \sqrt{m\Omega \omega_\pm\over 2}
x^2\nonumber\\
x_\pm&=&\sqrt{m\Omega\over 2\omega_\pm}x^1\mp{1\over \sqrt{2m\Omega\omega_\pm}}
p^2
\end{eqnarray}
where $\Omega=\sqrt{\omega^2+\frac{b^2}{4m^2}}$, and $\omega_\pm=\Omega\pm
\frac{b}{2m}$ are as in (\ref{frequencies}), separate $H$ into two distinct
harmonic oscillators of frequency $\omega_\pm$.
\vskip 1cm

A natural way to describe the lowest Landau level (LLL) projection is in terms
of coherent states \cite{jach,dunne2}. To see how these enter the picture,
consider the quantum mechanical Lagrangian, which includes a harmonic binding
term
\begin{equation}
L=\frac{1}{2}m\dot{x}_i^2+\frac{b}{2}\epsilon^{ij}\dot{x}_ix_j
-\frac{1}{2}m\omega^2 x_i^2
\end{equation}
This quantum mechanics problem can be solved exactly. Converting to polar
coordinates, the wavefunctions can be labelled by two integers, $N$ and $n$, 
with
\begin{equation}
<\vec{x}|N,n> =\sqrt{{N!\over \pi (N+|n|)!}}\,
\left(m\Omega\right)^{\frac{1+|n|}{2}} r^{|n|} e^{in\theta}
e^{-\frac{1}{2}m\Omega r^2} L_N^{|n|}\left(m\Omega r^2\right)
\end{equation}
 where $L_N^{|n|}$ is an associated Laguerre polynomial,
$\Omega=\sqrt{\frac{b^2}{4m^2}+\omega^2}$, and the energy is
$E(N,n)=(2N+|n|+1)\Omega-\frac{b}{2m}n$. In the $m\to 0$ limit, the $N=0$ and
$n\geq 0$ states decouple from the rest, and the corresponding wavefunctions
behave as
\begin{eqnarray}
<\vec{x}|0,n>&=&{1\over \sqrt{\pi n!}} \left(m\Omega\right)^{\frac{1+n}{2}}
r^{n} e^{in\theta} e^{-\frac{1}{2}m\Omega r^2} \nonumber\\
&\to& \sqrt{b\over 2\pi}{z^n\over \sqrt{n!}} e^{-\frac{1}{2}|z|^2}
\label{pro}
\end{eqnarray}
where we have defined the complex coordinate $z=\sqrt{\frac{b}{2}}(x_1+ix_2)$.
The norms of these states transform under this LLL projection limit as
\begin{equation}
\int d^2x\, |<\vec{x}|0,n>|^2\to \int {dz dz^*\over 2\pi i}\, e^{-|z|^2}
\, |<z|n>|^2
\label{norm}
\end{equation}
We recognize the RHS as the norm in the coherent state representation of a
one-dimensional quantum system. Thus, the natural description of the LLL is in
terms of coherent state wavefunctions $<z|n>=z^n/\sqrt{n!}$. The exponential
factor $e^{-\frac{1}{2}m\Omega r^2}$ becomes, in the $m\to 0$ limit, part of
the coherent state measure factor. This explains how the original two-\diml
system reduces to a one-\diml system, and how it is possible to have $z$ and
$z^*$ being conjugate to one another, as required by the commutation relations
(\ref{noncomm}).

Another way to find the lowest Landau level wavefunctions is to express
the single-particle Hamiltonian as
\begin{equation}
H=-\frac{1}{2m}\left( D_1^2+D_2^2\right)
\label{sh}
\end{equation}
where $D_1=\partial_1+i\frac{b}{2}x_2$ and $D_2=\partial_2-i\frac{b}{2}x_1$.
Then, define the complex combinations $D_\pm=D_1\pm iD_2$ as:
\begin{equation}
D_+=2\partial_{\bar{z}}+\frac{b}{2} z, \qquad\qquad
D_-=2\partial_z-\frac{b}{2}\bar{z}
\label{dpm}
\end{equation}
The Hamiltonian (\ref{sh}) factorizes as
\begin{equation}
H=-\frac{1}{2m}D_- D_+ +\frac{b}{2m}
\label{hfac}
\end{equation}
so that the lowest Landau level states [which all have energy
$\frac{1}{2}\omega_c=\frac{b}{2m}$] satisfy
\begin{equation}
D_+\psi=0, \qquad {\rm or} \qquad \psi=f(z)e^{-\frac{b}{4}|z|^2}
\label{hol}
\end{equation}
We recognize the exponential factor (after absorbing $\sqrt{\frac{b}{2}}$ into
the definition of $z$ as before) as the factor in (\ref{pro}) which contributes
to the coherent state measure factor. Thus, the lowest Landau level Hilbert
space consists of holomorphic wavefunctions $f(z)$, with coherent state norm as
defined in (\ref{norm}) \cite{jach}. This is a standard feature of the analysis
of the fractional quantum Hall effect -- see Steve Girvin's lectures for
further applications.

When the original Landau Hamiltonian contains also a potential term, this leads
to interesting effects under the LLL projection. With finite $m$, a potential
$V(x_1,x_2)$ depends on two commuting coordinates. But in the LLL limit ({\it
i.e.}, $m\to 0$ limit) the coordinates become non-commuting [see
(\ref{noncomm})] and $V(x_1,x_2)$ becomes the projected Hamiltonian on the
projected phase space. Clearly, this leads to possible operator-ordering
problems. However, these can be resolved \cite{jach,dunne2} by insisting that
the projected Hamiltonian is ordered in such a way that the coherent state
matrix elements computed within the LLL agree with the $m\to 0$ limit of the
matrix elements of the potential, computed with $m$ nonzero.

\subsection{Canonical Quantization of Abelian Chern-Simons Theories}

Motivated by the coherent state formulation of the lowest Landau level
projection of the quantum mechanical systems in the previous Section, we now
formulate the canonical quantization of abelian Chern-Simons theories in terms of {\it
functional} coherent states. Begin with the Maxwell-Chern-Simons Lagrangian in the
$A_0=0$ gauge:
\begin{equation}
{\cal L}_{\rm
MCS}=\frac{1}{2e^2}\dot{A}_i^2+\frac{\kappa}{2}\epsilon^{ij}\dot{A}_i A_j
-\frac{1}{2e^2} (\epsilon^{ij}\partial_i A_j)^2
\label{mcsweyl}
\end{equation}
This is a quadratic Lagrangian, so we expect we can find the groundstate
wavefunctional. Physical states must also satisfy the Gauss law constraint:
$\vec{\nabla} \cdot\vec{\Pi}-\kappa B=0$. This Gauss law is satisfied by
functionals of the form
\begin{equation}
\Psi[A_1,A_2]=e^{-i\frac{\kappa}{2}\int\, B\lambda} \Psi[A_T]
\end{equation}
where we have decomposed $\vec{A}$ into its longitudinal and transverse parts:
$A_i=\partial_i\lambda+A_i^T$. Using the Hamiltonian (\ref{mcsham}), the ground
state wavefunctional is \cite{deser1}
\begin{equation}
\Psi_0[A_1,A_2]=e^{-i\frac{\kappa}{2}\int B\lambda} e^{-\frac{1}{2e^2}\int
A^T_i\sqrt{\kappa^2e^4-\nabla^2}A^T_i}
\end{equation}
The pure Chern-Simons limit corresponds to taking $e^2\to\infty$, so that the
wavefunctional becomes
\begin{equation}
\Psi_0[A_1,A_2]\to e^{-\frac{1}{2}\int A\frac{\partial_+}{\partial_-}A} \,
e^{-\frac{1}{2}\int |A|^2}
\label{csgs}
\end{equation}
where we have defined $A=\sqrt{\frac{\kappa}{2}}(A_1+iA_2)$ [in analogy to the
definition of $z$ in the previous section], and $\partial_\pm=(\partial_1\mp
i\partial_2)$. From this form of the groundstate wavefunctional we recognize
the functional coherent state measure factor $e^{-\frac{1}{2}\int |A|^2}$,
multiplying a functional $\Psi[A]=e^{-\frac{1}{2}\int
A\frac{\partial_+}{\partial_-}A}$ that depends only on $A$, and not on $A^*$.
This is the functional analogue of the fact that the LLL wavefunctions have the
form $\psi=f(z)e^{-\frac{1}{2}|z|^2}$, as in (\ref{pro}) and (\ref{hol}). The
fact that the Chern-Simons theory has a single ground state, rather than a highly
degenerate LLL, is a consequence of the Gauss law constraint, for which there
was no analogue in the quantum mechanical model. These pure Chern-Simons wavefunctionals
have a functional coherent state inner product [compare with (\ref{norm})]
\begin{equation}
<\Psi |\Phi>=\int {\cal D}A{\cal D}A^*\, e^{-\int |A|^2} \left(\Psi[A]\right)^*
\Phi[A]
\label{fip}
\end{equation}
Actually, we needn't have gone through the process of taking the $e^2\to\infty$
limit of the Maxwell-Chern-Simons theory. It is much more direct simply to adopt the
functional coherent state picture. This is like going directly to the lowest
Landau level using coherent states, instead of projecting down from the full
Hilbert space of {\it all} the Landau levels. The canonical commutation
relations (\ref{purecrs}) imply $[A(z),A^*(w)]=\delta(z-w)$, so that we can
represent $A^*$ as a functional derivative operator: $A^*=-{\delta\over \delta
A}$. Then the pure Chern-Simons Gauss law constraint $F_{12}=0$ acts on states as
\begin{equation}
\left(\partial_-{\delta\over \delta A}+\partial_+ A\right) \Psi[A]=0
\end{equation}
with solution
\begin{equation}
\Psi_0[A]=e^{-\frac{1}{2}\int A\frac{\partial_+}{\partial_-}A}
\label{abgs}
\end{equation}
as in (\ref{csgs}).

If this pure Chern-Simons theory is coupled to some charged matter fields with a
rotationally covariant current, then the physical state (\ref{abgs}) is an
eigenstate of the conserved angular momentum operator $M=-\frac{\kappa}{2}\int
x^i\epsilon^{ij}(A_j B+B A_j)$:
\begin{equation}
M\Psi_0[A]={Q^2\over 4\pi \kappa} \Psi_0[A]
\label{sp}
\end{equation}
where $Q=\int d^2x\,\rho$. Comparing with the Aharonov-Bohm exchange phase
$\Delta\theta=\frac{e^2}{4\pi\kappa}$ in (\ref{spinstat}) we see that the
statistics phase $s$ coincides with the spin eigenvalue $M$. This is the
essence of the generalized spin-statistics relation for extended (field
theoretic) anyons.

\subsection{Quantization on the Torus and Magnetic Translations}
\label{torus}

The quantization of pure Chern-Simons theories on the plane is somewhat boring because
there is just a unique physical state (\ref{abgs}). To make this more interesting
we could include external sources, which appear in the canonical formalism as
point delta-function sources on the fixed-time surface. The appearance of these
singularities makes the projection to flat connections satisfying Gauss's law
more intricate, and leads to important connections with knot theory and the
braid group. Alternatively, we could consider the spatial surface to have
nontrivial topology, rather than simply being the open plane ${\bf R}^2$. For
example, take the spatial manifold to be a Riemann surface $\Sigma$ of genus
$g$. This introduces extra degrees of freedom, associated with the nontrivial
closed loops around the handles of $\Sigma$
\cite{witten,poly,bos,elitzur,iengo}. Interestingly, the quantization of this
type of Chern-Simons theory reduces once again to an effective quantum mechanics
problem, with a new feature that has also been treated long ago in the solid
state literature under the name of the ``magnetic translation group''.

To begin, it is useful to reconsider the case of ${\bf R}^2$. To make
connection with the coherent state representation, we express the
longitudinal-transverse decomposition of the vector potential,
$A_i=\partial_i\omega+\epsilon_{ij}\partial_j\sigma$, in terms of the
holomorphic fields $A=\frac{1}{2}(A_1+iA_2)$ and $A^*=\frac{1}{2}(A_1-iA_2)$.
Thus, with $z=x^1+ix^2$ and $A_i dx^i=A^*dz+ A d\bar{z}$, we have
\begin{equation}
A=\partial_{\bar{z}} \chi, \qquad\qquad A^*=\partial_z\chi^*
\label{hodge}
\end{equation}
where $\chi=\omega-i\sigma$ is a complex field. If $\chi$ were real, then $A$
would be purely longitudinal - {\it i.e.} pure gauge. But with a complex field
$\chi$, this representation spans all fields. A gauge transformation is
realized as a shift in the real part of $\chi$: $\chi\to \chi+\lambda$, where
$\lambda$ is real.

\begin{figure}[htb]
\vspace{-1.5cm}
\centering{\epsfig{file=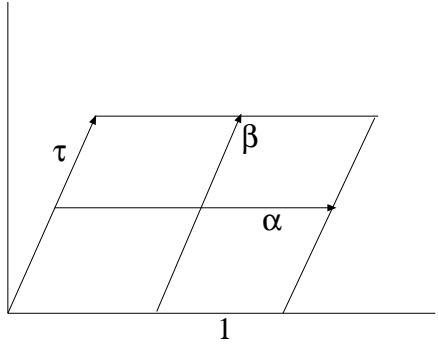,width=2.5in}}
\vspace{-1.5cm}
\caption{The torus can be parametrized as a parallelogram with sides $\tau$ and $1$. There are two cycles $\alpha$ and $\beta$ representing the two independent non-contractible loops on the surface.}
\label{tfig}
\end{figure}

On a nontrivial surface this type of longitudinal-tranverse decomposition is
not sufficient, as we know from elementary vector calculus on surfaces. The
gauge field is decomposed using a Hodge decomposition, which incorporates the
windings around the $2g$ independent noncontractible loops on $\Sigma$. For
simplicity, consider the $g=1$ case: {\it i.e.}, the torus. (The generalization
to higher genus is quite straightforward). The torus can be parametrized as a
parallelogram with sides $1$ and $\tau$, as illustrated in Figure \ref{tfig}. The area of the parallelogram is $Im(\tau)$, and the field $A$ can be expressed as
\begin{equation}
A=\partial_{\bar{z}}\chi+i {\pi \over Im(\tau)}\,\overline{\omega(z)}\, a
\label{torushodge}
\end{equation}
where $\omega(z)$ is a holomorphic one-form normalized according to $\int
|\omega(z)|^2=Im(\tau)$. This holomorphic form has integrals
$\oint_\alpha\omega=1$ and $\oint_\beta\omega=\tau$ around the homology basis
cycles $\alpha$ and $\beta$. For the torus, we can simpy take $\omega(z)=1$.

The complex parameter $a$ appearing in (\ref{torushodge}) is just a function of
time, independent of the spatial coordinates. Thus the $A_0=0$ gauge Chern-Simons
Lagrangian decouples into two pieces
\begin{equation}
L_{\rm CS} ={i\kappa\pi^2\over Im(\tau)} \left(\dot{a}a^*-\dot{a}^* a\right)+
i\kappa\int_\Sigma\left(\partial_{\bar{z}}\dot{\chi}\partial_z\chi^* -
\partial_z\dot{\chi}^*\partial_{\bar{z}}\chi\right)
\end{equation}
So the coherent state wavefunctionals factorize as $\Psi[A]=\Psi[\chi]
\psi(a)$, with the $\chi$ dependence exactly as discussed in the previous
section. On the other hand, the $a$ dependence corresponds exactly to a quantum
mechanical LLL problem, with ``magnetic field''
$B=\frac{4\pi^2\kappa}{Im\tau}$. So the quantum mechanical wavefunctions
$\psi(a)$ have inner product
\begin{equation}
<\psi|\phi>=\int da da^* e^{-{2\pi^2\kappa\over Im(\tau)}|a|^2}
(\psi(a))^*\phi(a)
\label{inn}
\end{equation}
But we have neglected the issue of gauge invariance. Small gauge
transformations, $\chi\to \chi+\lambda$, do not affect the $a$ variables. But
because of the nontrivial loops on the spatial manifold there are also
``large'' gauge transformations, which only affect the $a$'s:
\begin{equation}
a\to a+p+q\tau, \qquad\qquad p,q\in {\bf Z}
\label{lgts}
\end{equation}
To understand how these large gauge transformations act on the wavefunctions
$\psi(a)$, we recall the notion of the ``magnetic translation group''. That is,
in a uniform magnetic field, while the magnetic field is uniform, the
corresponding vector potential, which is what appears in the Hamiltonian, is
not! Take, for example,  $A_i=-\frac{B}{2}\epsilon_{ij}x^j$. Then there are
magnetic translation operators
\begin{equation}
T(\vec{R})\equiv e^{-i\vec{R}\cdot(\vec{p}-e\vec{A}) }
\label{magtrans}
\end{equation}
which commute with the particle Hamiltonian
$H=\frac{1}{2m}(\vec{p}+e\vec{A})^2$, but do not commute with one another:
\begin{equation}
T(\vec{R}_1)T(\vec{R}_2)=T(\vec{R}_2)T(\vec{R}_1)e^{-ie\vec{B}\cdot(\vec{R}_1
\times \vec{R}_2)}
\label{nc}
\end{equation}
The exponential factor here involves the magnetic flux through the
parallelogram spanned by $\vec{R}_1$ and $\vec{R}_2$. In solid state
applications, a crystal lattice establishes a periodic potential for the
electrons. If, in addition, there is a magnetic field, then we can ask how the
spectrum of Landau levels is modified by the periodic potential, or
alternatively we can ask how the Bloch band structure of the periodic potential
is modified by the presence of the magnetic field \cite{hofstadter}. The
important quantity in answering this question is the magnetic flux through one
unit cell of the periodic lattice. It is known \cite{zak,dubrovin} that the
magnetic translation group has finite dimensional representations if the
magnetic field is related to a primitive lattice vector $\vec{e}$ by
\begin{equation}
\vec{B} =2\pi  {1\over e\Omega} {N\over M}\,\vec{e}
\label{brown}
\end{equation}
where $\Omega$ is the area of the unit cell, and $N$ and $M$ are integers.
These representations are constructed by finding an invariant subgroup of
magnetic translation operators; the rationality condition arises because all
members of this invariant subgroup must commute, which places restrictions on
the phase factors in (\ref{nc}). Since we are considering a two-dimensional
system, with the magnetic field perpendicular to the two-dimensional surface,
the condition (\ref{brown}) simplifies to :
\begin{equation}
{eB\Omega\over 2\pi}={N\over M}
\label{brown2}
\end{equation}
The case $M=1$ is special; here the magnetic translations act as
one-dimensional ray representations on the Hilbert space, transforming the
wavefunction with a phase. Consistency of this ray representation gives the
number of states as $N=\frac{eB\Omega}{2\pi}$, which is just Landau's estimate
(\ref{landau}) of the degeneracy of the LLL. But when $\frac{N}{M}$ is
rational, we still have a consistent finite dimensional action of the magnetic
translation group on the wavefunctions. The invariant subgroup consists of
`superlattice' translations, where the superlattice is obtained by enlarging
each length dimension of the unit cell by a factor of $M$. This produces an
enlarged unit cell with effective flux $MN$ on which the magnetic translation
group acts one-dimensionally. Thus the total dimension is $MN$. Finally, if
$\frac{N}{M}$ is irrational, then the magnetic translation group has infinite
dimensional representations.

These results can be mapped directly to the quantization of the abelian Chern-Simons
theory on the torus. The quantum mechanical degrees of freedom, $a$, have a LLL
Lagrangian with magnetic field $eB=\frac{4\pi^2\kappa}{Im\tau}$. The large
gauge transformations (\ref{lgts}) are precisely magnetic translations across a
parallelogram unit cell. The area of the unit cell is $\Omega=Im\tau$, the area
of the torus. Thus 
\begin{equation}
{eB\Omega\over 2\pi}=
\frac{1}{2\pi}\left(\frac{4\pi^2\kappa}{Im\tau}\right)Im\tau =2\pi \kappa
\end{equation}
and the condition for finite dimensional representations of the
action of the large gauge transformations becomes
\begin{equation}
2\pi\kappa={N\over M}
\label{tc}
\end{equation}
If we require states to transform as a one-dimensional ray representation under
large gauge transformations then we must have $2\pi\kappa={\rm integer}$. But
if $2\pi\kappa$ is rational, then we still have a perfectly good quantization,
provided we identify the physical states with irreducible representations of
the global gauge transformations ({\it i.e.}, the magnetic translations). These
states transform according to a finite dimensional irreducible representation
of the global gauge transformations, and any element of a given irreducible
representation may be used to evaluate matrix elements of a gauge invariant
operator, because physical gauge invariant operators commute with the
generators of large gauge transformations. The dimension of the Hilbert space
is $MN$. If $2\pi \kappa$ is irrational, there is still nothing wrong with the
Chern-Simons theory -- it simply means that there are an infinite number of states in
the Hilbert space. These results are consistent with the connection between
abelian Chern-Simons theories and two \diml conformal field theories. Chern-Simons theories with
rational $2\pi\kappa$ correspond to what are known as ``rational CFT's'', which
have a finite number of conformal blocks, and these conformal blocks are in
one-to-one correspondence with the Hilbert space of the Chern-Simons theory
\cite{witten,poly,bos,elitzur}.

\subsection{Canonical Quantization of Nonabelian Chern-Simons Theories}
\label{qnab}

The canonical quantization of the nonabelian Chern-Simons theory with Lagrangian
(\ref{cslag}) is similar in spirit to the abelian case discussed in the
previous Section. There are, however, some interesting new features
\cite{witten,dunne1,bos,elitzur,labastida}. As before, we specialize to the
case where space-time has the form ${\bf R}\times \Sigma$, where $\Sigma$ is a
torus. With $\Sigma=T^2$, the spatial manifold has two noncontractible loops
and these provide gauge invariant holonomies. The problem reduces to an
effective quantum mechanics problem for these holonomies. Just as in the
abelian case, it is also possible to treat holonomies due to sources (which
carry a representation of the gauge algebra), and to consider spatial manifolds
with boundaries. These two approaches lead to deep connections with two-\diml
conformal field theories, which are beyond the scope of these lectures -- the
interested reader is referred to \cite{witten,bos,elitzur,labastida} for
details.

We begin as in the abelian case by choosing a functional coherent state
representation for the holomorphic wavefunctionals $\Psi=\Psi[A]$, where
$A=\frac{1}{2}(A_1+iA_2)$. The coherent state inner product is
\begin{equation}
<\Psi|\Phi>=\int {\cal D}A {\cal D}A^* e^{4\kappa\int \tr(AA^*)}
(\Psi[A])^* \Phi[A]
\label{nabinner}
\end{equation}
Note that with our Lie algebra conventions (see Section \ref{nabcs})
$\tr(AA^*)=-\frac{1}{2}A^a(A^a)^*$. Physical states are annihilated by the
Gauss law generator $F_{12}=-2i F_{z\bar{z}}$. Remarkably, we can solve this
constraint explicitly using the properties of the Wess-Zumino-Witten (WZW)
functionals:
\begin{equation}
S^\pm[g]=\frac{1}{2\pi}\int_\Sigma 
\tr(g^{-1}\partial_z g g^{-1}\partial_{\bar{z}}g)
\pm \frac{i}{12\pi}\int_{(3)} \epsilon^{\mu\nu\rho}\tr(g^{-1}\partial_\mu g
g^{-1}\partial_\nu g g^{-1}\partial_\rho g)
\label{wzw}
\end{equation}
where in the second term the integral is over a three \diml manifold with a two
\diml boundary equal to the two \diml space $\Sigma$.
\vskip 1cm
{\bf Exercise 3.5.1 :} Show that the WZW functionals (\ref{wzw}) have the
fundamental variations
\begin{equation}
\delta S^\pm[g]=\left\{\matrix{-\frac{1}{\pi}\int\tr(g^{-1}\delta g\partial_z
(g^{-1}\partial_{\bar{z}}g)]\cr -\frac{1}{\pi}\int\tr(g^{-1}\delta
g\partial_{\bar{z}}(g^{-1}
\partial_z g)]}\right.
\end{equation}
\vskip .5cm
Consider first of all quantization on the spatial manifold $\Sigma={\bf R}^2$.
To solve the Gauss law constraint we express the holomorphic field $A$, using
Yang's representation \cite{yang}, as
\begin{equation}
A=-\partial_{\bar{z}}U\,U^{-1}, \qquad\qquad U\in {\cal G}^{\bf C}
\label{yang}
\end{equation}
This is the nonabelian analogue of the complexified
longitudinal-transverse decomposition (\ref{hodge}) $A=\partial_{\bar{z}}\chi$
for the abelian theory on the plane. $U$ belongs to the complexification of the
gauge group, which, roughly speaking, is the exponentiation of the gauge
algebra, with complex parameters.

With $A$ parametrized in this manner, the Gauss law constraint
$F_{z\bar{z}}\Psi=0$ is solved by the functional
\begin{equation}
\Psi_0[A]=e^{-4\pi\kappa S^-[U]}
\label{nabgs}
\end{equation}
To verify this, note that the results of Exercise 3.5.1 imply that
\begin{equation}
\delta\Psi_0=4\kappa\left[\int\tr(\delta A\, \partial_zU U^{-1})\right]\Psi_0
\label{vari}
\end{equation}
{}From the canonical commutation relations (\ref{purecrs}), the field
$A^a_z=\frac{1}{2}(A_1^a-iA_2^a)$ acts on a wavefunctional $\Psi[A]$ as a
functional derivative operator
\begin{equation}
A_z^a=\frac{1}{2\kappa}{\delta\over \delta A^a}
\end{equation}
Thus, acting on the state $\Psi_0[A]$ in (\ref{nabgs}):
\begin{equation}
A_z^a\Psi_0[A]=-(\partial_zUU^{-1})^a\Psi_0[A]
\label{act}
\end{equation}
Since $A^a_{\bar{z}}$ acts on $\Psi_0$ by multiplication, it immediately follows
that $F_{z\bar{z}}\Psi_0[A]=0$, as required.

The physical state (\ref{nabgs}) transforms with a cocycle phase factor under a
gauge transformation. We could determine this cocycle from the variation
(\ref{change}) of the nonabelian Lagrangian \cite{dunne1}. But a more direct
way here is to use the fundamental Polyakov-Wiegmann transformation property
\cite{polyakov} of the WZW functionals:
\begin{equation}
S[g_1 g_2]=S[g_1]+S[g_2]+\frac{1}{\pi}\int\tr(g_1^{-1}\partial_z g_1
\partial_{\bar{z}} g_2 g_2^{-1})
\label{pw}
\end{equation}
With the representation $A=-\partial_{\bar{z}}U\,U^{-1}$ of the holomorphic
field $A$, the gauge transformation $A\to
A^g=g^{-1}Ag+g^{-1}\partial_{\bar{z}}g$ is implemented by $U\to g^{-1}U$, with
$g$ in the gauge group. Then
\begin{eqnarray}
\Psi_0[A^g]&=&e^{-4\pi\kappa S^-[g^{-1}U]}\nonumber\\
&=&e^{-4\pi\kappa S^+[g]-4\kappa\int\tr(A\partial_z g g^{-1})}\Psi_0[A]
\label{cocycle}
\end{eqnarray}
\vskip 1cm
{\bf Exercise 3.5.2 :} Check that the transformation law (\ref{cocycle}) is
consistent under composition, and that it combines properly with the measure
factor to make the coherent state inner product (\ref{nabinner}) gauge
invariant. 
\vskip 1cm

Furthermore, note that the WZW factors in (\ref{nabgs}) and
(\ref{cocycle}) are only well defined provided $4\pi\kappa={\rm integer}$. This
is the origin of the discreteness condition (\ref{dirac}) on the Chern-Simons
coefficient in canonical quantization.

This describes the quantum pure Chern-Simons theory with spatial manifold being the open
plane ${\bf R}^2$. There is a unique physical state (\ref{nabgs}). To make
things more interesting we can introduce sources, boundaries, or handles on the
spatial surface. As in the abelian case, here we just consider the effect of
higher genus spatial surfaces, and for simplicity we concentrate on the torus.
Then the nonabelian analogue of the abelian Hodge decomposition
(\ref{torushodge}) is \cite{bos,elitzur}
\begin{equation}
A=-\partial_{\bar{z}}U U^{-1}+{i\pi\over Im\tau}Ua U^{-1}
\label{nabtorushodge}
\end{equation}
where $U\in {\cal G}^{\bf C}$, and $a$ can be chosen to be in the Cartan
subalgebra of the gauge Lie algebra. This is the nonabelian generalization of
the abelian torus Hodge decomposition (\ref{torushodge}). To motivate this
decomposition, we note that when $U\in {\cal G}$ [{\it not} ${\cal G}^{\bf
C}$!], this is the most general pure gauge (flat connection) on the torus. The
$a$ degrees of freedom represent the nontrivial content of $A$ that cannot be
gauged away, due to the noncontractible loops on the spatial manifold. By a
gauge transformation, $a$ can be taken in the Cartan subalgebra (indeed, there
is further redundancy due to the action of Weyl reflections on the Cartan
subalgebra). Then, extending $U$ from ${\cal G}$ to ${\cal G}^{\bf C}$, the
representation (\ref{nabtorushodge}) spans out to cover all connections, just
as in Yang's representation (\ref{yang}) on ${\bf R}^2$.

Combining the representation (\ref{nabtorushodge}) with the transformation law
(\ref{cocycle}), we see that the physical state wavefunctionals on the torus
are
\begin{equation}
\Psi[A]=e^{-4\pi\kappa S^-[U]+{4\pi i\kappa\over Im\tau} \int\tr(a U^{-1}
\partial_z U)}\psi(a)
\end{equation}
In the inner product (\ref{nabinner}), we can change field variables from $A$
to $U$ and $a$. But this introduces nontrivial Jacobian factors
\cite{bos,elitzur}. The corresponding determinant is another Polyakov-Weigmann
factor \cite{kup}, with a coefficient $c$ arising from the adjoint
representation normalization ($c$ is called the dual Coxeter number of the
gauge algebra, and for $SU(N)$ it is $N$). The remaining functional integral
over the gauge invariant combination $U^\dagger U$ may be performed (it is the
generating functional of the gauged WZW model on the torus \cite{kup}). The
final result is an effective quantum mechanical model with coherent state inner
product
\begin{equation}
<\Psi|\Phi>=\int da da^* e^{\frac{\pi}{Im\tau}(4\pi\kappa+c) \tr(aa^*)}
(\psi(a))^* \phi(a)
\label{reduced}
\end{equation}
This looks like the abelian case, except for the shift of the Chern-Simons coefficient
$\kappa$ by $4\pi\kappa\to 4\pi\kappa+c$. In fact, we can represent the quantum
mechanical Cartan subalgebra degrees of freedom as $r$-component vectors
$\vec{a}$, where $r$ is the rank of the gauge algebra. Then large gauge
transformations act on these vectors as $\vec{a}\to \vec{a}+\vec{m}+\tau
\vec{n}$, where $\vec{m}$ and $\vec{n}$ belong to the root lattice
$\Lambda_{\rm R}$ of the gauge algebra. The wavefunction with the correct
transformation properties under these large gauge transformation shifts is a
generalized theta function, which is labelled by an element $\vec{\lambda}$ of
the weight lattice $\Lambda_{\rm W}$ of the algebra. These are identified under
translations by root vectors, and also under Weyl reflections. Thus the
physical Hilbert space of the nonabelian Chern-Simons theory on the torus corresponds to
\begin{equation}
{\Lambda_{\rm W}\over W\times (4\pi\kappa+c)\Lambda_{\rm R}}
\end{equation}
This parametrization of states is a familiar construction in the theory of
Kac-Moody algebras and conformal field theories.

\subsection{Chern-Simons Theories with Boundary}
\label{csb}

We conclude this review of basic facts about the canonical structure of Chern-Simons
theories by commenting briefly on the manifestation of boundary degrees of
freedom in Chern-Simons theories defined on spatial manifolds which have a boundary. We
have seen in the previous Sections that the canonical quantization of pure Chern-Simons
theory on the space-time
$\Sigma\times {\bf R}$, where $\Sigma$ is a compact Riemann surface, leads to a
Hilbert space that is in one-to-one correspondence with the conformal blocks of
a conformal field theory defined on $\Sigma$. But there is another important
connection between Chern-Simons theories and CFT -- namely, if the spatial manifold
$\Sigma$ has a boundary $\partial\Sigma$, then the Hilbert space of the Chern-Simons
theory is infinite dimensional, and provides a representation of the chiral
current algebra of the CFT defined on $\partial\Sigma\times {\bf R}$
\cite{witten,bos,elitzur}.

The source of these boundary effects is the fact that when we checked the
variation of the Chern-Simons action in (\ref{var}) we dropped a surface term. Retaining
the surface term, the variation of the Chern-Simons action splits naturally into a bulk
and a surface piece \cite{witten,elitzur}:
\begin{equation}
\delta S_{\rm CS}=\kappa\int d^3x \epsilon^{\mu\nu\rho}\tr(\delta A_\mu
F_{\nu\rho})+ \kappa\int d^3x\partial_\nu\left[\epsilon^{\mu\nu\rho}\tr(A_\mu
\delta A_\rho)\right]
\label{surfvar}
\end{equation}
The boundary conditions must be such that $\int_{\rm bndy}\tr(A\,\delta A)=0$.
When it is the spatial manifold $\Sigma$ that has a boundary $\partial\Sigma$,
we can impose the boundary condition that $A_0=0$. The remaining local symmetry
corresponds to gauge transformations that reduce to the identity on
$\partial\Sigma$, while the time independent gauge transformations on the
boundary are global gauge transformations.

With this boundary condition we can write
\begin{equation}
S_{\rm CS}=-\kappa\int_{\Sigma\times{\bf R}}
d^3x\,\epsilon^{ij}\tr(A_i\dot{A}_j) +\kappa\int_{\Sigma\times{\bf R}}
d^3x\,\epsilon^{ij}\tr(A_0 F_{ij})
\label{bred}
\end{equation}
Variation with respect to the Lagrange multiplier field $A_0$ imposes the
constraint $F_{ij}=0$, which has as its solution the pure gauges
$A_i=g^{-1}\partial_i g$. Then it follows that the Chern-Simons action becomes
\begin{equation}
S=-\kappa\int_{\partial\Sigma\times{\bf R}}
d\theta\,dt\,\tr(g^{-1}\partial_\theta g g^{-1}\partial_0 g)+ \frac{\kappa}{3}
\int_{\Sigma\times{\bf R}} \epsilon^{\mu\nu\rho} \tr \left( g^{-1}\partial_\mu
g g^{-1}\partial_\nu g g^{-1}\partial_\rho g \right)
\end{equation}
This is the chiral WZW action. The quantization of this system leads to a
chiral current algerba of the gauge group, with the boundary values of the
gauge field $A_\theta=g^{-1}\partial_\theta g$ being identified with the chiral
Kac-Moody currents. This relation gives another important connection between
Chern-Simons theories (here, defined on a manifold with a spatial boundary) and
conformal field theories \cite{witten,elitzur}.

Boundary effects also play an important role in the theory of the quantum Hall
effect \cite{wen,stone}, where there are gapless edge excitations which are
crucial for explaining the conduction properties of a quantum Hall liquid.
Consider the variation of the abelian Chern-Simons action
\begin{equation}
\delta\left(\int d^3x \epsilon^{\mu\nu\rho} A_\mu\partial_\nu A_\rho\right)
=2\int d^3x \epsilon^{\mu\nu\rho} \delta A_\mu \partial_\nu A_\rho+ \int d^3x
\epsilon^{\mu\nu\rho} \partial_\nu \left(A_\mu \delta A_\rho\right)
\end{equation}
For an infinitesimal gauge variation, $\delta A_\mu=\partial_\mu \lambda$, this
becomes a purely surface term
\begin{equation}
\delta\left(\int d^3x \epsilon^{\mu\nu\rho} A_\mu\partial_\nu A_\rho\right)
=\int d^3x \epsilon^{\mu\nu\rho} \partial_\mu \left(\lambda\partial_\nu
A_\rho\right)
\end{equation}
For a space-time ${\bf D}\times {\bf R}$, where ${\bf D}$ is a disc with
boundary $S^1$
\begin{equation}
\delta\left(\int d^3x \epsilon^{\mu\nu\rho} A_\mu\partial_\nu A_\rho\right)
=\int_{S^1\times{\bf R}} \lambda(\partial_0 A_\theta-\partial_\theta A_0)
\label{anom}
\end{equation}
Thus, the Chern-Simons action is not gauge invariant. Another way to say this is that
the current $J^\mu=\frac{\delta S_{\rm CS}}{\delta A_\mu} =
\frac{\kappa}{2}\epsilon^{\mu\nu\rho}F_{\nu\rho}$ is conserved within the bulk,
but not on the boundary. For a disc-like spatial surface, this noninvariance
leads to an accumulation of charge density at the boundary at a rate given by
the radial current :
\begin{equation}
J_{\rm r}=\kappa E_{\theta}
\end{equation}
where $E_\theta$ is the tangential electric field at the boundary. However, we
recognize this noninvariance as exactly that of a $1+1$ \diml Weyl fermion
theory defined on the boundary $S^1\times{\bf R}$. Due to the $1+1$ \diml
chiral anomaly, an electric field (which must of course point along the
boundary) leads to the anomalous creation of charge at the rate (with $n$
flavours of fermions) :
\begin{equation}
\frac{\partial}{\partial t} Q=\frac{n}{2\pi} E
\end{equation}
Therefore, when $2\pi \kappa$ is an integer [recall the abelian
discreteness condition (\ref{tc})] the noninvariance of the Chern-Simons theory matches
precisely the noninvariance of the anomalous boundary chiral fermion theory.
This corresponds to a flow of charge from the bulk to the edge and vice versa.
This gives a beautiful picture of a quantum Hall droplet, with integer filling
fraction, as an actual physical realization of the chiral anomaly phenomenon.
Indeed, when
$2\pi\kappa=n$, we can view the Hall droplet as an actual coordinate space
realization of the Dirac sea of the edge fermions \cite{stone2}. This also
provides a simple effective description of the integer quantum Hall effect as a
quantized flow of charge onto the edge of the Hall droplet. For the fractional
quantum Hall effect we need more sophisticated treatments on the edge, such as
bosonization of the $1+1$ \diml chiral fermion edge theory in terms of chiral
boson fields
\cite{wen}, or representations of $W_{1+\infty}$, the quantum algebra of area
preserving diffeomorphisms associated with the incompressibility of the quantum
Hall droplet \cite{ctz}.

\section{Chern-Simons Vortices}
\label{vortex}

Chern-Simons models acquire dynamics via coupling to other fields. In this Section we consider the dynamical consequences of coupling Chern-Simons fields to scalar fields that have either relativistic or nonrelativistic dynamics. These theories have vortex solutions, similar to (in some respects) but different from (in other respects) familiar vortex models such as arise in Landau-Ginzburg theory or the Abelian Higgs model. The notion of Bogomol'nyi self-duality is ubiquitous, with some interesting new features owing to the Chern-Simons charge-flux relation $\rho=\kappa B$. 

\subsection{Abelian-Higgs Model and Abrikosov-Nielsen-Olesen Vortices}
\label{abhiggs}

I begin by reviewing briefly the Abelian-Higgs model in $2+1$ \dim. This model
describes a charged scalar field interacting with a $U(1)$ gauge field, and
exhibits vortex solutions carrying magnetic flux, but no electric charge. These
vortex solutions are important in the Landau-Ginzburg theory of
superconductivity because the static energy functional [see (\ref{ahenergy})
below] for the relativistic Abelian-Higgs model coincides with the
nonrelativistic Landau-Ginzburg free energy in the theory of type II
superconductors, for which vortex solutions were first studied by Abrikosov
\cite{abrikosov}.

\begin{figure}[htb]
\vspace{-2in}
\centering{\epsfig{file=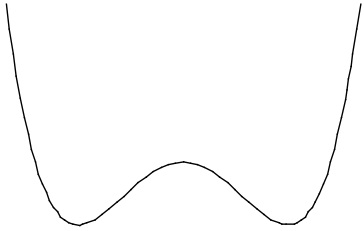,width=4in}}
\vspace{-1.75in}
\caption{The self-dual quartic potential $\frac{\lambda}{4}\left(|\phi|^2-v^2\right)^2$ for the Abelian-Higgs model. The vacuum manifold is $|\phi|=v$.}
\label{qqq}
\end{figure}

Consider the Abelian-Higgs Lagrangian \cite{nielsen}
\begin{equation}
{\cal L}_{\rm AH}=-\frac{1}{4}F_{\mu\nu}F^{\mu\nu}+|D_\mu \phi |^2-
\frac{\lambda}{4}\left(|\phi|^2-v^2\right)^2
\label{ahlag}
\end{equation}
where the covariant derivative is $D_\mu\phi=\partial_\mu\phi +ie A_\mu\phi$,
and the quartic potential has the standard symmetry breaking form as shown in Figure \ref{qqq}.

The static energy functional of the Abelian-Higgs model is
\begin{equation}
{\cal E}_{\rm AH}=\int d^2x\left[\frac{1}{2}B^2 + |\vec{D}\phi|^2 +
\frac{\lambda}{4} \left(|\phi|^2-v^2\right)^2\right]
\label{ahenergy}
\end{equation}
where $B=F_{12}$. The potential minimum has constant solutions
$\phi=e^{i\alpha} v$, where $\alpha$ is a real phase. Thus the vacuum manifold
is isomorphic to the circle $S^1$. Furthermore, any finite energy solution must
have $\phi(\vec{x})$ tending to an element of this vacuum manifold at infinity.
Therefore, finite energy solutions are classified by their winding number or
vorticity $N$, which counts the number of times the phase of $\phi$ winds
around the circle at spatial infinity:
\begin{equation}
\phi(\vec{x})|_{|\vec{x}|=\infty}=v\, e^{iN\theta}
\label{vort}
\end{equation}

The vorticity is also related to the magnetic flux because finite energy
solutions also require $|\vec{D}\phi|\to 0$ as $|\vec{x}|\to\infty$. This
implies that
\begin{equation}
e A_i\sim -i\partial_i \ln\phi \sim N\partial_i\theta \qquad {\rm as} \quad
|\vec{x}|\to \infty
\end{equation}
Therefore, the dimensionless magnetic flux is
\begin{equation}
\Phi=e\int d^2x B=e\oint_{|\vec{x}|=\infty}A_i dx^i = 2\pi N
\label{flux}
\end{equation}

A brute-force approach to vortex solutions would be to make, for example in the
1-vortex case, a radial ansatz:
\begin{equation}
\phi(\vec{x})=f(r)e^{i\theta}, \qquad\qquad \vec{A}(\vec{x})=a(r)\hat{\theta}
\label{radial}
\end{equation}
The field equations then reduce to coupled nonlinear ordinary differential
equations for $f(r)$ and $a(r)$. One can seek numerical solutions with the
appropriate boundary conditions: $f(r)\to v$ and $a(r)\to \frac{1}{er}$ as
$r\to \infty$; and $f(r)\to 0$ and $a(r)\to 0$ as $r\to 0$. No exact solutions
are known, but approximate solutions can be found numerically. The solutions
are localized vortices in the sense that the fields approach their asymptotic
vacuum values exponentially, with characteristic decay lengths set by the mass
scales of the theory. Note that $\lambda$, $e^2$ and $v^2$ each has dimensions
of mass; and the Lagrangian (\ref{ahlag}) has a Higgs phase with a massive
gauge field of mass $m_g=\sqrt{2} ev$, together with a massive real scalar
field of mass $m_s=\sqrt{\lambda}v$. In general, these two mass scales are
independent, but the Abelian-Higgs model displays very different behavior
depending on the relative magnitude of these two mass scales. Numerically, it
has been shown that two vortices (or two antivortices) repel if $m_s > m_g$, but
attract if $m_s < m_g$. When the masses are equal
\begin{equation}
m_s=m_g
\label{ahmassdeg}
\end{equation}
then the forces betwen vortices vanish and it is possible to find stable static
{\it multivortex} configurations. When translated back into the Landau-Ginzburg
model for superconductivity, this critical point, $m_s=m_g$, corresponds to the
boundary between type-I and type-II superconductivity. In terms of the
Abelian-Higgs model (\ref{ahlag}), this critical point is known as the
Bogomol'nyi \cite{bog} self-dual point where
\begin{equation}
\lambda=2e^2
\label{ahbog}
\end{equation}
With this relation between the charge $e$ and the potential strength $\lambda$,
special things happen.

To proceed, we need a fundamental identity -- one that will appear many times
throughout our study of vortex solutions in planar gauge theories.
\begin{equation}
|\vec{D}\phi|^2=|(D_1\pm i D_2)\phi|^2\mp e B|\phi|^2\pm
\epsilon^{ij}\partial_i J_j
\label{bochner}
\end{equation}
where $J_j=\frac{1}{2i}[\phi^* D_j \phi-\phi (D_j\phi)^*]$.
Using this identity, the energy functional (\ref{ahenergy}) becomes \cite{bog}
\begin{equation}
{\cal E}_{\rm AH}=\int d^2x\left[\frac{1}{2} \left(B\mp
e(|\phi|^2-v^2)\right)^2 + |D_\pm \phi|^2+ (\frac{\lambda}{4}-\frac{e^2}{2})
\left(|\phi|^2-v^2\right)^2 \mp ev^2 B\right]
\label{ahfactor}
\end{equation}
where $D_\pm\equiv(D_1\pm i D_2)$, and we have dropped a surface term. At the
self-dual point (\ref{ahbog}) the potential terms cancel, and we see that the
energy is bounded below by a multiple of the magnitude of the magnetic flux
(for positive flux we choose the lower signs, and for negative flux we choose
the upper signs):
\begin{equation}
{\cal E}_{\rm AH}\geq v^2 |\Phi |
\label{ahbound}
\end{equation}
This bound is saturated by fields satisfying the first-order Bogomol'nyi
self-duality equations \cite{bog}:
\begin{eqnarray}
D_\pm \phi&=&0\cr
B&=&\pm e(|\phi|^2-v^2)
\label{ahsd}
\end{eqnarray}

The self-dual point (\ref{ahbog}) is also the point at which the $2+1$ \diml
Abelian-Higgs model (\ref{ahlag}) can be extended to an $N=2$ supersymmetric
(SUSY) model \cite{llm,edelstein}. That is, first construct an $N=1$ SUSY
Lagrangian of which (\ref{ahlag}) is the bosonic part. This SUSY can then be
extended to $N=2$ SUSY only when the $\phi$ potential is of the form in
(\ref{ahlag}) and the self-duality condition (\ref{ahbog}) is satisfied. This
is clearly related to the mass degeneracy condition (\ref{ahmassdeg}) because
for $N=2$ SUSY we need {\it pairs} of bosonic particles with equal masses (in
fact, the extension to $N=2$ SUSY requires an additional neutral scalar field
to pair with the gauge field $A_\mu$). This feature of $N=2$ SUSY corresponding
to the self-dual point is a generic property of self-dual models
\cite{olive,hlousek}, and we will see it again in our study of Chern-Simons vortices.

The self-duality equations (\ref{ahsd}) are not solvable, or even integrable,
but a great deal is known about the solutions. To bring them to a  more
manageable form, we decompose the scalar field $\phi$ into its phase and
magnitude:
\begin{equation}
\phi=e^{i\omega} \, \rho^{\frac{1}{2}}
\label{phidecomp}
\end{equation}
Then the first of the self-duality equations (\ref{ahsd}) determines the gauge
field
\begin{equation}
e A_i=-\partial_i\omega \mp \frac{1}{2}\epsilon_{ij}\partial_j \ln \rho
\label{sdgauge}
\end{equation}
everywhere away from the zeros of the scalar field. The second self-duality
equation in (\ref{ahsd}) then reduces to a nonlinear elliptic equation for the
scalar field density $\rho$:
\begin{equation}
\nabla^2 \ln \rho=2 e^2\left( \rho - v^2\right)
\label{ahelliptic}
\end{equation}
No exact solutions are known for this equation, even when reduced to an
ordinary differential equation by the condition of radial symmetry. However, it
is easy to find (numerically) vortex-like solutions with $\phi=f(r)e^{\pm i N
\theta}$ where $f(r)$ satisfies
\begin{equation}
\frac{1}{r}\frac{d}{dr}\left(r \frac{d}{dr} f^2(r)\right)=2 e^2 (f^2-v^2)
\label{ahradial}
\end{equation}

Many interesting theorems have been proved concerning the general solutions to
the self-dual Abelian-Higgs equations (\ref{ahsd}). These are paraphrased
below. Readers interested in all the fine-print should consult \cite{jaffe} and
the original papers.

\bigskip

{\it Existence and Uniqueness}: Let $(\phi,\vec{A})$ be a smooth finite energy
solution to the Abelian-Higgs self-duality equations (\ref{ahsd}). Then

(i) $\phi$ has a finite number of zeros $z_1$, $\dots$ , $z_m$;

(ii) around each zero, $\phi\sim(z-z_k)^{n_k}h_k(z)$, where $h_k(z)$ is smooth
and $h_k(z_k)\neq 0$;

(iii) the vorticity is given by the net multiplicity of zeros: $N=\sum_{k=1}^m
n_k$;

(iv) given any set of zeros, $z_1$, $\dots$, $z_m$, the solution is unique, up
to gauge equivalence;

(v) $|\phi|< v$ on ${\bf R}^2$.

\bigskip

Furthermore, it has been shown that all finite energy solutions to the full
second-order static equations of motion are solutions to the first-order
self-duality equations. Thus, the solutions described in the above theorem
cover {\it all} finite energy static solutions.

These results mean that the moduli space of static multivortex solutions is
$2N$ dimensional, and these $2N$ parameters can be associated with the
locations of the zeros of the Higgs field $\phi$. This counting is confirmed by
an index-theorem fluctuation analysis \cite{weinberg}. We shall return to this
moduli space later in Section \ref{dynamics} when we discuss the dynamics of
vortices.

To conclude this review of the Abelian-Higgs model I mention that this model
has also been studied on spatial manifolds that are compact Riemann surfaces.
This is of interest for making comparisons with numerical simulations, which
are necessarily finite, and also for studying the thermodynamics of vortices
\cite{manton}. The main new feature is that there is an upper limit, known as
Bradlow's bound \cite{bradlow}, on the vorticity for a given area of the
surface. The appearance of such a bound is easy to see by integrating the
second of the self-duality equations (\ref{ahsd}) over the surface (assuming
positive flux, we take the lower signs):
\begin{equation}
\int d^2 x eB= e^2v^2\int d^2 x -e^2 \int d^2x |\phi|^2
\label{brad}
\end{equation}
Since $\int d^2 x eB=2\pi N$, and $\int d^2x |\phi|^2$ is positive, this
implies that
\begin{equation}
N\leq {e^2 v^2\over 2\pi} \, {\rm area}
\label{bradlow}
\end{equation}
[In the mathematics literature $v^2$ and $\lambda$ are usually scaled to 1, so
that the self-dual value of $e^2$ is $\frac{1}{2}$, in which case the bound
reads: $4\pi N\leq {\rm area}$.] A similar bound applies when considering the
Abelian-Higgs vortex solutions with periodic `t Hooft boundary conditions
defined on a unit cell of finite area \cite{yisong1}.

\subsection{Relativistic Chern-Simons Vortices}
\label{rcsv}

A natural generalization of the Abelian-Higgs model of the previous section is
to consider the effect of taking the gauge field to be governed by a Chern-Simons
Lagrangian rather than a Maxwell Lagrangian. The name ``relativistic'' Chern-Simons
vortices comes from the fact that a Chern-Simons gauge field inherits its dynamics from
the matter fields to which it is coupled, and here it is coupled to a
relativistic scalar field -- later we shall consider vortices arising from a
Chern-Simons gauge field coupled to matter fields with nonrelativistic dynamics.
Numerous studies were made of vortex solutions in models with Chern-Simons and/or
Maxwell terms, with symmetry breaking scalar field potentials
\cite{devega,khare1}. However, no analogue of the Bogomol'nyi self-dual
structure of the Abelian-Higgs model was found until a particular sixth-order
scalar potential was chosen in a model with a pure Chern-Simons term \cite{hong,jw}.

Consider the Lagrangian
\begin{equation}
{\cal L}_{\rm RCS}=\frac{\kappa}{2}\epsilon^{\mu\nu\rho}A_\mu \partial_\nu
A_\rho + |D_\mu \phi |^2-V(|\phi|)
\label{rcslag}
\end{equation}
where $V(|\phi|)$ is the scalar field potential, to be specified below. The
associated energy functional is
\begin{equation}
{\cal E}_{\rm RCS}=\int d^2x\left[|D_0\phi|^2+|\vec{D}\phi|^2+V(|\phi|)\right]
\label{rcsenergy}
\end{equation}
Before looking for self-dual vortices we note a fundamental difference between
vortices in a Chern-Simons model and those in the Abelian-Higgs model, where the gauge
field is governed by a Maxwell term. The Abelian-Higgs vortices carry magnetic
flux but are electrically neutral. In contrast, in a Chern-Simons model the Chern-Simons Gauss
law constraint relates the magnetic field $B$ to the conserved $U(1)$ charge
density as
\begin{equation}
B=\frac{1}{\kappa}J_0= \frac{i}{\kappa}\left(\phi^* D_0\phi-
(D_0\phi)^*\phi\right)
\label{rcsgauss}
\end{equation}
Thus, if there is magnetic flux there is also electric charge:
\begin{equation}
Q=\int d^2x J^0=\kappa\int d^2x B =\kappa \Phi
\label{csflux}
\end{equation}
So solutions of vorticity $N$ necessarily carry both magnetic flux $\Phi$ and
electric charge $Q$. They are therefore excellent candidates for {\it anyons}.

To uncover the Bogomol'nyi-style self-duality, we use the factorization
identity (\ref{bochner}), together with the Chern-Simons Gauss law constraint
(\ref{rcsgauss}), to express the energy functional as
\begin{equation}
{\cal E}_{\rm RCS}=\int d^2x \left[|D_0\phi\pm \frac{i}{\kappa}
(|\phi|^2-v^2)\phi|^2+|D_\pm\phi|^2+V(|\phi|)-
\frac{1}{\kappa^2}|\phi|^2(|\phi|^2-v^2)^2 \mp v^2 B\right]
\label{rcsfactor}
\end{equation}
Thus, if the potential is chosen to take the self-dual form
\begin{equation}
V(|\phi|)=\frac{1}{\kappa^2}|\phi|^2(|\phi|^2-v^2)^2
\label{rcspot}
\end{equation}
then the energy is bounded below [choosing signs depending on the sign of the
flux]
\begin{equation}
{\cal E}_{\rm RCS}\geq v^2 |\Phi|
\label{bound}
\end{equation}

\begin{figure}[htb]
\vspace{-4cm}
\centering{\epsfig{file=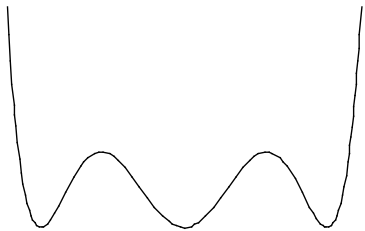,width=4in}}
\vspace{-5cm}
\caption{The self-dual potential $\frac{1}{\kappa^2}|\phi|^2(|\phi|^2-v^2)^2$ for the relativistic self-dual Chern-Simons system. Note the existence of two degenerate vacua: $\phi=0$ and $|\phi|=v$.}
\label{sss}
\end{figure}

The bound (\ref{bound}) is saturated by solutions to the first-order equations
\begin{equation}
D_\pm \phi=0,\qquad\qquad D_0\phi=\mp \frac{i}{\kappa} (|\phi|^2-v^2)\phi
\end{equation}
which, when combined with the Gauss law constraint (\ref{rcsgauss}), become the
self-duality equations:
\begin{eqnarray}
D_\pm \phi&=&0\cr
B&=&\pm \frac{2}{\kappa^2} |\phi|^2 (|\phi|^2-v^2)
\label{rcssd}
\end{eqnarray}

These are clearly very similar to the self-duality equations (\ref{ahsd})
obtained in the Abelian-Higgs model. However, there are some significant
differences. Before discussing the properties of solutions, a few comments are
in order. First, as is illustrated in Figure \ref{sss}, the self-dual potential (\ref{rcspot}) is sixth-order, rather than the more commonly considered case of fourth-order. 

Such a potential is still power-counting renormalizable in $2+1$ dimensions. Furthermore, the potential is such that the minima at $\phi=0$ and at $|\phi|^2=v^2$ are degenerate. Correspondingly, there are domain wall solutions that interpolate between the two vacua \cite{jlw}. In the Higgs vacuum, the Chern-Simons-Higgs mechanism leads to a massive gauge field [recall (\ref{cshprop})] and a massive real scalar field. With the particular form of the self-dual potential (\ref{rcspot}) these masses are {\it equal}:
\begin{equation}
m_s={2v^2\over \kappa}=m_g
\label{rcsmass}
\end{equation}
Just as in the Abelian-Higgs case, the relativistic Chern-Simons vortex model has an
associated $N=2$ SUSY, in the sense that the Lagrangian (\ref{rcslag}), with
scalar potential (\ref{rcspot}), is the bosonic part of a SUSY model with
extended $N=2$ SUSY \cite{llw}.
\vskip 1cm
{\bf Exercise 4.2.1 :} The $N=2$ SUSY extension of the relativistic Chern-Simons vortex
system (\ref{rcslag}) has Lagrangian
\begin{equation}
{\cal L}_{\rm SUSY}=\frac{\kappa}{2}\epsilon^{\mu\nu\rho}A_\mu \partial_\nu
A_\rho + |D_\mu \phi |^2+i\bar{\psi}\Dslash \psi
-\frac{1}{\kappa^2}|\phi|^2(|\phi|^2-v^2)^2+
\frac{1}{\kappa}(3|\phi|^2-v^2)\bar{\psi}\psi
\label{susyrcs}
\end{equation}
Show that there are pairs of bosonic fields degenerate with pairs of fermionic
fields, in both the symmetric and asymmetric phases.
\vskip 1cm

To investigate vortex solutions, we decompose the scalar field $\phi$ into its
magnitude $\sqrt{\rho}$ and phase $\omega$ as in (\ref{phidecomp}). The gauge
field is once again determined by the first self-duality equation to be
$A_i=-\partial_i\omega \mp \frac{1}{2}\epsilon_{ij}\partial_j \ln \rho$, as in
(\ref{sdgauge}), away from the zeros of the scalar field. The second
self-duality equation then reduces to a nonlinear elliptic equation:
\begin{equation}
\nabla^2 \ln |\phi|^2=\frac{4}{\kappa^2}|\phi|^2(|\phi|^2-v^2)
\label{rcselliptic}
\end{equation}

Just as in the Abelian-Higgs case (\ref{ahelliptic}), this equation is neither
solvable nor integrable. However, numerical solutions can be found using a
radial vortex-like ansatz. A significant difference from the Abelian-Higgs case
is that while the Abelian-Higgs vortices have magnetic flux strings located at
the zeros of the scalar field $\phi$, in the Chern-Simons case we see from (\ref{rcssd})
that the magnetic field {\it vanishes} at the zeros of $\phi$. The magnetic
field actually forms rings centred on the zeros of $\phi$.

Numerical studies lead to two different types of solutions, distinguished by
their behaviour at spatial infinity:

\bigskip

{\bf 1.} {\it Topological solutions}: $|\phi|\to v$ as $|\vec{x}|\to\infty$.

\bigskip

{\bf 2.} {\it Nontopological solutions}: $|\phi|\to 0$ as $|\vec{x}|\to\infty$.

\bigskip

In case 1, the solutions are topologically stable because they interpolate
between the unbroken vacuum $\phi=0$ at the origin and the broken vacuum
$|\phi|=v$ at infinity. For these solutions, existence has been proven using
similar complex analytic and variational techniques to those used for the
Ablian-Higgs model \cite{wang}.

\bigskip

{\it Existence}: There exist smooth finite energy solutions $(\phi,\vec{A})$ to
the relativistic Chern-Simons self-duality equations (\ref{rcssd}) such that

(i) $|\phi|\to v$ as $|\vec{x}|\to\infty$;

(ii) $\phi$ has a finite number of zeros $z_1$, $\dots$, $z_m$;

(iii) around each zero, $\phi\sim(z-z_k)^{n_k}h_k(z)$, where $h_k(z)$ is smooth
and $h_k(z_k)\neq 0$;

(iv) the vorticity is given by the net multiplicity of zeros: $N=\sum_{k=1}^m
n_k$.

\bigskip

Interestingly, the uniqueness of these solutions has not been rigorously
proved. Nor has the equivalence of these self-dual solutions to all finite
energy solutions of the full second-order equations of motion.

The topological vortex solutions have flux, charge, energy:
\begin{equation}
\Phi=2\pi N, \qquad Q=\kappa \Phi, \qquad {\cal E}=v^2|\Phi|
\label{top}
\end{equation}
Furthermore, they have nonzero angular momentum. For $N$ superimposed vortices,
the angular momentum can be evaluated as
\begin{equation}
J=-\pi \kappa N^2=-{Q^2\over 4\pi \kappa}
\label{ang}
\end{equation}
which is the anyonic relation (\ref{spinstat}).

The nontopological solutions, with asymptotic behaviour $|\phi|\to 0$ as
$|\vec{x}|\to\infty$,  are more complicated. The only existence proof so far is
for superimposed solutions \cite{spruck}. However, numerical studies are quite
convincing, and show that \cite{jlw}
\begin{equation}
\Phi=2\pi (N+\alpha), \qquad Q=\kappa \Phi, \qquad {\cal E}=v^2|\Phi|
\label{nontop}
\end{equation}
where $\alpha$ is a continuous parameter. They have nonzero angular momentum,
and for $N$ superimposed vortices
\begin{equation}
J=-\pi \kappa (N^2-\alpha^2)=-{Q\over 4\pi \kappa^2}+NQ
\label{ntang}
\end{equation}

There is an analogue of Bradlow's bound (\ref{bradlow}) for the relativistic
Chern-Simons vortices. Integrating the second self-duality equation in (\ref{rcssd}), we
get
\begin{equation}
\int d^2x B=\frac{v^4}{2\kappa^2}\int d^2x -\frac{2}{\kappa^2}\int
d^2x\left(|\phi|^2-\frac{v^2}{2}\right)^2
\end{equation}
which implies that the vorticity is bounded above by
\begin{equation}
N\leq {v^4\over 4\pi\kappa^2}\;{\rm area}
\end{equation}
A related bound has been found in the study of periodic solutions to the
relativistic Chern-Simons equations \cite{caff,gabriella}.

\vskip 1cm
{\bf Exercise 4.2.2 :} The self-dual model (\ref{rcslag}) may be generalized to
include also a Maxwell term for the gauge field, but this requires an
additional neutral scalar field ${\cal N}$ \cite{llm}:
\begin{equation}
{\cal L}_{\rm MCS}=-\frac{1}{4e^2}F_{\mu\nu}F^{\mu\nu} +
\frac{\kappa}{2}\epsilon^{\mu\nu\rho}A_\mu \partial_\nu A_\rho +
|D_\mu \phi |^2 +\frac{1}{2e^2}(\partial_\mu {\cal N})^2 -V(|\phi|,{\cal N})
\label{mcslag}
\end{equation}
with self-dual potential
\begin{equation}
V=|\phi|^2({\cal N}-\frac{v^2}{\kappa})^2+\frac{e^2}{2}(|\phi|^2-\kappa {\cal
N})^2
\end{equation}
Show that in the symmetric phase the neutral scalar field ${\cal N}$ is
degenerate with the massive gauge field. Show that in the asymmetric phase the
${\cal N}$ field and the real part of $\phi$ have masses equal to the two
masses of the gauge field. Check that

(i) the limit $e^2\to\infty$ reduces to the relativistic Chern-Simons vortex model of
(\ref{rcslag})

(ii) the limit $\kappa\to 0$ reduces to the Abelian-Higgs model (\ref{ahlag}).

\subsection{Nonabelian Relativistic Chern-Simons Vortices}
\label{nabrcsv}

The self-dual Chern-Simons vortex systems studied in the previous section can be
generalized to incorporate nonabelian local gauge symmetry
\cite{kimyeong,dunneb}. This can be done with the matter fields and gauge
fields in different representations, but the most natural and interesting case
seems to be with adjoint coupling, with the matter fields and gauge fields in
the same Lie algebra representation. Then the gauge covariant derivative is
$D_\mu \phi=\partial_\mu \phi+[A_\mu , \phi]$ and the Lagrangian is
\begin{equation}
{\cal L}= \kappa\epsilon^{\mu\nu\rho} tr \left ( A_\mu \partial_\nu A_\rho
+{2\over 3} A_\mu A_\nu A_\rho \right ) +\tr\left(|D_\mu \phi|^2\right)
-\frac{1}{4\kappa^2}\tr\left(|[[\phi,\phi^\dagger ],\phi]-v^2\phi|^2\right)
\label{nabrcslag}
\end{equation}
where we have used the short-hand notation $|D_\mu\phi|^2=(D_\mu\phi)^\dagger
D^\mu\phi$. There is a nonabelian version of the factorization identity
(\ref{bochner}) which with adjoint coupling reads
\begin{equation}
\tr\left(|\vec{D}\phi|^2\right)= \tr\left(|D_\pm \phi|^2\right) \pm i\,
\tr\left(\phi^\dagger [F_{12},\phi]\right) \pm \epsilon_{ij}\partial_i
\,\tr\left(\phi^\dagger D_j\phi -(D_j\phi)^\dagger \phi\right)
\label{nabbochner}
\end{equation}
By the same argument as in the abelian case, we can show that with the
potential as in (\ref{nabrcslag}), the associated energy functional is bounded
below by an abelian magnetic flux. This Bogomol'nyi bound is saturated by
solutions to the nonabelian self-duality equations
\begin{eqnarray}
D_\pm\phi&=&0\cr
F_{+-}&=&\frac{1}{\kappa^2}[v^2\phi-[[\phi,\phi^\dagger],\phi],\phi^\dagger]
\end{eqnarray}
Once again, the self-dual point is the point at which the model becomes the
bosonic part of an $N=2$ SUSY model. The self-dual potential has an intricate
pattern of degenerate minima, given by solutions of the embedding equation
\begin{equation}
[[\phi,\phi^\dagger],\phi]=v^2\phi
\label{embedding}
\end{equation}
This equation describes the embedding of $SU(2)$ into the gauge Lie algebra, as
can be seen by making the identifications:
\begin{equation}
\phi=\frac{1}{\sqrt{v}}J_+;\qquad \phi^\dagger=\frac{1}{\sqrt{v}}J_- ;\qquad
[\phi,\phi^\dagger]=\frac{1}{v}[J_+,J_-]=\frac{1}{v}J_3
\end{equation}
in which case the vacuum condition (\ref{embedding}) reduces to the standard
$SU(2)$ commutation relations. Therefore, for $SU(N)$, the number of gauge
inequivalent vacua is given by the number of inequivalent ways of
embedding $SU(2)$ into $SU(N)$. This number is in fact equal to the number
$P(N)$ of partitions of the integer $N$. In each of these vacua, the masses of
the gauge and scalar fields pair up in degenerate pairs, reflecting the $N=2$
SUSY of the extended model including fermions. The masses are given by universal
formulae in terms of the exponents of the gauge algebra
\cite{dunneb}.

Not many rigorous mathematical results are known concerning solutions to the
nonabelian self-duality equations, although partial results have been found
\cite{yisong3}. Physically, we expect many different classes of solutions, with
asymptotic behaviour of the solutions corresponding to the various gauge
inequivalent vacua.

\subsection{Nonrelativistic Chern-Simons Vortices : Jackiw-Pi Model}
\label{jp}

As mentioned before, Chern-Simons gauge fields acquire their dynamics from the matter
fields to which they couple, and so they can be coupled to either relativistic
or nonrelativistic matter fields. The nonrelativistic couplings discussed in
this and subsequent sections are presumably more immediately relevant for
applications in condensed matter systems. We shall see that Bogomol'nyi
self-duality is still realizable in the nonrelativistic systems.

We begin with the abelian Jackiw-Pi model \cite{jp1}
\begin{equation}
{\cal L}_{\rm JP}=\frac{\kappa}{2}\epsilon^{\mu\nu\rho}A_\mu \partial_\nu
A_\rho  +i\psi^* D_0\psi-\frac{1}{2m}|\vec{D}\psi|^2+\frac{g}{2}|\psi|^4
\label{jplag}
\end{equation}
The quartic term represents a self-coupling contact term of the type commonly
found in nonlinear Schr\"odinger systems. The Euler-Lagrange equations are
\begin{eqnarray}
iD_0 \psi&=&-{1\over 2m}\vec{D}^2\psi-g\left|\psi\right|^2\psi\cr
F_{\mu\nu}&=&{1\over \kappa} \epsilon_{\mu\nu\rho} J^{\rho}
\end{eqnarray}
where $J^{\mu}\equiv(\rho, \vec{J})$ is a Lorentz covariant short-hand notation
for the conserved nonrelativistic charge and current densities:
$\rho=\left|\psi\right|^2$, and $J^j=-{i\over 2m} \left(\psi^*
D^j\psi-\left(D^j\psi\right)^*\psi\right)$. This system is Galilean invariant,
and there are corresponding conserved quantities: energy, momentum, angular
momentum and Galilean boost generators. There is, in fact, an addition
dynamical symmetry \cite{jp1} involving dilations, with generator
\begin{equation}
D=t E -\frac{1}{2}\int d^2x\, \vec{x}\cdot\vec{{\cal P}}
\label{dil}
\end{equation}
and special conformal transformations, with generator
\begin{equation}
K=-t^2 E+2tD+\frac{m}{2}\int d^2x\, \vec{x}^2 \rho
\label{spec}
\end{equation}
Here $E$ is the energy and $\vec{{\cal P}}$ is the momentum density.

The static energy functional for the Jackiw-Pi Lagrangian (\ref{jplag}) is
\begin{equation}
{\cal E}_{\rm JP}=\int d^2x
\left[\frac{1}{2m}|\vec{D}\psi|^2-\frac{g}{2}|\psi|^4\right]
\end{equation}
Using the factorization identity (\ref{bochner}), together with the Chern-Simons Gauss
law constraint $F_{12}=\frac{1}{\kappa}|\psi|^2$, the energy becomes
\begin{equation}
{\cal E}_{\rm JP}=\int d^2x \left[\frac{1}{2m}|D_\pm\psi|^2-
\left(\frac{g}{2}\pm\frac{1}{2m\kappa}\right)|\psi|^4\right]
\end{equation}
Thus, with the self-dual coupling
\begin{equation}
g=\mp \frac{1}{m\kappa}
\label{jpcoupling}
\end{equation}
the energy is bounded below by zero, and this lower bound is saturated by
solutions to the first-order self-duality equations
\begin{eqnarray}
D_\pm\psi&=&0\cr
B&=&\frac{1}{\kappa}|\psi|^2
\label{jpsd}
\end{eqnarray}
Note that with the self-dual coupling (\ref{jpcoupling}), the original quartic
interaction term, $-\frac{g}{2}|\psi |^4=\pm\frac{1}{2m\kappa} |\psi |^4$, can be
understood as a Pauli interaction term $\pm \frac{B}{2m} |\psi|^2$, owing to the
Chern-Simons constraint $|\psi|^2 =\kappa B$.

The self-duality equations (\ref{jpsd}) can be disentangled as before, by
decomposing the scalar field $\psi$ into a phase and a magnitude
(\ref{phidecomp}), resulting in a nonlinear elliptic equation for the density
$\rho$:
\begin{equation}
\nabla^2 \ln \rho=\pm \frac{2}{\kappa}\rho
\label{liouville}
\end{equation}
Surprisingly [unlike the previous nonlinear elliptic equations
(\ref{ahelliptic},\ref{rcselliptic}) in the Abelian-Higgs and relativistic Chern-Simons
vortex models], this elliptic equation is exactly solvable! It is known as the
Liouville equation \cite{liouville}, and has the general real solution
\begin{equation}
\rho=\kappa \nabla^2\ln\left(1+|f|^2\right)
\label{liouvillesol}
\end{equation}
where $f=f(z)$ is a holomorphic function of $z=x^1+ix^2$ only.
\vskip 1cm
{\bf Exercise 4.4.1 :} Verify that the density $\rho$ in (\ref{liouvillesol})
satisfies the Liouville equation (\ref{liouville}). Show that only one sign is
allowed for physical solutions, and show that this corresponds to an attractive
quartic potential in the original Lagrangian (\ref{jplag}).
\vskip 1cm

As a consequence of the Chern-Simons Gauss law, these vortices carry both magnetic and
electric charge: $Q=\kappa \Phi$. The net matter charge $Q$ is
\begin{equation}
Q=\kappa \int d^2x\,\nabla^2 \ln\left(1+|f|^2\right)
=2\pi \kappa \left[ r{d\over dr}\ln\left(1+|f|^2\right)\right]_0^\infty
\label{qq}
\end{equation}

Explicit radially symmetric solutions may be obtained by taking
$f(z)=(\frac{z_0}{z})^{N}$. The corresponding charge density is
\begin{equation}
\rho={4\kappa N^2\over r_0^2}{\left(\frac{r}{r_0}\right)^{2(N-1)}\over
\left(1+\left(\frac{r}{r_0}\right)^{2N}\right)^2}
\label{radialsol}
\end{equation}

\begin{figure}[htb]
\vspace{-1in}
\centering{\epsfig{file=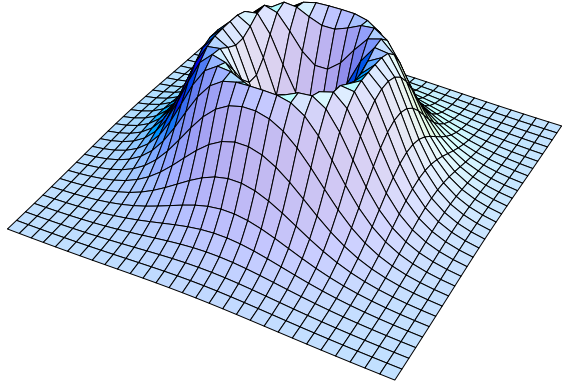,width=5in}}
\vspace{-3in}
\caption{Density $\rho$ for a radially symmetric solution (\protect{\ref{radialsol}}) representing one vortex with $N=2$.}
\label{1v}
\end{figure}

As $r\to 0$, the charge density behaves as $\rho \sim r^{2(N-1)}$, while as
$r\to\infty$, $\rho \sim r^{-2-2N}$. At the origin, the vector potential
behaves as $A_i(r)\sim -\partial_i\omega \mp (N-1)\epsilon_{ij}{x^j\over r^2}$.
We can therefore avoid singularities in the the vector potential at the
origin if we choose the phase of $\psi$ to be $\omega=\pm (N-1)\theta$. Thus
the self-dual $\psi$ field is
\begin{equation}
\psi= {2N\sqrt{\kappa} \over r_0} \left({\left(\frac{r}{r_0}\right)^{N-1}\over
1+\left(\frac{r}{r_0}\right)^{2N}}\right) \; e^{\pm i(N-1)\theta}
\label{sdpsi}
\end{equation}
Requiring that $\psi$ be single-valued we find that $N$ must be an
integer, and for $\rho$ to decay at infinity we require that $N$ be
positive. For $N>1$ the $\psi$ solution has vorticity $N-1$ at the origin and
$\rho$ goes to zero at the origin. See Figure \ref{1v} for a plot of the density for the $N=2$ case. Note the ring-like form of the magnetic field for these Chern-Simons vortices, as the magnetic field is proportional to $\rho$ and so $B$ vanishes where the field $\psi$ does.

For the radial solution (\ref{radialsol}) the net matter charge is $Q=\int
d^2x\, \rho = 4\pi \kappa N$; and the corresponding flux is $\Phi=4\pi N$,
which represents an {\it even} number of flux units. This quantized character
of the flux is a general feature and is not particular to the radially
symmetric solutions.

The radial solution (\ref{radialsol}) arose from choosing the
holomorphic function $f(z)=(\frac{z_0}{z})^{N}$, and corresponds to $N$
vortices superimposed at the origin. A solution corresponding to $N$ {\it
separated} vortices may be obtained by taking
\begin{equation}
f(z)=\sum_{a=1}^N {c_a\over z-z_a}
\label{separate}
\end{equation}

\begin{figure}[htb]
\vspace{-1in}
\centering{\epsfig{file=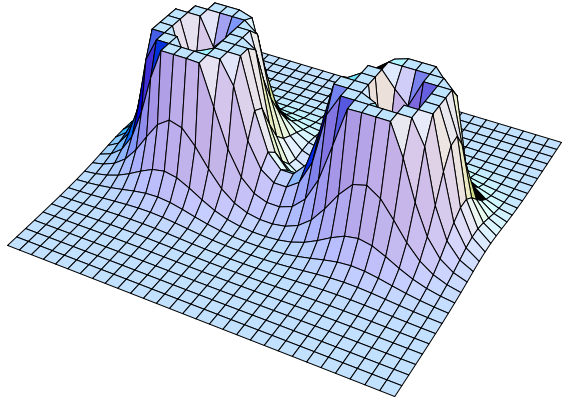,width=5in}}
\vspace{-3in}
\caption{Density $\rho$ for a solution (\protect{\ref{separate}}) representing two separated vortices.}
\label{2v}
\end{figure}

There are $4N$ real parameters involved in this solution : $2N$
real parameters $z_a\,\, (a=1,\dots N)$ describing the locations of the
vortices, and $2N$ real parameters $c_a\,\, (a=1,\dots N)$ corresponding
to the scale and phase of each vortex. See Figure \ref{2v} for a plot of the two vortex case. The solution in (\ref{separate}) is in fact the most general finite multi-soliton solution on the plane. Solutions with a {\it periodic} matter density $\rho$ may be obtained by choosing the function $f$ in
(\ref{liouvillesol}) to be a doubly periodic function \cite{olesen}.

I conclude this section by noting that the dynamical symmetry of the Jackiw-Pi
system guarantees that static solutions are necessarily self-dual. This follows
from the generators (\ref{dil}) and (\ref{spec}). Consider the dilation
generator $D$ in (\ref{dil}). It is conserved, but so is $\vec{{\cal P}}$ for
static solutions. This implies that $E$ must vanish, which is only true for
self-dual solutions.

\subsection{Nonabelian Nonrelativistic Chern-Simons Vortices}
\label{nabnrcsv}

Just as the relativistic Chern-Simons vortices of Section \ref{rcsv} could be
generalized to incorporate local nonabelian gauge symmetry, so too can the
nonrelativistic models discussed in the previous section. We consider the case
of adjoint coupling, with Lagrangian
\begin{equation}
{\cal L}=\kappa \epsilon^{\mu\nu\rho}\tr\left(A_\mu\partial_\nu A_\rho+{2\over
3}A_\mu A_\nu A_\rho \right)
 +i\, tr \left ( \psi^\dag D_0 \psi \right ) -
{1\over 2m} tr \left ( | \vec{D} \psi |^2 \right ) +
{1\over 4 m \kappa} tr \left ( [ \psi , \psi^\dag ]^2 \right )
\label{nabnrlag}
\end{equation}
Using the nonabelian factorization identity (\ref{nabbochner}), together with
the Gauss law constraint, $F_{+-}=\frac{1}{\kappa}[\psi,\psi^\dagger]$, the
static energy functional can be written as
\begin{equation}
{\cal E}=\frac{1}{2m}\int d^2x\, \tr\left(|D_\pm\psi|^2\right)
\end{equation}
which is clearly bounded below by $0$. The solutions saturating this lower
bound satisfy the first-order self-duality equations
\begin{eqnarray}
D_\pm \psi&=&0\cr
F_{+-}&=&\frac{1}{\kappa}[\psi,\psi^\dagger]
\label{nabnrsd}
\end{eqnarray}
These self-duality equations have been studied before in a different context,
as they are the dimensional reduction of the four-\diml self-dual Yang-Mills
equations
\begin{equation}
F^{\mu\nu}=\pm\frac{1}{2}\epsilon^{\mu\nu\rho\sigma}F_{\rho\sigma}
\label{sdym}
\end{equation}
\vskip .5cm
{\bf Exercise 4.5.1 :} Show that the self-dual Yang-Mills equations, with
signature
$(2,2)$, reduce to the self-dual Chern-Simons equations (\ref{nabnrsd}) if we take
fields independent of two of the coordinates, say $x^3$ and $x^4$, and combine
the gauge fields $A_3$ and $A_4$ to form the fields $\psi$ and $\psi^\dagger$.
\vskip 1cm

The self-duality equations (\ref{nabnrsd}) are integrable, as they can be
expressed as a zero curvature condition in the following way. Define a spectral
connection [with spectral paramter $\lambda$]
\begin{equation}
{\cal A}_+=A_+-\lambda\sqrt{\frac{1}{\kappa}}\psi,\qquad
{\cal A}_-=A_-+\frac{1}{\lambda}\sqrt{\frac{1}{\kappa}}\psi^\dagger
\label{spectral}
\end{equation}
Then the corresponding curvature is
\begin{eqnarray}
{\cal F}_{+-}&=&\partial_+{\cal A}_- -\partial_-{\cal A}_+ + [{\cal A}_+,
{\cal A}_-]\cr
&=&\left\{F_{+-}-\frac{1}{\kappa}[\psi,\psi^\dagger] \right\}
+\sqrt{\frac{1}{\kappa}} \lambda
D_-\psi-\sqrt{\frac{1}{\kappa}}\frac{1}{\lambda}D_+\psi^\dagger
\label{zeroc}
\end{eqnarray}
Therefore, the condition of zero curvature, ${\cal F}_{+-}=0$, for arbitrary
spectral parameter $\lambda$, encodes the self-dual Chern-Simons equations
(\ref{nabnrsd}). Explicit exact solutions can also be obtained by making
simplifying algebraic ans\"atze which reduce the self-duality equations to the
Toda equations, which are coupled analogues of the Liouville equation
(\ref{liouville}) and which are still integrable \cite{djpt,grossman}.

In fact, {\it all} finite charge solutions can be found by mapping the
self-duality equations (\ref{nabnrsd}) into the chiral model equations, which
can then be integrated exactly in terms of {\it unitons}. To see this, set the
spectral parameter $\lambda=1$ in (\ref{spectral}) and use the zero curvature
${\cal F}_{+-}=0$ to define
\begin{equation}
{\cal A}_\pm=g^{-1}\partial_\pm g
\end{equation}
Then the conjugation $\chi=\frac{1}{\sqrt{\kappa}}g\psi g^{-1}$ transforms the
self-duality equations (\ref{nabnrsd}) into a single eqaution
\begin{equation}
\partial_-\chi=[\chi^\dagger, \chi]
\label{single}
\end{equation}
Furthermore, if we define $\chi=\frac{1}{2}h^{-1}\partial_+h$, with $h$ in the
gauge group, then (\ref{single}) becomes the chiral model equation
\begin{equation}
\partial_+(h^{-1}\partial_- h)+ \partial_-(h^{-1}\partial_+ h)=0
\label{chiralmodel}
\end{equation}
All solutions to the chiral model equations with finite
$\int\tr(h^{-1}\partial_- h h^{-1}\partial_+ h)$ can be constructed in terms of
Uhlenbeck's unitons \cite{uhlenbeck,ward}. These are solutions of the form
\begin{equation}
h=2p-{\bf 1}
\label{uniton}
\end{equation}
where $p$ is a holomorphic projector   satisfying: (i) $p^\dagger=p$, (ii)
$p^2=p$, and (iii) $(1-p)\partial_+p=0$. This means that all finite charge
solutions of the self-dual Chern-Simons vortex equations (\ref{nabnrsd}) can be
constructed in terms of unitons \cite{dunneb}.

\vskip 1cm
{\bf Exercise 4.5.2 :} Show that a holomorphic projector $p$ can be expressed as
$p=M(M^\dagger M)^{-1}M^\dagger$, where $M=M(x^-)$ is any rectangular matrix.
For $SU(2)$ show that the uniton solution leads to a charge density $[\psi,
\psi^\dagger]$ which, when diagonalized, is just the Liouville solution
(\ref{liouvillesol}) times the Pauli matrix $\sigma^3$.

\subsection{Vortices in the Zhang-Hansson-Kivelson Model for FQHE}
\label{zhkm}

There have been many applications of Chern-Simons theories to the description of the quantum Hall effect, and the fractional quantum Hall effect in particular (see e.g. \cite{zee,wen,zhang,stone,shankar}). In this section I describe one such model, and show how it is related to our discussion of Chern-Simons vortices.

Zhang, Hansson and Kivelson \cite{zhk} reformulated the problem of interacting
fermions in an external magnetic field as a problem of interacting bosons with
an extra Chern-Simons interaction describing the statistical transmutation of the
fermions into bosons. This transmutation requires a particular choice for the
Chern-Simons coupling constant, as we shall see below. The Chern-Simons coupling is such that an odd number of flux quanta are `tied' to the fermions [recall Figure \ref{fluxes}]; thus the fermions acquire an additional statistics parameter [given by (\ref{spinstat})] and so become effective bosons. The ZHK model is basically a Landau-Ginzburg effective field theory description of these boson fields, coupled to a Chern-Simons field that takes care of the statistical transmutation. It looks like a fairly innocent variation on the Jackiw-Pi model of Section \ref{jp}, but the minor change makes a big difference to the vortex solutions. The ZHK Lagrangian is
\begin{eqnarray}
{\cal L}_{\rm ZHK}&=&-\frac{\kappa}{2}\epsilon^{\mu\nu\rho}a_\mu \partial_\nu
a_\rho  +i\psi^* \left(\partial_0+i a_0\right) \psi
-\frac{1}{2m}|\left(\partial_i+i (a_i+A_i^{ext})\right)\psi|^2 \cr\cr
&&\qquad\qquad-\frac{1}{2}\int d^2x^\prime \left(|\psi(\vec{x})|^2-n\right)
V(\vec{x}-\vec{x}^\prime ) \left(|\psi(\vec{x}^\prime)|^2-n\right)
\label{zhklag}
\end{eqnarray}
where we have adopted the notation that the statistical Chern-Simons gauge field is
$a_\mu$, while the external gauge field descrbing the external magnetic field
is $A_i^{ext}$. We have also, for convenience in some of the subsequent
equations, written the Chern-Simons coupling as $-\kappa$. The constant $n$ appearing in
the potential term denotes a uniform condensate charge density.

Normally a complex scalar field $\psi$ is used to describe bosons. But when the Chern-Simons coupling takes the values
\begin{equation}
\kappa={1\over 2\pi (2k-1)}; \qquad\qquad k\geq 1
\label{trans}
\end{equation}
the anyonic statistics phase (\ref{spinstat}) of the $\psi$ fields is $\frac{1}{2\kappa}=(2k-1)\pi$; that is, the fields are antisymmetric under interchange. Thus the fields are actually fermionic. We can alternatively view this as the condensing of the fundamental fermionic fields into bosons by the attachment of an odd number of fluxes through the Chern-Simons coupling \cite{zhk}. Consider a delta-function contact interaction with
\begin{equation}
V(\vec{x}-\vec{x}^\prime )={1\over m\kappa} \delta(\vec{x}-\vec{x}^\prime )
\label{contact}
\end{equation}
in which case we can simply express the potential as
\begin{equation}
V(\rho)={1\over 2m\kappa}(\rho-n)^2
\label{zhkpot}
\end{equation}
The static energy functional for this model is
\begin{equation}
{\cal E}_{\rm ZHK}=\int d^2x\left[\frac{1}{2m}|\left(\partial_i+i
(a_i+A_i^{ext})\right)\psi|^2 +{1\over 2m\kappa}(\rho-n)^2\right]
\label{zhkenergy}
\end{equation}
Clearly, the minimum energy solution corresponds to the constant field
solutions
\begin{equation}
\psi=\sqrt{n}, \qquad\qquad a_i=-A_i^{ext}, \qquad\qquad a_0=0
\end{equation}
for which the Chern-Simons gauge field opposes and cancels the external field.
Since the Chern-Simons constraint is $b=-\frac{1}{\kappa} \rho$, we learn that these
minimum energy solutions have density
\begin{equation}
\rho =n=\kappa B^{ext}
\label{laughlin}
\end{equation}
With the values of $\kappa$ in (\ref{trans}), these are exactly the conditions
for the uniform Laughlin states of filling fraction
\begin{equation}
\nu={1\over 2k-1}
\label{filling}
\end{equation}

To describe excitations about these ground states, we re-express the energy
using the factorization identity (\ref{bochner}).
\begin{eqnarray}
{\cal E}_{\rm ZHK}&=&\int d^2x\left[ \frac{1}{2m}|D_\pm \psi|^2
\mp \frac{1}{2m}\left(B^{ext}-\frac{1}{\kappa}\rho\right)\rho +
{1\over 2m\kappa}(\rho-n)^2\right] \cr
&=&\int d^2x\left[ \frac{1}{2m}|D_\pm \psi|^2
\pm\frac{1}{2m\kappa}\left(\rho-\kappa B^{ext}\right)^2 \mp \frac{\kappa}{2m}
B^{ext} B  +{1\over 2m\kappa}(\rho-n)^2  \right]\cr
&=&\int d^2x\left[ \frac{1}{2m}|D_- \psi|^2 +\frac{\kappa}{2m} B^{ext} B\right]
\end{eqnarray}
In the last step we have chosen the lower sign, and used the relation $n=\kappa
B^{ext}$ to cancel the potential terms. Note that in the last line, $B$ is the
total magnetic field $B=B^{ext}+b$, where $b$ is the Chern-Simons magnetic field.

Thus, the energy is bounded below by a multiple of the total magnetic flux.
This bound is saturated by solutions to the first-order equations
\begin{eqnarray}
D_-\psi&=&0\cr
B&=&B^{ext}-\frac{1}{\kappa}\rho
\label{zhksd}
\end{eqnarray}
As before, the first equation allows us to express the total gauge field
$A_i=a_i+A_i^{ext}$ in terms of the phase and the density, and the second
equation reduces to a nonlinear elliptic equation for the density:
\begin{equation}
\nabla^2 \ln \rho=\frac{2}{\kappa}(\rho-n)
\label{zhkelliptic}
\end{equation}
Comparing this with the corresponding equation (\ref{liouville}) in the
Jackiw-Pi model, we see that the effect of the external field and the modified
potential (\ref{zhkpot}) is to include a constant term on the RHS. But this
converts the Liouville equation back into the vortex equation
(\ref{ahelliptic}) for the Abelian-Higgs model! This can be viewed as both good
and bad news -- bad in the sense that we no longer have the explicit exact
solutions to the Liouville equation (\ref{liouville}) that we had in the
Jackiw-Pi model, but good because we know a great deal about the Abelian-Higgs
models vortices, even though we do not have any explicit exact solutions.
First, we learn that there are indeed well-behaved vortex solutions in the ZHK
model, and that their magnetic charge is related to their vorticity. But now,
because of the Chern-Simons relation, these vortices also have electric charge,
proportional to their magnetic charge. In particular these vortices have the
correct quantum numbers for the quasi-particles in the Laughlin model for the
FQHE \cite{zhk}.
\vskip 1cm
{\bf Exercise 4.6.1 :} Show that if we modify the Jackiw-Pi model by including a
background charge density $\rho_0$ (instead of an external magnetic field)
\cite{barashenkov}
\begin{equation}
{\cal L}=\frac{\kappa}{2}\epsilon^{\mu\nu\rho}A_\mu \partial_\nu A_\rho
+i\psi^* D_0\psi-\frac{1}{2m}|\vec{D}\psi|^2-V(\rho)+\rho_0 A_0
\label{bhlag}
\end{equation}
then with the potential $V(\rho)=\frac{1}{2m\kappa}(\rho-\rho_0)^2$, the
self-dual vortex equations also reduce to a nonlinear elliptic equation of the
Abelian-Higgs form (\ref{ahelliptic}):
\begin{equation}
\nabla^2 \ln \rho=\frac{2}{\kappa}(\rho-\rho_0)
\label{bhelliptic}
\end{equation}

\subsection{Vortex Dynamics}
\label{dynamics}

So far, we have only dealt with {\it static} properties of vortices in various
$2+1$-\diml field theories. However, the more interesting question concerns
their dynamics; and beyond that, we are ultimately interested in their
quantization. Various different approaches have been developed over the years
for studying vortex dynamics. Particle physicists and field theorists,
motivated largely by Manton's work \cite{manton1} on the low energy dynamics of
solitons (of which these planar Bogomol'nyi vortices are an example), have
studied the dynamics of vortices in the Abelian-Higgs model, which is governed
by relativistic dynamics for the scalar field. Condensed matter physicists have
developed techniques for studying vortices in superconductors and in Helium
systems, where the dynamics is nonrelativistic \cite{ao}. The Chern-Simons
vortices are particularly interesting, because in addition to introducing the
new feature of anyon statistics of vortices, they appear to require methods
from both the particle physics and condensed matter physics approaches. Having
said that, there is, as yet, no clear and detailed understanding of the
dynamics of Chern-Simons vortices. This is a major unsolved problem in the
field.

Consider first of all the dynamics of vortices in the Abelian-Higgs model of
Section \ref{abhiggs}. Since no exact vortex solutions are known, even for the
static case, we must be content with approximate analytic work and/or numerical
simulations. As mentioned earlier, it is known from numerical work
\cite{jacobs} that the vortices in the Abelian-Higgs model repel one another
when the scalar mass exceeds the gauge mass, and attract when the gauge mass
exceeds the scalar mass. When these two mass scales are equal (\ref{ahmassdeg})
we are in the self-dual case, and there are no forces between static vortices.
Manton's approach to the dynamics of solitons provides an effective description
of the dynamics at low energies when most of the field theoretic degrees of
freedom are frozen out. Suppose we have static multi-soliton solutions
parametrized by a finite dimensional `moduli space' -- the space consisting of
the minima of the static energy functional (\ref{ahenergy}). We assume that the
true dynamics of the full field theory is in some sense ``close to'' this
moduli space of static solutions. Then the full dynamics should be approximated
well by a projection onto a finite dimensional problem of dynamics on the
moduli space. This is an adiabatic approximation in which one assumes that at
each moment of time the field is a static solution, but that the parameters of
the static solution [in the vortex case we can loosely think of these
parameters as the locations of the vortices] vary slowly with time.

This approach has been applied successfully to the Abelian-Higgs vortices
\cite{samols}, with the $N$-vortex parameters taken to be the zeros $z_1$,
$\dots$, $z_N$ of the scalar $\phi$ field  [recall the theorem in Section
\ref{abhiggs}]. For well separated zeros we can think of these zeros as
specifying the locations of the vortices. Indeed, the exponential approach of
the fields to their asymptotic values motivates and supports the approximation
of well separated vortices as a superposition of single vortices, with only
exponentially small errors. (Actually, to be a bit more precise, the N-vortex
moduli space is not really ${\bf C}^N$; we need to take into account the
identical nature of the vortices and factor out by the permutation group ${\cal
S}_N$. Thus the true N-vortex moduli space is ${\bf C}^N/{\cal S}_N$, for which
a good set of global coordinates is given by the symmetric polynomials in the
zeros $z_1$, $\dots$, $z_N$.)

The total energy functional is
\begin{equation}
H=T+V
\end{equation}
where the kinetic energy is
\begin{equation}
T=\int d^2x\left[\frac{1}{2}\dot{A}_i\dot{A}_i +|\dot{\phi}|^2\right]
\label{kinetic}
\end{equation}
and the potential energy $V$ is the static energy functional (\ref{ahenergy}).
There is also the Gauss law constraint, $\vec{\nabla}\cdot\vec{E}=J^0$, to be
imposed. In the adiabatic approximation, the potential energy remains fixed at
$v^2|\Phi|$, given by the saturated Bogomol'nyi bound (\ref{ahbound}). But when
the moduli space parameters become time dependent, we can insert these
adiabatic fields
\begin{equation}
\phi=\phi(\vec{x}; z_1(t), \dots , z_N(t)), \qquad \qquad
\vec{A}=\vec{A}(\vec{x}; z_1(t), \dots , z_N(t))
\label{ad}
\end{equation}
into the kinetic energy (\ref{kinetic}), integrate over position $\vec{x}$, and
obtain an effective kinetic energy for the moduli parameters $z_a(t)$, for
$a=1, \dots N$. In terms of real coordinates $\vec{x}_a$ on the plane, this
kinetic energy takes the form
\begin{equation}
T=\frac{1}{2} g_{ab}\; \dot{\vec{x}}_a\cdot \dot{\vec{x}}_b
\label{effkinetic}
\end{equation}
where the metric $g_{ab}$ is a (complicated) function depending on the
positions and properties of all the vortices. Samols \cite{samols} has shown
that this construction has a beautiful geometric interpretation, with the
metric $g_{ab}$ being hermitean and K\"ahler. Furthermore, the dynamics of the
slowly moving vortices corresponds to geodesic motion on the moduli space.
While the metric cannot be derived in closed form, much is known about it, and
it can be expressed solely  in terms of the local properties of the vortices.
It should be mentioned that the step of performing the spatial integrations to
reduce the field theoretic kinetic energy (\ref{kinetic}) to the finite
dimensional moduli space kinetic energy (\ref{effkinetic}) involves some
careful manipulations due to the nature of vortex solutions in the
neighbourhood of the zeros of the scalar field $\phi$. The essential procedure
is first to excise small discs surrounding the zeros. The contributions from
the interior of the discs can be shown to be negligible as the size of the disc
shrinks to zero. The contribution from the outside of the discs can be
projected onto a line integral around each disc, using Stokes's theorem and the
linearized Bogomol'nyi self-duality equations. These line integrals may then be
expressed in terms of the local data of the scalar fields in the neighbourhood
of each disc :
\begin{equation}
\ln|\phi|^2\approx
\ln|z-z_k|^2+a_k+\frac{1}{2}\left\{b_k(z-z_k)+b_k^*(z^*-z_k^*)\right\}+\dots
\label{local}
\end{equation}

There are several important differences complicating the direct application of
this `geodesic approximation' to the dynamics of the relativistic Chern-Simons vortices
described in Section \ref{rcsv}. While it is still true that the static
multi-vortex solutions can be characterized by the zeros of the scalar field
[although no rigorous proof of uniqueness has been given so far], the fact that
the vortices appear to be anyonic means that we cannot simply factor out by the
symmetric group ${\cal S}_N$ to obtain the true moduli space. Presumably the
true moduli space would need to account for braidings of the vortex zeros.
Second, the gauge field makes no contribution to the kinetic energy in the case
of Chern-Simons vortices -- all the dynamics comes from the scalar field.
Correspondingly, even though there are no repulsive or attractive forces
between the static self-dual vortices, there may still be velocity dependent
forces that we do not see in the completely static limit. Thus, it is more
convenient to consider the effective action (rather than the energy) for motion
on the moduli space. Both these considerations suggest that we should expect a
term {\it linear} in the velocities, in addition to a quadratic kinetic term
like that in (\ref{effkinetic}).

To see how these velocity dependent forces might arise, consider the
relativistic Chern-Simons vortex model (\ref{rcslag}):
\begin{equation}
{\cal L}_{\rm RCS}=|D_0\phi|^2 +\kappa A_0 B-\frac{\kappa}{2} \epsilon^{ij}A_i
\dot{A}_j -|\vec{D}\phi|^2-\frac{1}{\kappa^2}|\phi|^2(|\phi|^2-v^2)^2
\label{lll}
\end{equation}
Decomposing $\phi=\sqrt{\rho} e^{i\omega}$ as in (\ref{phidecomp}), Gauss's law
determines the nondynamical field $A_0$ to be: $A_0=-\dot{\omega}-\frac{\kappa
B}{2\rho}$. Then the Lagrangian (\ref{lll}) can be re-expressed as
\begin{equation}
{\cal L}_{\rm RCS}=\frac{1}{4}\frac{\dot{\rho}^2}{\rho} -\kappa B \dot{\omega}
-\frac{\kappa}{2} \epsilon^{ij}A_i \dot{A}_j
-\left[|D_\pm\phi|^2+\frac{\kappa^2}{4\rho}\left(B\mp
\frac{2}{\kappa^2}\rho(\rho-v^2)\right)^2\right] \pm v^2 B
\label{lll2}
\end{equation}
To implement Manton's procedure, we take fields that solve the static self-dual
equations (\ref{rcssd}), but with adiabatically time-dependent parameters. As
moduli parameters we take the zeros $\vec{q}_a(t)$ of the $\phi$ field. Then
the term in the square brackets in (\ref{lll2}) vanishes for self-dual
solutions. Furthermore, for an N vortex solution the vorticity is such that
$\omega=\sum_{a=1}^N arg(\vec{x}-\vec{q}_a(t))$. Then we can integrate over
$\vec{x}$ to obtain an effective quantum mechanical Lagrangian for the vortex
zeros:
\begin{equation}
L(t)=\int d^2x{\cal L}=\frac{1}{2} g^{ij}_{ab}(q)\; \dot{q}^i_a \dot{q}^j_b +
\vec{{\cal A}}^i_a(q) \dot{q}^i_a \pm 2\pi v^2 N
\label{efflag}
\end{equation}
The term linear in the velocities comes from the $B \dot{\omega}$ term in
(\ref{lll2}), while the $\epsilon^{ij}A_i \dot{A}_j$ term integrates to zero
\cite{yk}. The coefficient of the linear term is
\begin{equation}
{\cal A}_a^i=2\pi\kappa \epsilon_{ij}\sum_{b\neq a} {q_a^j-q_b^j\over
|\vec{q}_a-\vec{q}_b|^2} +{\rm local}
\label{conn}
\end{equation}
where the first term is responsible for the anyonic nature of the vortices,
while the `local' term is only known approximately in terms of the local
expansion (\ref{local}) in the neighbourhood of each vortex, and is a
complicated function of the positions of all the vortices. The linear
coefficient ${\cal A}^i_a$ is interpreted as a linear connection on the moduli
space. But, despite a number of attempts \cite{kimmin,yk}, we still do not have
a good understanding of the quadratic metric term $g^{ij}$ in the effective
Lagrangian (\ref{efflag}). This is an interesting outstanding problem.

Another important problem concerns the implementation of this adiabatic
approximation for the description of vortex dynamics in {\it nonrelativistic}
Chern-Simons theories, such as the Jackiw-Pi model or the Zhang-Hansson-Kivelson model. In these cases the field Lagrangian has only first-order time derivatives, so the nature of the adiabatic approximation is somewhat different \cite{manton2,bak}.

\bigskip

\section{Induced Chern-Simons Terms}
\label{induced}

An important feature of Chern-Simons theories is that Chern-Simons terms can be induced by
radiative quantum effects, even if they are not present as bare terms in the
original Lagrangian. The simplest manifestation of this
phenomenon occurs in $2+1$ \diml QED, where a Chern-Simons term is induced in a simple
one-loop computation of the fermion effective action \cite{redlich}. Such a
term breaks parity and time-reversal symmetry, as does a fermion mass term
$m\bar{\psi}\psi$. There are two complementary ways to investigate this effective
action -- the first is a direct perturbative expansion in powers of the coupling
for an arbitrary background gauge field, and the second is based on a
Schwinger-style calculation of the induced current $<J^\mu>$ (from which the
form of the effective action may be deduced) in the presence of a special
background with constant field strength $F_{\mu\nu}$. Chern-Simons terms can also be
induced in gauge theories without fermions, and in the broken phases of
Chern-Simons-Higgs theories. Interesting new features arise when we consider induced Chern-Simons
terms at finite temperature.

\subsection{Perturbatively Induced Chern-Simons Terms : Fermion Loop}
\label{pertloop}
 
We begin with the perturbative effective action. To facilitate later comparison
with the finite temperature case, we work in Euclidean space. The one
fermion loop effective action is
\begin{equation} S_{\rm eff}[A,m]=N_f\, \log\det(i\partial \hskip-6pt /+ A \hskip -6pt /+m)
\label{eff}
\end{equation} where $m$ is a fermion mass. The physical significance of this fermion
mass
will be addressed below. We have also included the overall factor of $N_f$
corresponding to the number of fermion flavours. This, too, will be important
later. For now, simply regard $N_f$ and $m$ as parameters.

A straightforward perturbative expansion yields
\begin{equation} S_{\rm eff}[A,m]=N_f\, \tr\log(i\dslash+m)+N_f\, \tr\left({1\over
i\dslash+m}\aslash\right)+\frac{N_f}{2}\, \tr\left({1\over
i\dslash+m}\aslash{1\over i\dslash+m}\aslash\right)+\dots
\end{equation}

\begin{figure}[htb]
\vspace{-1in}
\centering{\epsfig{file=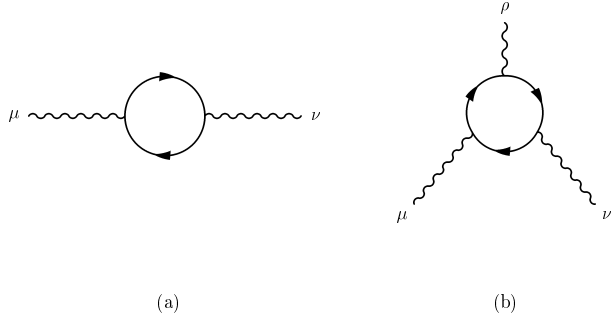,width=5in}}
\vspace{-4.25in}
\caption{The one-loop Feynman diagrams used in the calculation of the induced Chern-Simons term (at zero temperature). The self-energy diagram (a) is computed in (\protect{\ref{qkernel}}), while the three-photon-leg diagram (b) is treated in Exercise 5.1.1}
\label{feynman}
\end{figure}

The first term is just the free $(A=0)$ case, which is subtracted, while the
second term is the tadpole. Since we are seeking an
induced Chern-Simons term, and the abelian Chern-Simons term is quadratic in the gauge field
$A_\mu$, we restrict our attention to the quadratic term in the effective
action (interestingly, we shall see later that this step is not justified at
finite temperature)
\begin{equation}
S_{\rm eff}^{\rm quad}[A,m]=\frac{N_f}{2}\, \int 
{d^3p\over (2\pi)^3}\, \left[ A^\mu(-p)\Gamma^{\mu\nu}(p)A^\nu (p)\right]
\label{quadratic}
\end{equation}
where the kernel is
\begin{equation}
\Gamma^{\mu\nu}(p,m)=\int{d^3k\over (2\pi)^3} \, \tr\left[ \gamma^\mu\,
{\pslash+\kslash-m\over (p+k)^2+m^2}\, \gamma^\nu\, {\kslash-m\over k^2+m^2}
\right]
\label{qkernel}
\end{equation}
corresponding to the one-fermion-loop self-energy diagram shown in Figure
\ref{feynman} (a). Furthermore, since the Chern-Simons term involves the parity-odd Levi-Civita tensor $\epsilon^{\mu\nu\rho}$, we consider only the $\epsilon^{\mu\nu\rho}$ contribution to the fermion self-energy. This can arise because of the special property of the gamma matrices (here, Euclidean) in $2+1$ \dim
\begin{equation}
\tr(\gamma^\mu \gamma^\nu \gamma^\rho)=- 2\epsilon^{\mu\nu\rho}
\label{trodd}
\end{equation}
(Note that this may be somewhat unfamiliar because in $3+1$ \dim we are
used to the fact that the trace of an odd number of gamma matrices is zero). It
is then easy to see from (\ref{qkernel}) that the parity odd part of the kernel
has the form
\begin{equation}
\Gamma_{\rm odd}^{\mu\nu}(p,m)=\epsilon^{\mu\nu\rho}p_\rho
\Pi_{\rm odd}(p^2,m)
\label{form}
\end{equation} where
\begin{eqnarray}
\Pi_{\rm odd}(p^2,m)&=&2m\int{d^3k\over (2\pi)^3}
{1\over [(p+k)^2+m^2][k^2+m^2]}\nonumber\\
 &=&\frac{1}{2\pi}\frac{m}{|p|} \arcsin\left({|p|\over\sqrt{p^2+4m^2}}\right)
\label{arcs}
\end{eqnarray}
In the long wavelength ($p\to 0$) and large mass ($m\to\infty$) limit we
find
\begin{equation}
\Gamma_{\rm odd}^{\mu\nu}(p,m)\sim \,\frac{1}{4\pi}\,
\frac{m}{|m|}\epsilon^{\mu\nu\rho}p_\rho +O(\frac{p^2}{m^2})
\label{exp}
\end{equation}

Inserting the leading term into the quadratic effective action
(\ref{quadratic}) and returning to coordinate space, we find an induced Chern-Simons
term
\begin{equation}
S_{\rm eff}^{\rm CS}=-i \frac{N_f}{2}\frac{1}{4\pi}\frac{m}{|m|}\int d^3x
\epsilon^{\mu\nu\rho}A_\mu \partial_\nu A_\rho
\label{indquad}
\end{equation}

\vskip .5cm
{\bf Exercise 5.1.1 :} Consider the three-photon leg diagram in Figure \ref{feynman} (b), and show that in the large mass limit ($m\gg p_1, p_2$) :
\begin{equation}
\Gamma_{\rm odd}^{\mu\nu\rho}(p_1,p_2,m)\sim -i\,\frac{1}{4\pi}\,
\frac{m}{|m|}\epsilon^{\mu\nu\rho} +O(\frac{p^2}{m^2})
\label{threept}
\end{equation}
Hence show that in the nonabelian theory a nonabelian Chern-Simons term is induced
at one-loop (note that the Chern-Simons coefficient is imaginary in Euclidean space):
\begin{equation}
S_{\rm eff}^{\rm CS}=-i\frac{N_f}{2}\,\frac{1}{4\pi}\, \frac{m}{|m|} \int d^3x
\epsilon^{\mu\nu\rho} tr \left ( A_\mu \partial_\nu A_\rho +{2\over 3} A_\mu
A_\nu A_\rho \right )
\label{nabind}
\end{equation}
\vskip 1cm

We now come to the physical interpretation of these results \cite{redlich}.
Consider the evaluation of the QED effective action (\ref{eff}) at zero fermion
mass. The computation of $S_{\rm eff}[A,m=0]$ requires regularization because of
ultraviolet ($p\to\infty$) divergences. This regularization may be achieved,
for example, by the standard Pauli-Villars method:
\begin{equation}
S_{\rm eff}^{\rm reg}[A,m=0]=S_{\rm eff}[A,m=0]-
\lim_{M\to\infty} S_{\rm eff}[A,M]
\label{pv}
\end{equation}
The Pauli-Villars technique respects gauge invariance. But the $M\to\infty$
limit of the second term in (\ref{pv}) produces an induced Chern-Simons term, because of
the perturbative large mass result (\ref{exp}). Therefore, in the process of
maintaining gauge invariance we have broken parity symmetry -- this is
initiated by the introduction of the Pauli-Villars mass term  $M\bar{\psi}\psi$
which breaks parity, and survives the $M\to\infty$ limit in the form of an
induced Chern-Simons term (\ref{indquad}). This is the ``parity anomaly'' of $2+1$ \diml
QED \cite{redlich}. It is strongly reminiscent of the well known axial anomaly in
$3+1$ \dim, where we can maintain gauge invariance only at the expense of the
discrete axial symmetry. There, Pauli-Villars regularization introduces a
fermion mass which violates the axial symmetry. Recall that there is no
analogous notion of chirality in $2+1$ \dim because of the different Dirac
gamma matrix algebra; in particular, there is no ``$\gamma^5$'' matrix that
anticommutes with all the gamma matrices $\gamma^\mu$. Nevertheless, there is a
parity anomaly that is similar in many respects to the $3+1$ \diml axial
anomaly.

In the nonabelian case, the induced Chern-Simons term (\ref{nabind}) violates
parity but restores invariance under large gauge transformations. It is known
from a nonperturbative spectral flow argument \cite{redlich} that $S_{\rm
eff}[A,m=0]$ for a single flavour of fermion ($N_f=1$) is not gauge invariant,
because the determinant (of which
$S_{\rm eff}$ is the logarithm) changes by a factor $(-1)^N$ under a large
gauge transformation with winding number $N$. Thus $S_{\rm eff}[A,m=0]$ is
shifted by $N\pi i$. But the induced Chern-Simons term (\ref{nabind}) also shifts by
$N\pi i$, when $N_f=1$, under a large gauge transformation with winding number
$N$. These two shifts cancel, and the regulated effective action (\ref{pv}) is
gauge invariant. This is reminiscent of Witten's ``SU(2) anomaly'' in $3+1$
\dim \cite{witten2}. This is a situation where the chiral fermion
determinant changes sign under a large gauge transformation with odd
winding number, so that the corresponding effective action is not invariant
under such a gauge transformation. As is well known, this anomaly is avoided in
theories having an {\it even} number $N_f$ of fermion flavours, because the
shift in the effective action is $N_f\,N\,\pi i$, which is always an integer
multiple of $2\pi i$ if
$N_f$ is even (here $N$ is the integer winding number of the large gauge
transformation). The same is true here for the parity anomaly in the nonabelian
$2+1$ \diml case; if $N_f$ is even then both $S_{\rm eff}[A,m=0]$ and the
induced Chern-Simons term separately shift by a multiple of $2\pi i$ under any large
gauge transformation.

These results are from one-loop calculations. Nevertheless, owing to the
topological origin of the Chern-Simons term, there is a strong expectation that the
induced Chern-Simons terms should receive no further corrections at higher loops. This
expectation is based on the observation that in a nonabelian theory the Chern-Simons
coefficient must take discrete quantized values in order to preserve large
gauge invariance. At one loop we have seen that the induced coefficient is
$\frac{N_f}{2}$, which is an integer for even numbers of fermion flavours, and
reflects the parity anomaly in theories with an odd number of fermion flavours.
At higher loops, if there were further corrections they would necessarily
destroy the quantized nature of the one-loop coefficient. This suggests that
there should be no further corrections at higher loops. This expectation has
strong circumstantial evidence from various higher order calculations. Indeed,
an explicit calculation \cite{kao} of the two-loop induced Chern-Simons coefficient for
fermions showed that the two-loop contribution vanishes, in both the abelian
and nonabelian theories. This is a highly nontrivial result, with the zero
result arising from cancellations between different diagrams. This led to a
recursive diagrammatic proof by Coleman and Hill \cite{coleman} that in
the abelian theory there are no contributions to the induced Chern-Simons term beyond
those coming from the one fermion loop self-energy diagram. This has come to be
known as the `Coleman-Hill theorem'. There is, however, some important
fine-print -- the Coleman-Hill proof only applies to abelian theories (and zero
temperature) because it relies on manifest Lorentz invariance and the absence
of massless particles.

\subsection{Induced Currents and Chern-Simons Terms}
\label{ic}

Another way to compute the induced Chern-Simons term in the fermionic effective action
(\ref{eff}) is to use Schwinger's proper time method to calculate the induced
current $<J^\mu >$, and deduce information about the effective action from the
relation
\begin{equation} <J^\mu >={\delta\over \delta A_\mu}S_{\rm eff}[A]
\label{indcurrent}
\end{equation}
Schwinger's famous `proper-time' computation \cite{schwinger} showed that the
$3+1$ \diml QED effective action can be computed exactly for the special case
of a background gauge field $A_\mu$ whose corresponding field strength
$F_{\mu\nu}$ is constant. The corresponding calculation in $2+1$ \dim
\cite{redlich} is actually slightly easier because there is only a single
Lorentz invariant combination of $F_{\mu\nu}$, namely $F_{\mu\nu}F^{\mu\nu}$.
(In $3+1$ \dim there is also $F_{\mu\nu}\tilde{F}^{\mu\nu}$.) Schwinger's
`proper-time' technique is also well-suited for computing the induced current
$<J^\mu>$ in the presence of a constant background field strength.

A constant field strength may be represented
by a gauge field {\it linear} in the space-time coordinates:
$A_\mu=\frac{1}{2}x^\nu F_{\nu\mu}$, with $F_{\mu\nu}$ being the constant field
strength. Since $A$ is linear in $x$, finding the spectrum of the Dirac
operator $\dslash +i\aslash$ reduces to finding the spectrum of a harmonic
oscillator. This spectrum is simple and discrete, thereby permitting an
explicit exact solution. This computation does, however, require the
introduction of a regulator mass $m$ for the fermions. The result for the
induced current is \cite{redlich}
\begin{equation}
<J^\mu>=\frac{1}{2}\frac{m}{|m|}\frac{1}{4\pi}
\epsilon^{\mu\nu\rho}F_{\nu\rho}
\label{indc}
\end{equation}
By Lorentz invariance, we conclude that this result should hold
for nonconstant background fields, at least to leading order in a derivative
expansion. This is the result for a single flavour of fermions. For $N_f$
flavours the result is simply multiplied by $N_f$. Integrating back to get the
effective action, we deduce that the effective action must have the form
\begin{equation} S_{\rm eff}[A]=S_{\rm eff}^{\rm
NA}[A]+\frac{N_f}{2}\frac{m}{|m|}\frac{1}{4\pi} S_{\rm CS}
\label{sind}
\end{equation} where $S_{\rm eff}^{\rm NA}[A]$ is parity even but nonanalytic in the
background field. This agrees with the perturbative calculation described in
the previous section. Furthermore, we can also do this same computation of the
induced current for special nonabelian
backgrounds with constant field strength (note that a constant nonabelian
field strength, $F_{\mu\nu}=\partial_\mu A_\nu-\partial_\mu A_\nu+[A_\mu,
A_\nu]$, can be obtained by taking commuting gauge fields that are linear in
the space-time coordinates, as in the abelian case, or by taking constant but
non-commuting gauge fields).
\vskip .5cm
{\bf Exercise 5.2.1 :} Illustrate the appearance of terms in the $2+1$ \diml
effective action that are parity preserving but nonanalytic in the background
field strength, by computing the effective energy of $2+1$ \diml fermions in a
constant background magnetic field $B$. Make things explicitly parity
preserving by computing $\frac{1}{2}(S_{\rm eff}[B,m]+S_{\rm eff}[B,-m])$.

\subsection{Induced Chern-Simons Terms Without Fermions}
\label{pr}

The issue of induced Chern-Simons terms becomes even more interesting when bare Chern-Simons
terms are present in the original Lagrangian. Then Chern-Simons terms may be radiatively
induced even in theories without fermions. In a classic calculation, Pisarski
and Rao \cite{rao} showed that a gauge theory of $2+1$ \diml $SU(N)$
Yang-Mills coupled to a Chern-Simons term has, at one-loop order, a finite additive
renormalization of the bare Chern-Simons coupling coefficient:
\begin{equation}
4\pi\kappa_{\rm ren} =4\pi\kappa_{\rm bare}+N
\label{prshift}
\end{equation}
where the $N$ corrsponds to the $N$ of the $SU(N)$ gauge group.
This radiative correction is consistent with the
discretization condition [recall (\ref{dirac})] that the Chern-Simons coefficient
$4\pi\kappa$ must be an integer for consistency with large gauge invariance at
the quantum level. As such, this integer-valued finite shift is a startling
result, since it arises from a one-loop perturbative computation, which {\it a
priori} we would not expect to `know' anything about the nonperturbative large
gauge transformations.

Here I briefly outline the computation of the renormalized Chern-Simons coefficient in
such a Chern-Simons-Yang-Mills (CSYM) theory \cite{rao}. The Euclidean space bare
Lagrangian is
\begin{equation} {\cal L}_{\rm CSYM}=-\frac{1}{2}\tr(F_{\mu\nu} F^{\mu\nu})-i m\;
\epsilon^{\mu\nu\rho}\tr(A_\mu\partial_\nu A_\rho+
\frac{2}{3}e A_\mu A_\nu A_\rho)
\label{csymlag}
\end{equation}
where $F_{\mu\nu}=\partial_\mu A_\nu-\partial_\mu A_\nu+e[A_\mu, A_\nu]$.
Note that the Chern-Simons coefficient is imaginary in Euclidean space. The discreteness
condition (\ref{dirac}) requires
\begin{equation} 4\pi\, \frac{m}{e^2}={\rm integer}
\label{disc}
\end{equation}
where $m$ is the mass of the gauge field. The bare gauge propagator (with
covariant gauge fixing) is
\begin{equation}
\Delta_{\mu\nu}^{\rm bare}(p)={1\over p^2+m^2}\left(\delta_{\mu\nu}-{p_\mu
p_\nu\over p^2}-m \epsilon_{\mu\nu\rho}{p^\rho\over p^2}\right)+\xi{p_\mu
p_\nu\over (p^2)^2}
\label{bareprop}
\end{equation}
The gauge self-energy $\Pi_{\mu\nu}$ comes from the relation
$\Delta^{-1}_{\mu\nu}=(\Delta_{\mu\nu}^{\rm bare})^{-1}+\Pi_{\mu\nu}$, and may
be decomposed as
\begin{equation}
\Pi_{\mu\nu}(p)=(\delta_{\mu\nu} p^2-p_\mu p_\nu)\Pi_{\rm even}(p^2)+m
\epsilon_{\mu\nu\rho}p^\rho \Pi_{\rm odd}(p^2)
\label{gse}
\end{equation}
Then the renormalized gauge propagator is defined as
\begin{equation}
\Delta_{\mu\nu}(p)={1\over Z(p^2)[p^2+m_{\rm ren}^2(p^2)] }
\left(\delta_{\mu\nu}-{p_\mu p_\nu\over p^2}-m_{\rm ren}(p^2)
\epsilon_{\mu\nu\rho}{p^\rho\over p^2}\right)+\xi{p_\mu p_\nu\over (p^2)^2}
\label{renprop}
\end{equation}
where $Z(p^2)$ is a wavefunction renormalization factor and the renormalized
mass is
\begin{equation} m_{\rm ren}(p^2)={Z_m(p^2)\over Z(p^2)} m
\label{renmass}
\end{equation}
with
\begin{equation} Z(p^2)=1+\Pi_{\rm even}(p^2), \qquad\qquad Z_m(p^2)=1+\Pi_{\rm odd}(p^2)
\label{zeds}
\end{equation}
The important divergences are in the infrared ($p^2\to 0$), and we define
the renormalized Chern-Simons mass to be
\begin{equation} m_{\rm ren}=m_{\rm ren}(0)={Z_m(0)\over Z(0)}m
\label{mren}
\end{equation}
There is also, of course, charge renormalization to be considered. The
renormalized charge is
\begin{equation} e^2_{\rm ren}={e^2\over Z(0) (\tilde{Z}(0))^2}
\label{eren}
\end{equation}
where $\tilde{Z}(p^2)$ comes from the renormalization of the ghost
propagator. In writing this expression for the renormalized charge we have used
the standard perturbative Ward-Takahashi identities for (infinitesimal) gauge
invariance (note, however, that the Chern-Simons term introduces new vertices; but this
is only a minor change). The important thing is that none of
the Ward-Takahashi identities places any constraint on $Z_m(0)$, which comes
from the odd part of the gauge self-energy at zero external momentum
(\ref{zeds}). The renormalization factors $Z(0)$ and $\tilde{Z}(0)$ are finite
in Landau gauge, and a straightforward (but messy) one-loop calculation
\cite{rao} leads to the results
\begin{equation} 
Z_m(0)=1+{7 \over 12\pi} N \frac{e^2}{m},\qquad\qquad \tilde{Z}(0)=1-
{1\over 6\pi}N\frac{e^2}{m}
\label{zedres}
\end{equation} 
Putting these together with the renormalized mass (\ref{mren}) and charge
(\ref{eren}) we find that, to one-loop order:
\begin{equation}
\left(\frac{m}{e^2}\right)_{\rm ren}
=\left(\frac{m}{e^2}\right)Z_m(0)(\tilde{Z}(0))^2
=\left(\frac{m}{e^2}\right)\left\{1+(\frac{7}{12\pi}-\frac{1}{3\pi})
N\frac{e^2}{m}\right\}
=\left(\frac{m}{e^2}\right)+\frac{N}{4\pi}
\end{equation}
But this is just the claimed result:
\begin{equation}
4\pi\kappa_{\rm ren}=4\pi\kappa_{\rm bare}+N
\label{prresult}
\end{equation}

It is widely believed that this is in fact an all-orders result, although no
rigorous proof has been given. This expectation is motivated by the
observation that if there {\it were} further contributions to $Z_m(0)$ and
$\tilde{Z}(0)$ at two loops, for example,
\begin{equation}
\left(\frac{m}{e^2}\right)_{\rm ren}=\left(\frac{m}{e^2}\right)+
\frac{N}{4\pi} +\alpha {N^2\over (m/e^2)}
\end{equation}
(where $\alpha$ is some numerical coefficient) then the renormalized
combination $4\pi(\frac{m}{e^2})_{\rm ren}$ could no longer be an integer.
Explicit two-loop calculations have shown that there is indeed no two-loop
contribution \cite{wchen}, and there has been much work done (too much to
review here) investigating this finite renormalization shift to all orders.
Nevertheless, from the point of view of perturbation theory, the result $4\pi
\kappa_{\rm ren} =4\pi\kappa_{\rm bare}+N$ seems almost {\it too} good. We will
acquire a deeper appreciation of the significance of this result when we
consider the computation of induced Chern-Simons terms using finite temperature
perturbation theory in Section \ref{ft}. I should also mention that there are
nontrivial subtleties concerning regularization schemes in renormalizing these
Chern-Simons theories \cite{asorey}, in part due to the presence of the antisymmetric
$\epsilon^{\mu\nu\rho}$ tensor which does not yield easily to dimensional
regularization. These issues are particularly acute in the renormalization of
{\it pure} Chern-Simons theories (no Yang-Mills term).

The story of induced Chern-Simons terms becomes even more interesting when we include
scalar (Higgs) fields and spontaneous symmetry breaking. In a theory with a
Higgs scalar coupled to a gauge field with a bare Chern-Simons term, there is a
radiatively induced Chern-Simons term at one loop. If this Higgs theory has a nonabelian
symmetry that is completely broken, say $SU(2)\to U(1)$, then the computation
of the zero momentum limit of the odd
part of the gauge self-energy suggests the shift
\begin{equation}
4\pi\kappa_{\rm ren}=4\pi \kappa_{\rm bare}+
f\left({m_{\rm Higgs}\over m_{\rm CS}}\right)
\label{higgsshift}
\end{equation}
where $f$ is some complicated (noninteger!) function of the dimensionless
ratio of the Higgs and Chern-Simons masses \cite{khleb}. So $4\pi\kappa_{\rm ren}$ is
not
integer valued. But this is not a problem here because there is no residual
nonabelian symmetry in the broken phase, since the $SU(2)$ symmetry has been
completely broken. However, consider instead a {\it partial} breaking of the
original nonabelian symmetry [say from $SU(3)$ to $SU(2)$] so that the broken
phase does have a residual nonabelian symmetry. Then, remarkably, we find
\cite{chen,khare2} that
the complicated function $f$ reduces to an integer: $4\pi \kappa_{\rm ren} =
4\pi \kappa_{\rm bare}+2$, (the $2$ corresponds to the residual $SU(2)$
symmetry in this case). This result indicates a surprising robustness at the
perturbative level of the nonperturbative discreteness condition on the Chern-Simons
coefficient, when there is a nonabelian symmetry present.

Actually, in the case of complete symmetry breaking, the shift
(\ref{higgsshift}) should really be interpreted as the appearance of ``would
be'' Chern-Simons terms in the effective action. For example, a term
$\epsilon^{\mu\nu\rho}\tr(D_\mu\phi F_{\nu\rho})$ in the effective action is
gauge invariant, and in the Higgs phase in which $\phi\to<\phi>$ at large
distances, this term looks exactly like a Chern-Simons term. This is because we
extracted the Chern-Simons coefficient in the large distance ($p^2\to 0$) limit where
$\phi$ could be replaced by its asymptotic expectation value $<\phi>$. This
observation has led to an interesting extension of the Coleman-Hill theorem to
include the case of spontaneous symmetry breaking \cite{khare3}. However, in
the partial symmetry breaking case no such terms can be written down with the
appropriate symmetry behaviour, so this effect does not apply in a phase with
residual nonabelian symmetry. Correspondingly, we find that the integer shift
property does hold in such a phase.

\subsection{A Finite Temperature Puzzle}
\label{ft}

In this section we turn to the question of induced Chern-Simons terms at nonzero
temperature. All the results mentioned above are for $T=0$. The case of $T>0$
turns out to be significant both for practical and fundamental reasons. In the
study of anyon superconductivity \cite{weiss} one of the key steps involves a
cancellation between the bare Chern-Simons term and an induced Chern-Simons term. While this
cancellation was demonstrated at $T=0$, it was soon realised that at $T>0$ this
same cancellation does not work because the finite $T$ induced Chern-Simons coefficient
is temperature dependent. The resolution of this puzzle is not immediately
obvious. This strange $T$ dependent Chern-Simons coefficient has also caused significant
confusion regarding the Chern-Simons discreteness condition: $4\pi\kappa={\rm integer}$.
It seems impossible for a temperature dependent Chern-Simons coefficient $\kappa(T)$ to
satisfy this consistency condition. However, recent work
\cite{dll,deser3,rossini,aitchison} has led to a new understanding and appreciation of this issue, with some important lessons about finite temperature perturbation theory in general.

We concentrate on the induced Chern-Simons terms arising from the fermion loop, as
discussed in Sections \ref{pertloop} and \ref{ic}, but now generalized to
nonzero temperature. Recall from (\ref{arcs}) and (\ref{exp}) that the induced
Chern-Simons coefficient is essentially determined by
\begin{eqnarray}
\kappa_{\rm ind}&=&\frac{N_f}{2}\Pi_{\rm odd}(p^2=0,m)\nonumber\\
&=&\frac{N_f}{2}\int{d^3k\over (2\pi)^3}{2m\over (k^2+m^2)^2}\nonumber\\
&=&\frac{N_f}{2}\frac{1}{4\pi}\frac{m}{|m|}
\label{kind}
\end{eqnarray}

If we simply generalize this one loop calculation to finite temperature (using
the imaginary time formalism) then we arrive at
\begin{equation}
\kappa_{\rm ind}^{(T)}=\frac{N_f}{2}\, T\, \sum_{n=-\infty}^\infty\; \int
{d^2k\over (2\pi)^2}{2m\over [((2n+1)\pi T)^2+\vec{k}^2+m^2]^2}
\label{ktind1}
\end{equation}
where we have used the fact that at finite temperature the `energy' $k_0$ takes
discrete values $(2n+1)\pi T$, for all integers $n\in {\bf Z}$. 

\vskip 1cm {\bf Exercise 5.4.1 :} Take the expression (\ref{ktind1}) and do the
$\vec{k}$ integrals and then the $k_0$ summation, to show that
\begin{eqnarray}
\kappa_{\rm ind}^{(T)}&=&\frac{N_f}{2}\frac{T}{4\pi} \sum_{n=-\infty}^\infty\;
{2m\over [((2n+1)\pi T)^2+m^2]}\nonumber\\
&=&\frac{N_f}{2}\frac{1}{4\pi}\, \tanh(\frac{\beta m}{2})\nonumber\\
&=& \frac{N_f}{2}\frac{1}{4\pi}\frac{m}{|m|}\, \tanh(\frac{\beta |m|}{2})
\label{ktind2}
\end{eqnarray}
where $\beta=\frac{1}{T}$.
\vskip 1cm

Thus, it looks as though the induced Chern-Simons coefficient is temperature dependent.
Note that the result (\ref{ktind2}) reduces correctly to the
zero $T$ result (\ref{kind}) because $\tanh(\frac{\beta |m|}{2})\to 1$ as $T\to
0$ ({\it i.e.}, as $\beta\to\infty$). Indeed, the $T>0$ result is just the
$T=0$ result multiplied by the smooth function $\tanh(\frac{\beta |m|}{2})$.
This result has been derived in many different ways \cite{babu}, in both
abelian and nonabelian theories, and in both the real time and imaginary time
formulations of finite temperature field theory. The essence of the calculation
is as summarized above.

On the face of it, a temperature dependent induced Chern-Simons term would seem
to violate large gauge invariance. However, the nonperturbative (spectral flow)
argument for the response of the fermion determinant to large gauge
transformations at zero $T$ \cite{redlich} is unchanged when generalized to
$T>0$. The same is true for the hamiltonian argument for the discreteness of
$4\pi\kappa$ in the canonical formalism. Thus the puzzle. Is large gauge
invariance really broken at finite $T$, or is there something wrong with the
application of finite $T$ perturbation theory? We answer these questions in the
next Sections. The essential new feature is that at finite temperature, other
parity violating terms (other than the Chern-Simons term) can and do appear in
the effective action; and if one takes into account all such terms to all orders
(in the field variable) correctly, the full effective action maintains gauge
invariance even though it contains a Chern-Simons term with a temperature
dependent coefficient. In fact, it is clear that if there are higher order terms
present (which are not individually gauge invariant), one cannot ignore them in
discussing the question of invariance of the effective action under a large
gauge transformation. Remarkably, this mechanism requires the existence of
nonextensive terms ({\it i.e.}, terms that are not simply space-time integrals
of a density) in the finite temperature effective action, although only
extensive terms survive in the zero temperature limit.

\subsection{Quantum Mechanical Finite Temperature Model}
\label{qm}

The key to understanding this finite temperature puzzle can be illustrated with
a simple exactly solvable $0+1$ \diml Chern-Simons theory \cite{dll}. This is a quantum
mechanical model, which at first sight might seem to be a drastic
over-simplification, but in fact it captures the essential points of the $2+1$
\diml computation. Moreover, since it is solvable we can test various
perturbative approaches precisely.

Consider a $0+1$ dimensional field theory with $N_f$ flavours of fermions
$\psi_j$, $j=1\dots N_f$, minimally coupled to a $U(1)$ gauge field $A$. It is
not possible to write a Maxwell-like kinetic term for the gauge field in $0+1$
dimensions, but we can write a Chern-Simons term - it is linear in $A$. [Recall
that it is possible to define a Chern-Simons term in any odd dimensional
space-time]. We formulate the theory in Euclidean space ({\it i.e.}, imaginary
time $\tau$, with $\tau\in [0,\beta]$) so that we can go smoothly between
nonzero and zero temperature using the imaginary time formalism. The Lagrangian
is
\begin{equation}
{\cal L}=\sum_{j=1}^{N_f}\psi^\dagger_j
\left(\partial_\tau-iA+m\right)\psi_j -i\kappa A
\label{qmlag}
\end{equation}
There are many similarities between this model and the $2+1$ dimensional model
of fermions coupled to a nonabelian Chern-Simons gauge field. First, this model
supports gauge transformations with nontrivial winding number. This may look
peculiar since it is an {\it abelian} theory, but under the $U(1)$ gauge
transformation $\psi\to e^{i \lambda}\psi$, $ A\to A+\partial_\tau
\lambda$, the Lagrange density changes by a total derivative and the action
changes by
\begin{equation}
\Delta S=-i\kappa\int_0^\beta d \tau\,\partial_\tau \lambda=-2\pi i\kappa N
\label{wind}
\end{equation}
where $N\equiv\frac{1}{2\pi}\int_0^\beta d\tau \partial_\tau \lambda$ is
the integer-valued winding number of the topologically nontrivial gauge
transformation.
\vskip 1 cm
{\bf Exercise 5.5.1 :} Show that, in the imaginary time formalism, such a
nontrivial gauge transformation is
$\lambda(\tau)=\frac{2N\pi}{\beta}(\tau-\frac{\beta}{2})$; while, in real time,
a nontrivial gauge transformation is $\lambda(t)=2N\,\arctan(t)$. In each case,
explain why the winding number $N$ must be an integer.
\vskip 1cm

{}From (\ref{wind}) we see that choosing $\kappa$ to be an integer, the
action changes by an integer multiple of $2\pi i$, so that the Euclidean 
quantum path integral $e^{-S}$ is invariant. This is the analogue of the
discreteness condition (\ref{dirac}) on the Chern-Simons coefficient in three
dimensional nonabelian Chern-Simons theories. (The extra $4\pi$ factor in the
$2+1$ \diml case is simply a solid angle normalization factor.)

Another important similarity of this quantum mechanical model to its three
\diml counterpart is its behaviour under discrete symmetries. Under naive
charge conjugation $C$ : $\psi\to\psi^\dagger$, $A\to -A$, both the fermion
mass term and the Chern-Simons term change sign. This mirrors the situation in
three \dim where the fermion mass term and the Chern-Simons term each change
sign under a discrete parity transformation. In that case, introducing an equal
number of fermions of opposite sign mass, the fermion mass term can be made
invariant under a generalized parity transformation. Similarly, with an equal
number of fermion fields of opposite sign mass, one can generalize charge
conjugation to make the mass term invariant in our $0+1$ dimensional model.

Induced Chern-Simons terms appear when we compute the fermion effective action
for this theory:
\begin{equation}
S[A]=\log\left[{\det\left(\partial_\tau-i A+m\right)\over
\det\left(\partial_\tau+m\right)}\right]^{N_f}
\label{effective}
\end{equation}
The eigenvalues of the operator $\partial_\tau-iA+m$ are fixed by imposing the
boundary condition that the fermion fields be antiperiodic on the imaginary
time interval, $\psi(0)=-\psi(\beta)$, as is standard at finite temperature.
Since the eigenfunctions are
\begin{equation}
\psi(\tau)=e^{(\Lambda-m)\tau+i\int^\tau A(\tau^\prime)d\tau^\prime}
\end{equation}
the antiperiodicity condition determines the eigenvalues to be
\begin{equation}
\Lambda_n=m-i{a\over \beta}+{(2n-1)\pi i\over \beta}, \qquad\qquad
n=-\infty,\dots,+\infty
\label{eigenvalues}
\end{equation}
where we have defined
\begin{equation}
a\equiv \int_0^\beta d\tau A(\tau)
\label{acs}
\end{equation}
which is just the $0+1$ \diml Chern-Simons term.

Given the eigenvalues (\ref{eigenvalues}), the determinants in
(\ref{effective}) are simply
\begin{eqnarray}
{\det\left(\partial_\tau-i A+m\right)\over
\det\left(\partial_\tau+m\right)} =\prod_{n=-\infty}^\infty\left[
{m-i{a\over \beta}+{(2n-1)\pi i\over \beta}\over m+
{(2n-1)\pi i\over \beta}}\right] = {\cosh\left(\frac{\beta m}{2}
-i \frac{a}{2}\right)\over \cosh\left(\frac{\beta m}{2}\right)}
\label{determinant}
\end{eqnarray}
where we have used the standard infinite product representation of the $\cosh$
function. Thus, the {\it exact} finite temperature effective action is
\begin{equation}
S[A]=N_f \log\left[\cos\left(\frac{a}{2}\right)-i
\tanh\left(\frac{\beta m}{2}\right) \sin\left(\frac{a}{2}\right)\right]
\label{answer}
\end{equation}
Several comments are in order. First, notice that the effective action
$S[A]$ is not an extensive quantity ({\it i.e.}, it is not an integral of a
density). Rather, it is a complicated function of the Chern-Simons action:
$a=\int d\tau \,A$. We will have more to say about this later. Second, in the
zero temperature limit, the effective action
reduces to
\begin{equation}
S[A]_{T=0}=-i\frac{N_f}{2}{m\over |m|} \int d\tau\, A(\tau)
\label{zero}
\end{equation}
which [compare with (\ref{nabind})] is an induced Chern-Simons term, with
coefficient $\pm \frac{N_f}{2}$. This mirrors precisely the zero $T$ result
(\ref{indquad}) for the induced Chern-Simons term in three dimensions [the factor of
$\frac{1}{4\pi}$ is irrelevant because with our $2+1$ \diml normalizations it
is $4\pi\kappa$ that should be an integer, while in the $0+1$ \diml model it is
$\kappa$ itself that should be an integer. This extra $4\pi$ is just a solid
angle factor].

At nonzero temperature the effective action is much more complicated. A formal
perturbative expansion of the exact result (\ref{answer}) in powers of the
gauge field yields
\begin{equation}
S[A]=-i\frac{N_f}{2}\left[\tanh\left(\frac{\beta m}{2}\right) a-
\frac{i}{4}{\rm sech}^2\left(\frac{\beta m}{2}\right) a^2
+\frac{1}{12} \tanh\left(\frac{\beta m}{2} \right) {\rm
sech}^2\left(\frac{\beta m}{2}\right) a^3
+\dots\right]
\label{expansion}
\end{equation}
The first term in this perturbative expansion
\begin{equation}
S^{(1)}[A]=-i\frac{N_f}{2}\tanh(\frac{\beta m}{2}) \int A
\label{first}
\end{equation}
is precisely the Chern-Simons action, but with a temperature dependent
coefficient. Moreover, this $T$ dependent coefficient is simply the zero $T$
coefficient from (\ref{zero}), multiplied by the smooth function
$\tanh(\frac{\beta |m|}{2})$. Once again, this mirrors exactly what we found in
the $2+1$ dimensional case in the previous Section -- see (\ref{kind}) and
(\ref{ktind2}).

If the computation stopped here, then we would arrive at the apparent
contradiction mentioned earlier -- namely, the ``renormalized'' Chern-Simons
coefficent
\begin{equation}
\kappa_{\rm ren}=\kappa_{\rm bare}-\frac{N_f}{2}\tanh(\frac{\beta m}{2})
\label{ren}
\end{equation}
would be temperature dependent, and so could not take discrete values. Thus, it
would seem that the effective action cannot be invariant under large gauge
transformations.

The flaw in this argument is clear. At nonzero temperature there are other
terms in the effective action, besides the Chern-Simons term, which cannot be
ignored; and these must all be taken into account when considering the question
of the large gauge invariance of the effective action. Indeed, it is easy to
check that the exact effective action (\ref{answer}) shifts by $(N_f\, N)\pi
i$, independent of the temperature, under a large gauge transformation, for
which $a\to a+2\pi N$. But if the perturbative expansion (\ref{expansion}) is
truncated to {\it any} order in perturbation theory, then the result cannot be
invariant under large gauge transformations: large gauge invariance is only
restored once we resum all orders. The important point is that the full finite
$T$ effective action transforms under a large gauge transformation {\it in
exactly the same way as} the zero $T$ effective action. When $N_f N$ is odd,
this is just the familiar global anomaly, which can be removed (for example) by
taking an even number of flavours, and is not directly related to the issue of
the temperature dependence of the Chern-Simons coefficient. The clearest way to
understand this global anomaly is through zeta function regularization of the
theory \cite{deser3}, as is illustrated in the following exercise.
\vskip 1cm
{\bf Exercise 5.5.2 :} Recall the zeta function regularization definition of the
fermion determinant, $\det({\cal O})=\exp(-\zeta^\prime(0))$, where the zeta
function $\zeta(s)$ for the operator ${\cal O}$ is
\begin{equation}
\zeta(s)=\sum_{\lambda}\, (\lambda)^{-s}
\label{zeta}
\end{equation}
where the sum is over the entire spectrum of ${\cal O}$. Using the eigenvalues in
(\ref{eigenvalues}), express this zeta function for the $0+1$ \diml Dirac
operator in terms of the Hurwitz zeta function $\zeta_H(s,v)\equiv
\sum_{n=0}^\infty (n+v)^{-s}$. Hence show that the zeta function regularized
effective action is
\begin{equation}
S_{\rm zeta}[A]=\pm i\frac{N_f}{2} a + N_f
\log\left[\cos\left(\frac{a}{2}\right)
-i \tanh\left(\frac{\beta m}{2}\right) \sin\left(\frac{a}{2}\right)\right]
\label{zz}
\end{equation}
[You will need the Hurwitz zeta function properties :
$\zeta_H(0,v)=\frac{1}{2}-v$, and
$\zeta^\prime_H(0,v)=\log\Gamma(v)-\frac{1}{2} \log(2\pi)$]. The sign ambiguity
in the first term corresponds to the ambiguity in defining $(\lambda)^{-s}$.
The effect of this additional term is that the zeta function regularized
effective action (\ref{zz}) changes by an integer multiple of $2\pi i$ under
the large gauge transformation $a\to a+2\pi N$, even when $N_f$ is odd. Show
that this is consistent with the fact that this large gauge transformation
simply permutes the eigenvalues in (\ref{eigenvalues}) and so should not affect
the determinant. (Note that this explanation of the global anomaly \cite{frish} is independent of the temperature, so it is somewhat beside the point
for the resolution of the problem of an apparently $T$ dependent Chern-Simons
coefficient.)
\vskip 1cm

To conclude this section, note that only the first term in the perturbative
expansion (\ref{expansion}) survives in the zero temperature limit. The higher
order terms all vanish because they have factors of ${\rm sech}^2(\frac{\beta
m}{2})$. This is significant because all these higher order terms are
nonextensive -- they are powers of the Chern-Simons action. We therefore do not
expect to see them at zero temperature. Indeed, the corresponding Feynman
diagrams vanish identically at zero temperature. This
is usually understood by noting that they {\it must} vanish because there is no
gauge invariant (even under infinitesimal gauge transformations) term involving
more than one factor of $A(\tau)$ that can be written down. This argument,
however, assumes that we only look for {\it extensive} terms; at nonzero
temperature, this assumption breaks down and correspondingly we shall see that
our notion of perturbation theory must be enlarged to incorporate nonextensive
contributions to the effective action. For example, let us consider an action
quadratic in the gauge fields which can have the general form
\begin{equation}
S^{(2)}[A] = \frac{1}{2}\int d\tau_{1}\,d\tau_{2} \,
A(\tau_{1})F(\tau_{1}-\tau_{2})A(\tau_{2})
\label{q}
\end{equation}
where, by symmetry, $F(\tau_{1}-\tau_{2}) = F(\tau_{2}-\tau_{1})$. Under an
infinitesimal gauge transformation, $A\to A+\partial_\tau \lambda$, this action
changes by: $\delta S^{(2)}[A] = -\int
d\tau_{1}\,d\tau_{2}\,\lambda(\tau_{1})\partial_{\tau_{1}}
F(\tau_{1}-\tau_{2})A(\tau_{2})$. Clearly, the action (\ref{q}) will be
invariant under an infinitesimal gauge transformation if $F=0$. This
corresponds to excluding such a quadratic term from the effective action. But
the action can also be invariant under infinitesimal gauge transformations if
$F=constant$, which would make the quadratic action (\ref{q}) nonextensive,
and in fact proportional to the square of the Chern-Simons action. The origin
of such nonextensive terms will be discussed in more detail in Section \ref{pt}
in the context of finite temperature perturbation theory.

\subsection{Exact Finite Temperature $2+1$ Effective Actions}
\label{eft}

Based on the results for the $0+1$ \diml model described in the previous
Section, it is possible to compute exactly the parity violating part of the
$2+1$ \diml QED effective action when the backgound gauge field
$A_\mu(\vec{x},\tau)$ takes the following special form:
\begin{equation}
A_0(\vec{x},\tau)=A_0, \qquad\qquad \vec{A}(\vec{x},\tau)=\vec{A}(\vec{x})
\label{ans}
\end{equation}
and the static background vector potential $\vec{A}(\vec{x})$ has quantized
flux:
\begin{equation}
\int d^2x \epsilon^{ij}\partial_i A_j=\int d^2x\, B = 2\pi {\cal N},
\qquad\qquad {\cal N}\in{\bf Z}
\label{fl}
\end{equation}

Under these circumstances, the three dimensional finite temperature effective
action breaks up into an infinite sum of two dimensional effective actions for
the two dimensional background $\vec{A}(\vec{x})$. To see this, choose
Euclidean gamma matrices in three dimensions to be: $\gamma^0=i\sigma^3$,
$\gamma^1=i\sigma^1$, $\gamma^2=i\sigma^2$. Then the Dirac operator appearing
in the three dimensional effective action is
\begin{equation}
-i(\dslash-i\aslash)+m=\left(\matrix{\partial_0-iA_0+m &D_-\cr D_+ &
-\partial_0+iA_0+m}\right)
\end{equation}
where $D_\pm=D_1\pm iD_2$ are independent of $\tau$ by virtue of the ansatz
(\ref{ans}). Recalling that at finite $T$ the operator $\partial_0$ has
eigenvalues $\frac{(2n+1)\pi i}{\beta}$, for $n\in {\bf Z}$, we see that the
problem is reduced to an infinite set of Euclidean two dimensional problems.

To proceed, consider the eigenfunctions $\left(\matrix{f\cr g}\right)$, and
eigenvalues $\mu$, of the massless two dimensional Dirac operator
\begin{equation}
\left(\matrix{0&D_-\cr D_+&0}\right) \left(\matrix{f\cr g}\right) =\mu
\left(\matrix{f\cr g}\right)
\label{2d}
\end{equation}
It is a straightforward (but messy) algebraic exercise to show that given such
an eigenfunction corresponding to a {\it nonzero} eigenvalue, $\mu\neq 0$, it
is possible to construct two independent eigenfunctions $\phi_\pm$ of the three
dimensional Dirac operator \cite{deser3} :
\begin{equation}
\left(\matrix{[m-iA_0+\frac{(2n+1)\pi i}{\beta}]& D_-\cr
D_+& [m+iA_0-\frac{(2n+1)\pi i}{\beta}]}\right)\phi_\pm=\lambda_\pm\,\phi_\pm
\label{3d}
\end{equation}
where
\begin{equation}
\lambda_\pm=m\pm i\sqrt{\mu^2+(A_0-\frac{(2n+1)\pi i}{\beta})^2}
\label{3deig}
\end{equation}
and $\phi_\pm=\left(\matrix{f\cr \alpha_\pm g}\right)$, with
\begin{equation}
\alpha_\pm=\frac{i}{\mu}(A_0-\frac{(2n+1)\pi i}{\beta})\pm
i\sqrt{1+\frac{1}{4\mu^2} (A_0-\frac{(2n+1)\pi i}{\beta})^2}
\end{equation}
So, for each nonzero eigenvalue $\mu$ of the two dimensional problem, there are
two eigenvalues $\lambda_\pm$ of the three \diml Dirac operator. But from the
form (\ref{3deig}) of these eigenvalues, we see that their contribution to the
three \diml determinant is {\it even} in the mass $m$; and therefore these
eigenvalues (coming from {\it nonzero} eigenvalues of the two-\diml problem) do
not contribute to the parity odd part of the three \diml effective action.

In fact, the only contribution to the parity odd part comes from the zero
eigenvalues of the two \diml problem. From the work of Landau \cite{landau}
(and Aharonov and Casher \cite{aharonov}) we know that there are ${\cal N}
=\frac{1}{2\pi}\int d^2x B$ of these zero eigenvalues. This `lowest Landau
level' can be defined by the condition $D_-g=0$, so that the eigenfunctions of
the three \diml Dirac operator are
\begin{equation}
\phi_0=\left(\matrix{0\cr g}\right), \qquad {\rm where} \quad D_-g=0
\label{z}
\end{equation}
Thus the relevant eigenvalues of the three \diml Dirac operator are
\begin{equation}
\lambda_0^{(n)}=m+iA_0-\frac{(2n+1)\pi i}{\beta}, \qquad\qquad n\in {\bf Z}
\label{zeig}
\end{equation}
each with degeneracy ${\cal N}$.

There is no paired eigenvalue, so to compute the parity odd part of the finite
temperature three \diml effective action we simply trace over these
eigenvalues, and multiply by ${\cal N}$. But this is {\it exactly} the same
problem that we solved in the last section [see (\ref{eigenvalues})], with
${\cal N}$ playing the role of $N_f$, the number of fermion flavours. Thus, we
see immediately that
\begin{eqnarray}
S_{\rm eff}^{\rm odd}[A]&=&\frac{{\cal N}}{2}
\left( \log\left[\cos(\frac{a}{2}) -i
\tanh(\frac{\beta m}{2}) \sin(\frac{a}{2})\right]-
\log\left[\cos(\frac{a}{2})+i \tanh(\frac{\beta m}{2})
\sin(\frac{a}{2})\right]\right)\nonumber\\
&=&-i{\cal N} \, \arctan\left[\tanh(\frac{\beta m}{2})\,
\tan(\frac{a}{2})\right]
\label{dd}
\end{eqnarray}
where $a\equiv \beta A_0=\int_0^\beta A_0$. This is simply the imaginary part
of the $0+1$ exact effective action (\ref{answer}).

A more rigorous zeta function analysis of this problem has been given in
\cite{deser3}, along the lines outlined in the Exercise from the last section.
But the key idea is the same -- when the three dimensional gauge background has
the restricted static form of (\ref{ans}), the problem reduces to a set of two
\diml problems; and moreover, only the zero modes of this two \diml system
contribute to the parity odd part of the three \diml effective action. This can
also be phrased in terms of chiral Jacobians of the two \diml system
\cite{rossini}.

The background in (\ref{ans}) supports large gauge transformations at
finite temperature as a consequence of the $S^1$ of the Euclidean time
direction. So, if $\lambda(\tau)=\frac{2N\pi}{\beta}(\tau-\frac{\beta}{2})$,
independent of $\vec{x}$, then the gauge transformation $A_\mu\to
A_\mu+\partial_\mu \lambda$, does not affect $\vec{A}$, but $A_0\to A_0+
\frac{2N\pi}{\beta}$. In the notation of (\ref{dd}) this means $a\to a+2N\pi$.
Thus our discussion of these large gauge transformations reduces exactly to the
discussion of the previous section for the $0+1$ \diml model.

While this is a nice result, it is still a bit unsatisfying because these are
not the nonabelian large gauge transformations in three \dim that we were
originally considering. In fact, if we adopt the static ansatz (\ref{ans}) then
the abelian Chern-Simons term reduces to
\begin{equation}
\int d^3x \epsilon^{\mu\nu\rho}A_\mu \partial_\nu A_\rho = 4\pi{\cal N}
\int_0^\beta A_0
\end{equation}
which is just the $0+1$ \diml Chern-Simons term. So transformations that stay within
this ansatz are simply the nontrivial winding number transformations of the
$0+1$ \diml model. We can make a similar static ansatz in the nonabelian case.
For static fields, the nonabelian Chern-Simons term simplifies to
\begin{equation}
\int d^3x\, \epsilon^{\mu\nu\rho}\tr\left(A_\mu\partial_\nu
A_\rho+\frac{2}{3}A_\mu A_\nu A_\rho\right) =  \int_0^\beta d\tau\;
\tr\left[A_0\,\left(\int d^2x\, \epsilon^{ij}F_{ij}\right)\right]
\end{equation}
where $\epsilon^{ij}F_{ij}$ is the (Lie algebra valued) nonabelian covariant
anomaly in two dimensions. It is possible to make gauge transformations that
shift this Chern-Simons action by a constant, and by choosing $\vec{A}$ appropriately
(for example, in terms of unitons) this constant shift can be made integer
mulitple of $2\pi i$. But this constant shift is not due to the winding number
term in the change (\ref{change}) of the nonabelian Chern-Simons Lagrangian under a
gauge transformation -- rather, it is due to the total derivtive term.
Therefore, the simple nonabelian generalization of (\ref{dd}), with a static
nonabelian ansatz, does not really answer the question of what happens to the
discreteness condition (\ref{dirac}) at finite temperature.

\subsection{Finite Temperature Perturbation Theory and Chern-Simons Terms}
\label{pt}

These results for the finite temperature effective action contain some
interesting lessons concerning finite temperature perturbation theory. The
exact results of the previous sections are clearly very special. For general
$2+1$ \diml backgrounds we cannot compute the effective action exactly. Nor can
we do so in truly nonabelian backgrounds that support large gauge
transformations with nonvanishing winding number. Furthermore, Chern-Simons terms may be
induced not only in fermionic systems, but also in Chern-Simons-Yang-Mills
\cite{rao} and in gauge-Higgs models with spontaneous symmetry breaking
\cite{khleb,chen}. In such models there are no known exact results, even at
zero temperature. At finite $T$, perturbation theory is one of the few tools we
have.

An important lesson we learn is that there is an inherent incompatibility
between large gauge invariance and finite temperature perturbation theory. We
are accustomed to perturbation theory being gauge invariant order-by-order in
the coupling $e$, but this is not true for large gauge invariance at finite
temperature. We see this explicitly in the perturbative expansion
(\ref{expansion}) [note that since we had absorbed $e$ into the gauge field
$A$, the order of perturbation is effectively counted by the number of $A$
factors]. If we truncate this expansion at {\it any} finite order, then the
result is invariant under small gauge transformations, but it transforms under
a large gauge transformation in a $T$ dependent manner. It is only when we
re-sum {\it all orders}, to obtain the exact effective action (\ref{answer}),
that the response of the effective action to a large gauge transformation
becomes $T$ independent, as it should be. There is actually a simple way to
understand this breakdown of large gauge invariance at any finite order of
perturbation theory \cite{deser3}. A gauge transformation (with factors of $e$
restored) is
\begin{equation}
A_\mu\to A_\mu+\frac{1}{e}\partial_\mu \Lambda
\label{lgt}
\end{equation}
For an infinitesimal gauge transformation, the $\frac{1}{e}$ factor can be
absorbed harmlessly into a redefinition of the gauge function $\Lambda$. But
such a rescaling does not remove the $e$ dependence for a large gauge
transformation, because such a gauge transformation must satisfy special
boundary conditions at $\tau=0$ and $\tau=\beta$ (in the imaginary time
formalism). A rescaling of $\Lambda$ simply moves $e$ into the boundary
conditions. The effect is that a large gauge transformation can mix all orders
in a perturbative expansion in powers of $e$, thus destroying the large gauge
invariance order-by-order.

Diagrammatically, the appearance of higher order terms, other than the Chern-Simons
term, in the perturbative expansion (\ref{expansion}) means that at finite
temperature the diagrams with many external `photon' legs contribute to the
parity odd part of the effective action. This is in contrast to the case at
$T=0$ where only a single graph contributes -- in $0+1$ \dim it is the one-leg
graph, and in $2+1$ \dim it is the two-leg self-energy graph. Actually, these
higher-leg graphs are perfectly compatible with infinitesimal gauge invariance,
but they violate the zero temperature requirement of only have extensive
quantities in the effective action. In the $0+1$ \diml model, the standard Ward
identities for infinitesimal gauge invariance [$p_\mu \Gamma^{\mu\nu\dots}=0$,
etc ...] simplify (because there is no contraction of indices) to imply that
the diagram is proportional to a product of delta functions in the external
energies. In position space this simply means that each term is proportional to
a nonextensive term like $(\int A)^n$. But at zero temperature such
nonextensive terms are excluded for $n>1$, and indeed one finds, reassuringly,
that the corresponding diagrams vanish identically. At finite temperature we
cannot exclude terms that are nonextensive in time, and so these terms can
appear; and correspondingly we discover that these diagrams are indeed
nonvanishing at $T>0$. Accepting the possibility of nonextensive terms, the
requirement that the fermion determinant change by at most a sign under a large
gauge transformation, $a\to a+2\pi N$, leads to the general form:
\begin{equation}
\exp\left[ -\Gamma(a)/N_f\right] =i\sum_{j=0}^\infty \left( d_j \,
\cos(\frac{(2j+1)a}{2}) + f_j \, \sin(\frac{(2j+1)a}{2})\right)
\end{equation}
The actual answer (\ref{answer}) gives as the only nonzero
coefficients:
$d_0=1$ and $f_0=i\, \tanh(\frac{\beta m}{2})$. This fact can only be deduced
by
computation, not solely from gauge invariance requirements.

These same comments apply to the $2+1$ \diml case when the background is
restricted by the static ansatz (\ref{ans}). The static nature of the
background once again makes a multi-leg diagram proportional to a product of
delta functions in the external energies. For the answer to be extensive in
space (but possibly nonextensive in time) we can only have one external spatial
index, say $i$, and then invariance under infinitesimal static gauge
transformations requires this diagram to be proportional to $\epsilon^{ij}p_j$.
Factoring this out, the remaining diagrams are just like the multi-leg diagrams
of the $0+1$ \diml model, and can be computed exactly. [There is a slight
infrared subtlety due to the difficulty in Fourier transforming a finite flux
static background, but this is easily handled.] So, not surprisingly, the
perturbative computation in the static anstaz reduces to that of the $0+1$
case, just as happens in the exact evaluation.

As soon as we attempt to go beyond the static ansatz, or consider induced Chern-Simons
terms in non-fermionic theories, we strike some critical problems. The most
significant is that the zero momentum limit (\ref{exp}), via which we
identified the induced Chern-Simons terms, is no longer well defined at finite
temperature. This is a physics problem, not just a mathematical complication.
At finite $T$, Lorentz invariance is broken by the thermal bath and so a
self-energy function $\Pi(p)=\Pi(p^0,\vec{p})$ is separately a function of
energy $p^0$ and momentum $\vec{p}$. Thus, as is well known even in scalar
field theories \cite{weldon}, the limits of $p^0\to 0$ and $\vec{p}\to 0$ do
not commute. The original computations of the finite temperature induced Chern-Simons
coefficient [see (\ref{ktind2})]  explicitly employed the ``static limit''
\begin{equation}
\lim_{|\vec{p}|\to 0}\, \Pi(p^0=0,\vec{p})
\label{s}
\end{equation}
It is easy to see that the `opposite' limit $\lim_{p^0\to 0}
\,\Pi(p^0,|\vec{p}|=0)$ gives a different answer at finite $T$ \cite{kaoyang}.
This ambiguity simply does not arise in the $0+1$ \diml model, and the exact
$2+1$ \diml results of the previous section avoided this ambiguity because the
static ansatz (\ref{ans}) corresponds explicitly to the static limit (\ref{s}).

Finally, another important issue that is not addressed by our $0+1$ \diml
model, or the corresponding static $2+1$ \diml results, is the Coleman-Hill
theorem \cite{coleman}, which essentially states that only one-loop graphs
contribute to the induced Chern-Simons term. This is an explicitly zero
temperature result, as the proof assumes manifest Lorentz covariance. But the
question of higher loops does not even come up in the $0+1$ \diml model, or the
static $2+1$ \diml backgrounds, because the `photon' does not propagate; thus,
there are no higher loop diagrams to consider.

It would be interesting to learn more about finite temperature effective
actions whose zero temperature forms have induced Chern-Simons terms. There is
undoubtedly more to discover.

\bigskip\bigskip

\noindent{\bf Acknowledgement:} I thank the Les Houches organizers, A. Comtet, Th. Jolicoeur, S. Ouvry and F. David, for the opportunity to participate in this Summer School. This work has been supported by the U.S.
Department of Energy grant DE-FG02-92ER40716.00, and by the University of
Connecticut Research Foundation.

\end{document}